\documentclass[a4paper,11pt]{article}
\pdfoutput=1 

\usepackage{jcappub} 

\usepackage[utf8]{inputenc}
\usepackage[T1]{fontenc} 
\usepackage{mathtools}
\usepackage[dvipsnames]{xcolor}

\title{\boldmath A self-consistent analytical model for both the photoionization rate and reionization history}

\author[a]{Christopher Cain,}
\author[a]{Kevin S. Croker,}
\author[b]{Anson D'Aloisio,}
\author[c,d]{Ivelin Georgiev,}
\author[e]{Hurum Maksora Tohfa,}
\author[f]{Yongda Zhu,}
\author[a]{and Rogier Windhorst}

\affiliation[a]{School of Earth and Space exploration, Arizona State University, Tempe, AZ 85281, USA}
\affiliation[b]{Department of Physics and Astronomy, University of California, Riverside, CA 92521, USA}
\affiliation[c]{Department of Astronomy and Oskar Klein Centre, AlbaNova, Stockholm University, SE-10691 Stockholm, Sweden}
\affiliation[d]{ARCO (Astrophysics Research Center), Department of Natural Sciences, The Open University of Israel, 1 University Road, PO Box 808, Ra’anana 4353701, Israel} 
\affiliation[e]{Department of Astronomy, University of Washington, Seattle, WA 98195, USA }
\affiliation[f]{Steward Observatory, University of Arizona, 933 North Cherry Avenue, Tucson, AZ 85721, USA}

\emailAdd{clcain3@asu.edu}

\abstract{

Recent developments at the intersection of cosmology and astrophysics have highlighted the need for improved analytical models of observables that probe the Epoch of Reionization.  
With few exceptions, fast analytical treatments of reionization suitable for use in Bayesian inference have been limited to modeling the reionization history, $x_i(z)$.
Such models cannot take full advantage of observables that constrain $x_i$ indirectly. One such observable is the photoionization rate of neutral hydrogen, $\Gamma_{\rm HI}(z)$, which can be inferred from the mean transmission of the Lyman-$\alpha$ forest of high-redshift quasars and galaxies. It has been shown by several prior works that the evolution of $\Gamma_{\rm HI}$ at $5 \lesssim z \lesssim 6$ is highly sensitive to the tail end of reionization, potentially providing a tight astrophysical constraint on the reionization timeline. We present a new analytical formalism, based on the cosmological radiative transfer equation, that self-consistently predicts $x_i$ and $\Gamma_{\rm HI}$.  We test our model against detailed radiative transfer simulations and find it to be percent-level accurate in $x_i$ and $20-30\%$ accurate in $\Gamma_{\rm HI}$ at $z \lesssim 6$ - better than or comparable to existing observational uncertainties. Finally, we demonstrate that modest shifts in the ionizing photon output of high-redshift galaxies and/or the endpoint of reionization lead to differences in $\Gamma_{\rm HI}$ much larger that the model's intrinsic uncertainty, highlighting its utility for interpreting existing data.  We explore the origin of modeling uncertainty in $\Gamma_{\rm HI}$ and comment on future pathways for improvement.

}

\begin{document}
\maketitle
\flushbottom

\section{Introduction}
\label{sec:intro}

The Epoch of Reionization (EoR), during which the first UV sources ionized intergalactic hydrogen, plays a crucial role in our understanding of high-redshift astrophysics and cosmology.  It has long been recognized as key window into the properties of the first galaxies, black holes, and stars - believed to be its main drivers~\citep{Furlanetto2006,Robertson2015,Finkelstein2019}.  It is also an important ingredient in precision cosmology, most notably because it sets Thomson scattering optical depth to the CMB~\citep[][]{Hu2003,Gnedin2004,Sailer2026}, the least well-constrained cosmological parameter.  
In the past decade, the quantity and quality of data probing reionization has grown rapidly.  
The spectra of high-redshift quasars up to $z \sim 6$, including the Ly$\alpha$ forest and Lyman Continuum transmission, have placed tight constraints on the tail end of reionization~\citep{Fan2006,McGreer2015,Becker2015,Becker2021,Bosman2021,Gaikwad2023,Zhu2024,Spina2024}.  
At higher redshifts, Ly$\alpha$ damping wing absorption in the spectra of quasars~\citep{Davies2018,Greig2024,Hennawi2025} and galaxies~\citep{Umeda2023} and the statistics of Ly$\alpha$ emitting galaxies~\citep{Mason2018,Nakane2023,Kageura2025} have probed the neutral fraction up to $z \sim 10$.  
JWST has constrained the galaxy population up to even higher redshifts~\citep{Adams2024,Donnan2024,Finkelstein2024,Atek2026}, and has begun to characterize the statistics of galaxy ionizing properties during the EoR~\citep{Simmonds2024}.  
Constraints on small-scale CMB anisotropies have started to constrain reionization's duration~\citep{Reichardt2020,Raghunathan2024,Beringue2025,Chaubal2026}.  
Upper limits on the 21 cm power spectrum from reionization are closing in on the expected signal~\citep{HERA2021a,HERA2021b,HERA2022,Trott2025,Sims2025,HERA2026}.  

The astrophysical and cosmological implications of these data sets have only begun to be explored, largely thanks to the challenges inherent in modeling reionization.  
The production and emission of ionizing photons into the intergalactic medium (IGM) is governed by the sub-parsec-scale physics of the interstellar and circumgalactic media of galaxies~\citep{Trebitsch2017,Rosdahl2022}.  
After escaping their host galaxies, these photons can be absorbed by Jeans-scale structures in the intergalactic medium as small as a kiloparsec~\citep{Gnedin2000,Shapiro2004,Iliev2005,Park2016,DAloisio2020}.  
Finally, ionized ``bubbles'' growing around clusters of galaxies can reach tens to hundreds of Mpc, requiring simulated volumes as large as a Gpc to fully capture~\citep[][]{Alvarez2012,Iliev2014,Kaur2020}.  
The parameter space of reionization is also large, thanks our limited knowledge of the physics of its sources~\citep{Qin2024} and the IGM~\citep{Choudhury2025}.  
As such, the community must rely on a diverse set of tools, including computationally intensive numerical simulations~\citep[e.g.][]{Gnedin2014,Ocvirk2016,Rosdahl2018,Kannan2022}, fast analytical models~\citep{Madau1999,Furlanetto2004,Bolton2007,Madau2017,McQuinn2018,Chen2020} and everything in between~\citep[e.g.][]{Mesinger2007,Choudhury2018,Trac2021,Munoz2023}, to model reionization and the observables that probe it.  

Despite reionization's complexity, the relationship between the reionization history and the globally-averaged ionizing properties of the galaxy population is surprisingly simple.  
A typical ionizing photon emitted from a source during reionization survives for a very short time (that is, $\ll$ a Hubble time) before being absorbed, since its mean free path (MFP) is limited by Lyman-limit systems~\citep[LLSs,][]{McQuinn2011,Alvarez2012}, Damped Lyman$\alpha$ absorbers (DLAs)~\citep[DLAs,][]{Rahmati2013,Crighton2015}, and fully neutral patches of the IGM\footnote{Lyman Limit Systems are defined to be gas structures (typically associated with dark matter halos) that are optically thick to ionizing photons at the Lyman edge, $912\text{\AA}$.  Damped Lyman-$\alpha$ absorbers are systems massive enough to self-shield and retain significant neutral gas, which imprints a damping wing feature in their Ly$\alpha$ absorption spectra.  }. 
It follows that photons are absorbed by the IGM at nearly the same rate at which they are emitted.  
For a given ionizing photon emissivity $\dot{n}_{\rm ion}(z)$, the global reionization history can be computed by balancing the number of H atoms ionized with the number of photons produced minus those absorbed by Lyman Limit Systems and DLAs.  
An approximation to this type of ``photon-counting'' argument was popularized by Ref.~\cite{Madau1999} (see also Ref.~\cite{Madau2017}), and has since become perhaps the most widely-used reionization model in the literature~\citep[e.g.][]{Robertson2013,Robertson2015,Finkelstein2019}.  
It has been used extensively to assess the consistency of galaxy models and observations with astrophysical and cosmological probes of reionization~\citep[e.g.][]{Munoz2024}, and is suitably fast for use in Bayesian analyses~\citep[e.g.][]{Robertson2015,Finkelstein2019}.  

The photon counting argument is limited to predicting the global reionization history.  
However, in the past decade, the spectra of high-redshift quasars up to $z \sim 6$ have placed tight indirect constraints on the tail end of reionization.   
The evolution of the mean transmission in the Ly$\alpha$ forest, and its large-scale spatial fluctuations, have provided strong evidence that reionization ended somewhere between $z = 6$ and $5.3$~\citep{Kulkarni2019,Keating2019,Keating2020,Nasir2020,Qin2021,Choudhury2021,Bosman2021,Gaikwad2023,Cain2024b}.  
Supporting evidence has emerged from measurements of the ionizing photon mean free path~\citep{Prochaska2009,Worseck2014,Becker2021,Cain2021,Garaldi2022,Lewis2022,Zhu2023,Gaikwad2023} and the thermal history of the IGM~\citep{Becker2011,Walther2019,Boera2019,Gaikwad2020}.  
The Ly$\alpha$ forest can place direct constraints on the neutral fraction, via dark pixels~\citep{McGreer2015,Jin2023,Davies2026}, dark gaps~\citep{Zhu2022}, and signatures of damping wing absorption~\citep{Zhu2024,Spina2024,Becker2024}.  
However, much of its constraining power arises from its sensitivity to the ionization state of ionized IGM gas, which is itself highly sensitive to the details of how and when reionization ends~\citep{Keating2020,Cain2023}.  
This sensitivity motivates the development of analytical models that can go beyond predicting the reionization history.  

One such observable is the average photoionization rate of HI in ionized IGM gas - denoted $\Gamma_{\rm HI}$.  
It quantifies the intensity of the ionizing radiation field within ionized regions, and is sensitive to both the ionizing output of the galaxy population and the MFP~\citep{Bolton2007,Becker2013}.  
It is typically measured by re-scaling the Ly$\alpha$ optical depths in cosmological simulations of the high-redshift IGM to recover the measured mean transmission of the Ly$\alpha$ forest~\citep{Bosman2018,Bosman2021}.  
The error bars on $\Gamma_{\rm HI}$ measurements are driven mainly by uncertainties in the IGM thermal history\footnote{Additional uncertainty in $\Gamma_{\rm HI}$ include continuum estimation for quasar spectra, statistical uncertainty in the forest transmission measurements, and additional modeling uncertainty associated with modeling the IGM at the end of reionization.   }~\citep{Nasir2016,Boera2019,Gaikwad2020}. 
These measurements provide important physical insight into the physics at play in reionization's tail end at $5 < z < 6$.  
As ionized bubbles overlap at the end of reionization, the IGM transitions from a state where $\Gamma_{\rm HI}$ is regulated by the neutral IGM to one in which it is regulated by over-dense absorbers~\citep{Alvarez2012}.  
The timing of this transition is expected to be accompanied by a sudden decrease in $\Gamma_{\rm HI}$ (with increasing redshift), and indeed this has been found to be true in simulations~\citep[e.g.][]{Kulkarni2019,Cain2023,Asthana2024} and analytical models~\citep{Munoz2016}.  
As such, measurements of $\Gamma_{\rm HI}$ offer a key independent constraint on the endpoint of reionization.  

\begin{figure}[h!]
    \centering
    \includegraphics[scale=0.255]{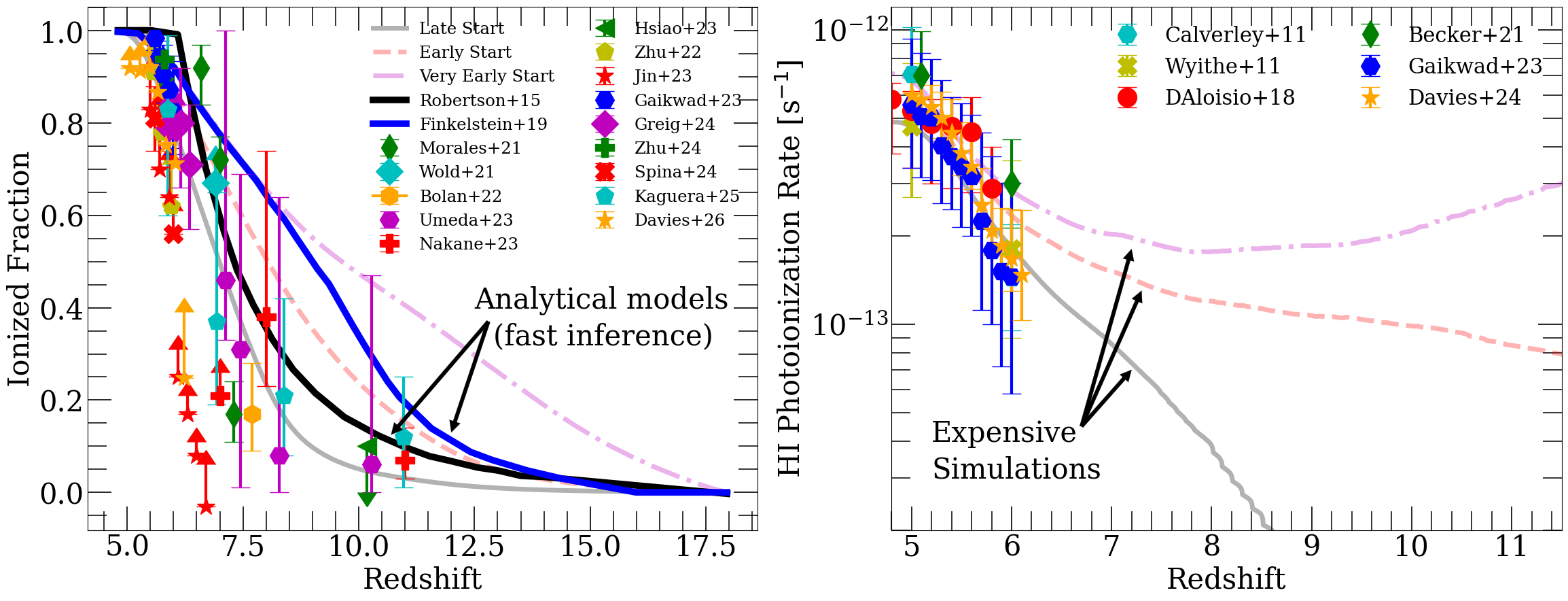}
    \caption{Summary of recent constraints on the reionization history and IGM photoionization rate from the Ly$\alpha$ forest.  The left panel shows the ionized fraction, and the right shows the IGM-averaged $\Gamma_{\rm HI}$.  Measurements of the ionized fraction at $z > 6$ are from Ly$\alpha$ emitters and Ly$\alpha$ damping wings in quasar and galaxy spectra.  The thin faded curves show three simulated reionization histories from Refs.~\cite{Cain2024b,Cohon2025} that are in agreement with forest data at $z \leq 6$.  The thick bold curves on the left show the best-fitting photon counting models from Refs.~\cite{Robertson2015,Finkelstein2019} constrained using galaxy and CMB data.  Note that no analytical models are shown on the right, because no analytical models exist that self-consistently predict the ionized fraction and $\Gamma_{\rm HI}$.  We show measurements of the ionized fraction from Refs.~\citep{Morales2021,Wold2022,Bolan2022,Zhu2022,Jin2023,Hsiao2023,Umeda2023,Nakane2023,Gaikwad2023,Greig2024,Zhu2024,Spina2024,Kageura2025,Davies2026}, and $\Gamma_{\rm HI}$ from Refs.~\citep{Calverley2011,Wyithe2011,DAloisio2018,Becker2021,Gaikwad2023,Davies2024}.  }
    \label{fig:sim_summary}
\end{figure}

In Figure~\ref{fig:sim_summary}, we summarize observational constraints on the ionized fraction and $\Gamma_{\rm HI}$ (left and right panels, respectively).   
The top left panel also includes the most recent constraints on the ionized fraction from Ly$\alpha$ emitters and quasar/galaxy damping wings at $z > 6$ (see caption for references).  
The faded curves show results from numerical simulations of three reionization histories that are compatible with the $z \lesssim 6$ quasar data (see \S\ref{subsec:methods} for details).  
On the left, we show the best-fit models from Refs.~\cite{Robertson2015,Finkelstein2019} (thick bold curves), which were inferred from galaxy and CMB data using analytical photon counting arguments to model the reionization history.  
However, as yet (to our knowledge) there are no analytical models capable of self-consistently modeling $\Gamma_{\rm HI}$ and the ionized fraction simultaneously - doing so presently requires expensive numerical simulations that are ill-suited for Bayesian analyses.   
{\bf As such, analyses aimed at linking galaxy properties and the reionization history using  existing photon counting arguments miss out on potential additional constraining power available from $\Gamma_{\rm HI}$ measurements.  }

The need for computationally efficient analytical reionization models has been accentuated by recent developments in observational cosmology.  
The latest results from the Dark Energy Spectroscopic Instrument (DESI) have displayed a preference for dynamical dark energy when combined with the latest CMB constraints from \textsl{Planck}~\citep{Planck2018,DESIDR22025}.  
These have also revealed at least one new tension in cosmology - a preference for formally negative neutrino masses when physically-motivated priors are omitted~\citep{Elbers2025a}.  
Among several proposed solutions to this problem~\citep{KumarAjith2025,Ahlen2025,ChenZaldarriaga2025}, Refs.~\cite{Sailer2026,Jhaveri2025} have argued that invoking a higher Thomson optical depth ($\tau$) to the CMB than inferred from \textsl{Planck} low-$\ell$ EE polarization~\citep{Tristram2024} can resolve the neutrino mass tension and restore $2\sigma$ consistency with $\Lambda$CDM.  
Several subsequent works~\citep{Cain2025c,Elbers2025,GarciaGallego2025,Kageura2026} have countered that constraints on the timing of reionization from complementary probes - including Ly$\alpha$ forest data - place tight limits on $\tau$ independent of \textsl{Planck}.  {\bf These findings motivate the development of fast analytical reionization models - suitable for use in high-dimensional cosmological parameter analyses - that can self-consistently leverage as much astrophysical EoR data as possible.  }

In this work, we develop a novel extension to the standard photon-counting formalism that can jointly predict both the reionization history and $\Gamma_{\rm HI}$.  
For the first time (to our knowledge), our formalism self-consistently models both quantities, requiring only one additional assumption about IGM conditions beyond that of standard photon counting arguments.  
We show that our model has the potential to accurately leverage additional constraining power from $\Gamma_{\rm HI}$ measurements in the context of astrophysical and cosmological analyses.  
The paper is organized as follows: \S\ref{sec:model} lays out our formalism, including a brief review of standard analytical treatments of the reionization history and $\Gamma_{\rm HI}$.    
In \S\ref{sec:testing}, we test our model against radiative transfer simulations.  
\S\ref{sec:usefulness} conducts a simple assessment of how useful our model is expected to be for leveraging $\Gamma_{\rm HI}$ measurements to constrain the reionization history and galaxy properties.  
In \S\ref{sec:inaccuracies}, we explore the accuracy of several assumptions underlying our formalism, and suggest pathways for forthcoming improvements.  
We conclude in \S\ref{sec:conc}. 
Throughout, we assume the following cosmological parameters: $\Omega_m = 0.305$, $\Omega_{\Lambda} = 1 - \Omega_m$, $\Omega_b = 0.048$, $h = 0.68$, $n_s = 0.9667$ and $\sigma_8 = 0.82$, consistent with Ref.~\cite{Planck2018} results. Distances are in co-moving units unless otherwise specified. 

\section{Formalism}
\label{sec:model}

In \S\ref{subsec:review}, we begin by reviewing previously-developed approaches to modeling the reionization history and $\Gamma_{\rm HI}$ that are based on the cosmological radiative transfer equation.  
We show why these approaches cannot model both quantities self-consistently during reionization, and why this substantially limits the constraining power of existing data.  
The remaining sections present our self-consistent formalism that jointly predicts the two quantities. 

\subsection{The cosmological RT equation: a brief review}
\label{subsec:review}

Our starting point is the isotropically and spatially-averaged cosmological radiative transfer equation, which reads~\cite{Peebles1993}, 
\begin{equation}
    \label{eq:cosmoRT_eq}
    \left(\frac{\partial}{\partial t} - \nu \frac{\dot{a}}{a}\frac{\partial}{\partial \nu} + 3 \frac{\dot{a}}{a}\right) J_{\nu} = - c \kappa(\nu) J_{\nu} + \frac{c}{4\pi}\epsilon_{\nu}
\end{equation}
where $J_{\nu}$ is the spatially-averaged radiation intensity per frequency $\nu$ per steradian, $\kappa(\nu)$ is the intergalactic absorption coefficient, $\epsilon_{\nu}$ is the emissivity of the ionizing source population, and the remaining parameters have their usual meanings.  Note that we have used the sub-script $\nu$ to indicate a spectrum per unit frequency.  
The solution is given by~\citep{Madau1999}, 
\begin{equation}
    \label{eq:Jnu}
    J_{\nu}(z) = \frac{c}{4 \pi} \int_{z}^{\infty} \frac{dz'}{(1+z')H(z')}\frac{(1+z)^3}{(1+z')^3} \epsilon_{\nu'}(z') e^{-\tau_{\rm eff}(\nu,\nu',z,z')},
\end{equation}
where $\nu' = \nu (1+z')/(1+z)$ is the frequency of radiation that redshifts to frequency $\nu$ between redshifts $z' > z$ and $z$, and the effective optical depth $\tau_{\rm eff}$ can be expressed as~\citep{McQuinn2016b}, 
\begin{equation}
    \label{eq:taueff_matt}
    \tau_{\rm eff}(\nu,\nu',z,z') = \int_{s(z')}^{s(z)}\frac{ds}{1+z(s)} \lambda(\nu'')^{-1} = \int_{t'}^{t} dt'' c \kappa(\nu''),
\end{equation}
where $\lambda(\nu) \equiv \kappa(\nu)^{-1}$, $s$ is co-moving distance, $t'$ is the cosmic time corresponding to $z'$ and $\nu'$, and double-primed variables are defined analogously to single-primed ones.   
It is noteworthy that Equation~\ref{eq:cosmoRT_eq} is agnostic to the physics that sets $\kappa(\nu)$ - that is, whether it arises due to recombinations in the diffuse IGM, absorptions by self-shielded mini-halos/Lyman limit systems\footnote{That is, systems that are optically thick to Lyman-Limit ($912\text{\AA}$, $13.6$eV) photons.  }~\citep{Shapiro2004,McQuinn2011}, and/or fully neutral patches of the IGM.  
As such, in principle it gives the complete solution for the average ionization state of the IGM, {\it including the contribution of regions that have yet to be re-ionized.  }

Equation~\ref{eq:cosmoRT_eq} is commonly applied in one of two ways.  The first is in the post-reionization limit, when the entire IGM is highly ionized, and neutral gas is found only in the ISM and CGM of galaxies (and perhaps in mini-halos~\citep[e.g.][]{Iliev2005}).  
In this limit ($z \lesssim 5$), the mean free path (MFP, $\lambda \equiv \kappa^{-1}$) has been measured to high precision~\citep{Prochaska2009,OMeara2013,Fumagalli2013,Worseck2014,Becker2021,Zhu2023,Gao2024}.  
As such, Equation~\ref{eq:Jnu} can be evaluated given an assumed $\epsilon_{\nu}$ and a physically-motivated assumption about the frequency dependence of $\kappa(\nu)$~\citep[e.g.][]{Haardt2012}.  
The latter can be estimated using the HI column density distribution, which has been measured at $z \lesssim 4$ from the Ly$\alpha$ forest~\citep{Rudie2013,Kim2021} and can be estimated by extrapolation or from simulations at higher redshifts~\cite{Rahmati2013}.
Equation~\ref{eq:Jnu} also be inverted to solve for $\epsilon_{\nu}$ given Ly$\alpha$ forest measurements of $\Gamma_{\rm HI}$ (which provides $J$).  
This approach has been used to measure the ionizing properties of the galaxy population in the post-reionization universe by several works~\citep[e.g.][]{Bolton2007,Becker2013,Bosman2024,Cain2025}.  

This approach has two important limitations when it is applied at redshifts when reionization is ongoing.  
First, measurements of the MFP from observations become increasingly uncertain at $z > 5$~\citep{Becker2021,Zhu2023,Gaikwad2023}, and disappear altogether at $z > 6$.  
As such, any application of this approach well into reionization must extrapolate the MFP to higher redshifts than it is measured, or make a prediction for the MFP based on some simulation or analytical model.  
A more serious problem, however, is that this approach does not self-consistently account for the dependence of $\Gamma_{\rm HI}$ on the reionization history itself.  
As mentioned earlier, the end of reionization has been found in simulations to be accompanied by a steep increase in $\Gamma_{\rm HI}$, which is driven by the overlap of ionized bubbles and the disappearance of the last fully neutral regions in the IGM.  
In other words, the $\kappa(\nu)$ that enters Equation~\ref{eq:cosmoRT_eq} depends on the reionization history, which is itself a function of $\epsilon_\nu$, and this dependence is not captured in existing formalism.  
Thus, {\bf to take full advantage of $\Gamma_{\rm HI}$ measurements at $z > 5$, modeling the coupling between $\Gamma_{\rm HI}$ and ionized fraction is crucial.}

The other case is the EoR limit, when we are most interested in solving for the reionization history.  
In this limit, we will invoke the assumption discussed previously - namely, that the MFP is very short and we can safely assume that every photon emitted is ``quickly'' absorbed.  
In this so-called ``local source approximation'', the LHS of Equation~\ref{eq:cosmoRT_eq} (which keeps track of how many photons are ``in transit'' between emission and absorption) can be set to $0$ - that is, we assert
\begin{equation}
    \label{eq:dJdt=0}
    \left(\frac{\partial}{\partial t} - \nu \frac{\dot{a}}{a}\frac{\partial}{\partial \nu} + 3 \frac{\dot{a}}{a}\right) J_{\nu} = \frac{dJ_{\nu}}{dt} =  0
\end{equation}
The rate at which photons are absorbed by the IGM per frequency is $\kappa(\nu) J_{\nu}$, and this can be divided into two contributions: absorptions inside the ionized regions (by Lyman Limit systems and DLAs), and absorptions by fully neutral gas at the edges of ionized regions (or, as they are commonly called, ``bubbles'').  
The latter make a net contribution to the reionization of the universe, because neutral atoms at the edges of bubbles are being (re-)ionized for the first time.  
Their co-moving ionization rate is given by 
\begin{equation}
    \label{eq:ndot_abs_neutral}
    \dot{n}_{\mathcal{A}}^{\rm neutral} = n_{\rm H}\frac{dx_i}{dt}
\end{equation}
where $n_{\rm H}$ is the co-moving hydrogen number density and $dx_i/dt$ is the time derivative of the mass-weighted ionized fraction $x_i$. 
Note that here and throughout the rest of the paper, the sub-script $\mathcal{A}$ denotes an absorption rate.  
By contrast, photons absorbed inside ionized bubbles do not make a net contribution to reionization, since they must re-ionize an H atom that occupies an already re-ionized region.  
It is commonly assumed that the rate of such ionizations is balanced exactly by hydrogen recombinations\footnote{More specifically, recombinations that do not produce another ionizing photon.  See Ref.~\cite{FaucherGiguere2009} for discussion of this nuance.  }.   
Under this approximation, the co-moving absorption rate within ionized regions can be expressed as
\begin{equation}
    \label{eq:ndot_abs_ionized}
    \dot{n}_{\mathcal{A}}^{\rm ionized} = \dot{n}_{\mathcal{R}}
\end{equation}
where $\dot{n}_{\mathcal{R}}$ is the co-moving (case B) recombination rate of ionized hydrogen.  
Note that this is not quite true, because ionized regions may contain self-shielded HI absorbers with neutral gas.  If these systems are being photo-evaporated or are accreting gas (which recombines at high densities), they will not be in photoionization equilibrium~\citep{Shapiro2004,Park2016,DAloisio2020,Chan2023} (we will return to this point later).  

The sum of these two absorption rates is equal to the total, which itself is just the integral over frequency of $\kappa(\nu) J_{\nu}$ - that is, 
\begin{equation}
    \label{eq:ndot_abs_total}
     \dot{n}_{\mathcal{A}}^{\rm total} = 4 \pi \int d\nu [\kappa(\nu) J_{\nu} /h_p \nu] = \dot{n}_{\mathcal{A}}^{\rm neutral} + \dot{n}_{\mathcal{A}}^{\rm ionized}
\end{equation}
where we have divided by $h_p\nu$ to count the number of photons being absorbed, and the factor of $4\pi$ comes from integrating $J$ over solid angle.  
Lastly, we re-cast the source term in terms of the ionizing photon emissivity $\dot{n}_{\rm ion}$, 
\begin{equation}
    \label{eq:ndot_ion}
    \int d\nu [\epsilon_\nu/h_p \nu] = \dot{n}_{\rm ion}
\end{equation}  
Finally, we divide both sides of Equation~\ref{eq:cosmoRT_eq} by $h_p\nu$, set the LHS to $0$ (under the local source approximation), integrate all remaining terms over frequency, and substitute Equations~\ref{eq:ndot_abs_neutral}-\ref{eq:ndot_ion}.  
It is a simple exercise to verify that this gives
\begin{equation}
    \label{eq:int_step}
    0 = -\left(n_{\rm H} \frac{dx_i}{dt} + \dot{n}_{\mathcal{R}}\right) + \dot{n}_{\rm ion}
\end{equation} 
which is simply a statement of photon conservation.  
We finally define the hydrogen recombination clumping factor\footnote{Note that $\dot{n}_{\mathcal{R}}$ is defined in Equation~\ref{eq:int_step} to be the recombination rate averaged over the whole IGM (including neutral regions where it is trivially $0$).  The factor of $x_i$ in the denominator of Equation~\ref{eq:CR} recovers the average recombination rate within ionized regions, which is how the clumping factor is typically defined.  }~\cite[following][]{Kaurov2015} to be
\begin{equation}
    \label{eq:CR}
    C_{\mathcal{R}} \equiv \dot{n}_{\mathcal{R}}/[\alpha_{\rm B}(T_{\rm ref}) (1+\chi) n_{\rm H}^2 x_i]
\end{equation}
where $\alpha_{\rm B}(T_{\rm ref})$ is the temperature-dependent case-B recombination coefficient evaluated at some reference temperature $T_{\rm ref}$, and the factor of $(1+\chi) = 1.082$ accounts for singly ionized helium.  
Putting Equation~\ref{eq:CR} into Equation~\ref{eq:int_step} and re-arranging terms yields a more familiar form: 
\begin{equation}
\label{eq:madau99}
    \frac{dx_i}{dt} = \frac{\dot{n}_{\rm ion}}{n_{\rm H}} - C_{\mathcal{R}} \alpha_{\rm B}(T) (1+\chi) n_{\rm H} x_i
\end{equation}
This is (essentially) the photon counting equation of Ref.~\cite{Madau1999}.  
The only difference is that we have used the mass-weighted ionized fraction $x_i$ rather\footnote{Note that Ref.~\cite{Chen2020} showed that $x_i$ should be used in place of $Q_{\rm HII}$ using a derivation based on the local ionization balance equation.  
We have confirmed their result here, this time starting directly from the cosmological RT equation.  } than the volume-filling factor of ionized hydrogen, $Q_{\rm HII}$.  
This demonstrates that the standard photon counting equation derives directly from the cosmological radiative transfer equation, assuming the local source approximation and photoionization equilibrium in highly ionized gas.  
One problem with this approach can be seen from it's first assumption (Equation~\ref{eq:dJdt=0}): setting $dJ/dt = 0$ explicitly discards information about $\Gamma_{\rm HI}$ from the outset.  
As such, one can only model $x_i$ with this formalism, and not $\Gamma_{\rm HI}$.  
Modeling both at once will require relaxing the $dJ/dt = 0$ assumption (which we do below).

We have reviewed existing treatments of $\Gamma_{\rm HI}$ and $x_i$, and noted why the assumptions the usual models for each ignore or preclude a self-consistent treatment of the other.  
In the rest of this section, we will provide the first analytical model (to the best of our knowledge) that unifies these two regimes.   

\subsection{A self-consistent model for $x_i$ and $\Gamma_{\rm HI}$}
\label{subsec:new_equations}

In this section, we will re-trace the arguments made in the previous section to derive Equation~\ref{eq:madau99}, {\it without assuming the local source approximation or photoionization equilibrium} (Equations~\ref{eq:dJdt=0} and~\ref{eq:ndot_abs_ionized}).  
To begin, we consider the properties of the radiation field and IGM ionization state at a single frequency.  
Dropping $\nu$-dependence from Equation~\ref{eq:cosmoRT_eq} and writing the operator on the LHS as a total time derivative gives
\begin{equation}
    \label{eq:cosmoRT_eq_nu0}
    \frac{dJ}{dt} = - c \kappa J + \frac{c}{4\pi}\epsilon
\end{equation}
If {\it all} photons have a single frequency $\nu_0$, we can see from Equation~\ref{eq:ndot_ion} that $\dot{n}_{\rm ion} = \epsilon /h_p \nu_0$.  
The photoionization rate is given by $\Gamma_{\rm HI} = 4\pi\sigma_{\rm HI}J/h_p\nu_0$, where $\sigma_{\rm HI}$ is the HI photoionization cross-section at frequency $\nu_0$.  
Then we have
\begin{equation}
    \label{eq:first_step}
    \frac{1}{c\sigma_{\rm HI}}\frac{d\Gamma_{\rm HI}}{dt} = -\frac{\kappa}{\sigma_{\rm HI}}\Gamma_{\rm HI} + \dot{n}_{\rm ion} 
\end{equation} 
This equation remains valid in the general multi-chromatic case, provided $\sigma_{\rm HI}$ and $\kappa$ are replaced by appropriately-defined weighted frequency averages of their frequency-dependent counterparts $\sigma_{\rm HI}(\nu)$ and $\kappa(\nu)$ (see Equation~\ref{eq:avg_sigma} of Appendix~\ref{app:multi_freq}).
\footnote{The technique of using a single effective frequency to approximate a spectrum is common in numerical radiative transfer simulations of hydrogen reionization~\cite[e.g.][]{Kulkarni2019,Cain2021}.  }

We next expand the absorption term on the RHS using Equations~\ref{eq:ndot_abs_neutral} and~\ref{eq:ndot_abs_total}, which yields
\begin{equation}
    \label{eq:second_step}
    \frac{\kappa}{\sigma_{\rm HI}}\Gamma_{\rm HI} = n_{\rm H} \frac{dx_i}{dt} + \dot{n}_{\mathcal{A}}^{\rm ionized}
\end{equation}
Note that we do not assume that $\dot{n}_{\mathcal{A}}^{\rm ionized} = \dot{n}_{\mathcal{R}}$, since this is not true if dense, self-shielded neutral gas is present within ionized regions.  
We instead define the ``absorption rate'' clumping factor (following Ref.~\cite{Cain2026}) to be
\begin{equation}
    \label{eq:C_A}
    C_{\mathcal{A}} \equiv \frac{\dot{n}_{\mathcal{A}}^{\rm ionized}}{\alpha_{\rm B}(T_{\rm ref}) (1+\chi) n_{\rm H}^2 x_i}
\end{equation}
Note that this is the same as Equation~\ref{eq:CR}, but with $\dot{n}_{\mathcal{R}}$ replaced by $\dot{n}_{\mathcal{A}}^{\rm ionized}$.
We note that this definition of the clumping factor is equivalent to the observationally measured one defined in Ref.~\cite{Davies2024c} in the limit that reionization is complete.  
Then we have (from Equations~\ref{eq:first_step}-\ref{eq:C_A}), 
\begin{equation}
    \label{eq:first_eq}
    \frac{1}{n_{\rm H} c \sigma_{\rm HI}}\frac{d\Gamma_{\rm HI}}{dt} + \frac{dx_i}{dt} = \frac{\dot{n}_{\rm ion}}{n_{\rm H}} - C_{\mathcal{A}} \alpha_{\rm B}(T_{\rm ref}) (1+\chi) n_{\rm H} x_i
\end{equation}
The most notable difference between Equations~\ref{eq:first_eq} and~\ref{eq:madau99} is the presence of the $d\Gamma_{\rm HI}/dt$ term on the LHS.  Because of this term, Equation~\ref{eq:first_eq} is a differential equation relating two unknowns - $\Gamma_{\rm HI}$ and $x_i$.  
We therefore require an additional constraint to solve for both of them.  
We can get this by noticing that the RHS of Equation~\ref{eq:second_step} consists of two contributions to $\kappa$ - one from absorptions by neutral gas (at bubble edges), and the other from Lyman Limit systems in ionized regions.  
This fact suggests the following definitions: 
\begin{equation}
    \label{eq:kappa_defs}
    \frac{\kappa_n}{\sigma_{\rm HI}}\Gamma_{\rm HI} \equiv n_{\rm H} \frac{dx_i}{dt} \hspace{1cm}
\frac{\kappa_i}{\sigma_{\rm HI}}\Gamma_{\rm HI} \equiv  \dot{n}_{\mathcal{A}}^{\rm ionized}
\end{equation}
where we have introduced the ``$i$'' and ``$n$'' subscripts to refer to absorptions by ionized gas within bubbles and neutral H atoms at the edges of bubbles, respectively\footnote{We use these subscripts instead of writing out ```ionized'' and ``neutral'' to make the notation more compact and readable.  }.  
We see that $\kappa_{n}$ and $\kappa_i$ are the effective absorption coefficients arising from these two types of absorptions.  
Lastly, we re-write the first of these equations in the suggestive form
\begin{equation}
    \label{eq:second_eq}
    \frac{dx_i}{dt} = \frac{\Gamma_{\rm HI}}{n_{\rm H}\sigma_{\rm HI}\lambda_n(x_i)}
\end{equation}
where we have defined $\lambda_n \equiv \kappa_n^{-1}$ and made the observation that one would, in general, expect this quantity to depend on $x_i$ (among other things).  
Equation~\ref{eq:second_eq} provides the additional independent constraint required to solve for $x_i$ and $\Gamma_{\rm HI}$.  
Of course, doing so requires that we know $\lambda_n(x_i)$.  We will describe a model for this quantity in the next section.  

\subsection{Modeling the opacity due to neutral regions}
\label{subsec:opacity_neutral}

Our goal in this section is to determine what $\lambda_n(x_i)$ is, and how we can model it.  
We will start making the crucial observation that, in the special case in which $\kappa_i \rightarrow 0$ (or, equivalently, $\lambda_i \rightarrow \infty$), the MFP to ionizing photons is equivalent to the average distance an emitted ionizing photon travels until it reaches neutral gas - that is, the nearest bubble edge along its trajectory.  
Thus, in this limit we may write
\begin{equation}
    \label{eq:lambda_n_lami=0}
    \lambda_t = \lambda_n = \int_{0}^{\infty} dR \left(R \frac{dP}{dR}\right) = \langle R \rangle
\end{equation}
where we have defined $\lambda_t$ to be the ``total'' MFP for the entire IGM, $R$ the nearest distance to the bubble edge for a randomly oriented sightline starting at a randomly selected location within a bubble, and $dP/dR$ is the distribution of $R$ for all such sightlines.  
We can then recognize that $dP/dR$ is the ``mean free path'' definition of the ionized bubble size distribution, introduced by Ref.~\cite{Mesinger2007}.   
This is a very convenient result because the ionized bubble size distribution has been extensively studied and  modeled over the last two decades~\citep[e.g.][]{Furlanetto2004,Furlanetto2005,Shin2008,Lin2016,Shimabukuro2020,Doussot2022,Wyatt2026}.  
It also agrees with prior work showing that $\Gamma_{\rm HI}$ should depend on the average size of ionized bubbles~\citep[e.g.][]{Zuo1992,Furlanetto2009}.  
Recently, Ref.~\cite{Umeda2023} reported the first observational constraints on $\langle R \rangle$ from galaxy damping wing observations.   
{\bf As such, our model not only extends the photon counting formalism to predict $\Gamma_{\rm HI}$ - it also does so by leveraging a quantity that has been extensively studied and characterized by the EoR community. }

We will now show that, under certain assumptions, the association of $\lambda_n$ with $\langle R \rangle$ holds in the more general case in which $\kappa_i > 0$.
First, define the ionizing transmission a distance $s$ along the $j$-th line of sight (LOS)
\begin{align}
T_j(s) &\equiv \exp\left[-\tau_j(s)\right] \\
\tau_j(s) &\equiv \int_0^s ds' \kappa_j(s')
\end{align}
where $\tau_j$ is the optical depth along the $j$-th LOS.  
Eventually this LOS will intercept a bubble edge at a distance $R$ from its starting point.
Transmission abruptly drops to $0$ here because the MFP through neutral gas for ionizing photons near the Lyman edge is negligibly short ($10$s of kpc).  
We may thus write,
\begin{align}
\label{eq:fofx}
T_j(s) =
\begin{cases}
      \exp\left[-\displaystyle\int_{0}^{s} \frac{ds}{\lambda_i(j, s')}\right] & \hspace{0.5cm} s\leq R_j \\
      0 & \hspace{0.5cm} s > R_j
\end{cases} 
\end{align}
where $R_j$ is the distance along the $j$-th LOS to the nearest bubble edge and $\lambda_i(j,s)$ is the MFP in the ionized gas at a distance $s < R_j$.
The sightline dependence in $\lambda_i(j,s)$ arises from local differences in the residual HI density (which sets $\lambda_{\rm i}$) along different lines of sight.  

By analogy to Equation~\ref{eq:lambda_n_lami=0}, we may write the average over some large number of lines of sight $N_{\rm sl}$ as,
\begin{equation}
    \label{eq:lambda_def}
    \lambda_t = -\frac{1}{N_{\rm sl}}\sum_{j=1}^{N_{\rm sl}} \int_0^\infty ds \left(s \frac{dT_j}{ds}\right) = \left<s\right>
\end{equation}
where we note that $-dT_j/ds$ gives the (normed) distribution of distances that photons travel along the $j$-th LOS prior to being absorbed (the minus sign accounts for the fact that $dT_j/ds < 0$).    
Note that this definition is equivalent to the one provided in Appendix C of Ref.~\cite{Chardin2015} (their Eq. C2).  
Integrate by parts and use that $sT_j(s) = 0$ at $s = 0$ and $\infty$, 
\begin{equation}
    \label{eq:lambda_t_2}
     \lambda_t = \frac{1}{N_{\rm sl}}\sum_{j=1}^{N_{\rm sl}} \int_0^\infty ds\,T_j(s)
\end{equation}
Using Equation~\ref{eq:fofx} gives
\begin{equation}
    \label{eq:lambda_def_int}
    \lambda_t = \frac{1}{N_{\rm sl}}\sum_{j = 1}^{N_{\rm sl}} \int_{0}^{R_j} ds \exp\left[-\int_{0}^{s} \frac{ds'}{\lambda_{\rm i}(j,s')}\right].
\end{equation}
We can now express this in the form of an expectation value over the distribution of bubble sizes,
\begin{equation}
    \label{eq:LambdadPdR}
    \lambda_t = \int_0^\infty dR\frac{dP}{dR} \int_{0}^{R} ds \left\{ \frac{1}{N_{\rm sl}(R)}\sum_{j = 1}^{N_{\rm sl}(R)} \exp\left[-\int_{0}^{s} \frac{ds'}{\lambda_{\rm i}(j,s')}\right]\right\}. 
\end{equation}
Here, $1 \ll N_{\rm sl}(R) \ll N_{\rm sl}$ is the number of lines of sight that travel a distance $R$ before reaching a bubble edge, and we have commuted the integral along the LOS with the sum over sightlines of fixed $R$. 
 
When measuring the MFP from stacks of observed quasar spectra~\citep{Prochaska2009,Worseck2014,Becker2021} or forward-modeling the MFP in numerical simulations~\citep{Cain2021,Lewis2022,Roth2023,Satyavolu2023,Chen2025}, it is common to assume that the stack of many transmission spectra can be well-approximated by an exponential function with a single attenuation length.  
Here, we will assume that this is true in the ionized gas at distances $s < R$.  
In other words, we will assume that we can write the stacked transmission profile over all sightlines with $R_j = R$ as 
\begin{align}
\label{eq:first_assumption}
\frac{1}{N_{\rm sl}(R)}\sum_{j = 1}^{N_{\rm sl}(R)} \exp\left[-\int_{0}^{s} \frac{ds'}{\lambda_{\rm i}(R)(j,s')}\right] \equiv
\begin{cases}
\exp(-s/\overline{\lambda}_i(R)) & s \leqslant R \\
0 & s > R
\end{cases}
\end{align}
where $\overline{\lambda}_i(R)$ is an effective attenuation length. 
This does not require $\lambda_{\rm i}(j,s)$ for individual sightlines to be locally independent of $s$ (which it certainly should not be), but it does assume that spatial fluctuations in $\lambda_i$ should be uncorrelated between different sightlines.   
We explore the validity of this assumption more carefully in \S\ref{sec:inaccuracies}. 
With this definition, Equation~\ref{eq:LambdadPdR} becomes
\begin{align}
    \label{eq:lam_t_2}
    \lambda_t = \int_0^\infty dR~\frac{dP}{dR} \int_{0}^{R} ds \exp\left(-s/\overline{\lambda}_i(R)\right). 
\end{align}
The LOS integration can be performed immediately,
\begin{align}
\lambda_t = \int_0^\infty dR~\frac{dP}{dR}~\overline{\lambda}_i(R) \left[1 - \exp\left(-R/\overline{\lambda}_i(R)\right)\right].
\end{align} 
Finally, we assume that $\overline{\lambda}_i(R)$ is independent of $R$ and equal to $\kappa_i^{-1}$ as defined in the previous section. 
It follows that
\begin{equation}
    \label{eq:first_simplification}
    \lambda_t =  \lambda_i \left[1 - \int_{0}^{\infty} dR\frac{dP}{dR} \exp(-R/\lambda_i)\right],
\end{equation}
where we have used the fact that $dP/dR$ is a normalized probability density.  

Because ionized bubbles have ``percolated'' late in reionization, and the distribution of neutral gas can be well-described by a collection of isolated ``islands''~\citep{Furlanetto2016,Lin2016,Xu2017}, we might expect
\begin{equation}
    \label{eq:dPdR_exp}
    \frac{dP}{dR} = \frac{1}{\langle R \rangle} \exp\left[-\frac{R}{\langle R \rangle}\right]
\end{equation}
to reasonably approximate bubble sizes, where $\langle R \rangle$ is (again) the average ionized bubble size. 
This turns out (\S\ref{sec:inaccuracies})
to be a reasonably good approximation during most of reionization in our simulations.  
Putting Equation~\ref{eq:dPdR_exp} into Equation~\ref{eq:first_simplification} and integrating/simplifying gives the desired result, 
\begin{equation}
    \label{eq:lambdat_model}
    \lambda_t^{-1} = {\lambda_i}^{-1} + \langle R \rangle^{-1}
\end{equation}
If we identify identify $\kappa = \lambda_t^{-1}$ and $\kappa_i = {\lambda}_{i}^{-1}$, we can see that $\kappa_n = \lambda_n^{-1} = \langle R \rangle^{-1}$ given the assumptions above, even in the presence of optically thick gas inside ionized bubbles.  
It follows that {\bf if one can write down a functional form for $\langle R \rangle(x_i)$ - by parameterizing it, predicting it from a model/simulation, or measuring it - Equations~\ref{eq:first_eq} and~\ref{eq:second_eq} can be solved.  }

\section{Testing the model against radiative transfer simulations}
\label{sec:testing}

\subsection{Simulations \& Test Methodology}
\label{subsec:methods}

In this section, we test our formalism against radiative transfer (RT) simulations of reionization run with the \textsc{FlexRT} code of Ref.~\cite{Cain2024c}.  
Details about the numerical setup and calibration procedure for these simulations is described in detail in Ref.~\cite{Cain2024b}, so we briefly summarize here.  
\textsc{FlexRT} is an adaptive ray-tracing RT code optimized for fast, self-consistent reionization simulations.  
It is equipped with a sub-grid model for the ionizing photon opacity in ionized gas~\citep{Cain2021,Cain2022b}, allowing it to run in post-processing on a coarse-grained grid of IGM densities.  
Our simulations are run in a $200$ $h^{-1}$Mpc box with $\Delta s_{\rm cell} = 1$ $h^{-1}$Mpc RT cells, and ionizing sources are identified with dark matter halos from an N-body simulation with matching large-scale structure. 
We assign UV luminosities to our halos by abundance-matching against the UV luminosity function of Ref.~\cite{Adams2024}.  
Our runs are calibrated to match the measured Ly$\alpha$ forest mean transmission at $5 < z < 6$ from Ref.~\cite{Bosman2021}.  
We use the \textsl{Late Start/Late End} and \textsl{Early Start/Late End} models from Ref.~\cite{Cain2024b}, which for brevity we refer to as \textsl{Late Start} and \textsl{Early Start}, respectively.  
We also use the \textsl{Very Early Start} model introduced in Ref.~\cite{Gangolli2024} (see also Ref.~\cite{Cohon2025}).  
The predictions from these models are shown by the faded curves in Figure~\ref{fig:sim_summary}.  

To test our model, we take the following approach.  
We first take the inputs required for the model - namely, $\dot{n}_{\rm ion}(z)$, $C_{\mathcal{A}}(z)$, and $\langle R \rangle(x_i)$ from each \textsc{FlexRT} simulation, then evaluate Equations~\ref{eq:first_eq}-\ref{eq:second_eq} and compare the predicted $x_i$ and $\Gamma_{\rm HI}$ to the \textsc{FlexRT} result.    
Since \textsc{FlexRT} also takes $\dot{n}_{\rm ion}(z)$ as a (calibrated) input, we simply assume the same $\dot{n}_{\rm ion}(z)$ input to \textsc{FlexRT}.  
The other two quantities are predicted by \textsc{FlexRT}, so we must extract them from the simulations.  
We compute $C_{\mathcal{A}}$ by directly counting the number of photons absorbed by ionized cells in the simulation\footnote{In practice, we do this by subtracting from $\dot{n}_{\rm ion}$ the net ionization rate ($n_{\rm H} dx_i/dt$) and the time derivative of the number density of ionizing photons in the ionized gas.  What remains must be the absorption rate in the ionized gas, since photons are conserved.  Note that this assumes photon conservation, which holds exactly for ray-tracing radiative transfer, the underlying algorithm of \textsc{FlexRT}.  }, $\dot{n}_{\mathcal{A}}^{\rm ionized}$, and then use this to evaluate Equation~\ref{eq:C_A}.  To estimate $\langle R \rangle$, we throw down $10^5$ randomly positioned and oriented sightlines (starting in ionized cells), compute the distance along them to the nearest neutral cell, and average the results.  
We use the same set of sightlines to evaluate Equations~\ref{eq:fofx}-\ref{eq:lambda_def} for $\lambda_t$, using the value of $\lambda_i$ in each RT cell predicted by \textsc{FlexRT}'s simulation-calibrate sub-grid opacity algorithm.  

When $\langle R \rangle$ approaches $\Delta s_{\rm cell}$, our MFP estimate will become sensitive to how partially ionized cells affect the value of $R$ we compute for each sightline.  
We handle this in the following way.  
First, we assume that photons emitted in partially ionized cells ``escape'' without encountering neutral gas with probability equal to the cell's ionized fraction, $x_i^{\rm cell}$.  
When selecting starting points for each sightline, we draw a random number $x$ between $0$ and $1$, and use that sightline only if the starting cell has $x > x_i^{\rm cell}$.  
When a sightlines terminate in a partially neutral cell (which they almost always do), we assume that it travels a distance $x_i^{\rm cell} \Delta s_{\rm cell}^{\rm sightline}$ through that cell before encountering neutral gas, where $\Delta s_{\rm cell}^{\rm sightline}$ is the total path length of the sightline through the terminating cell\footnote{We note that this procedure matches the treatment assumed in \textsc{FlexRT}'s two-zone ionization model when rays encounter partially ionized cells.  }.  
We note that regardless of how carefully we treat partially ionized cells, some inaccuracy (in the model and the simulations themselves) is inevitable as $\lambda_t \rightarrow \Delta s_{\rm cell}$.  
Fortunately, as we will see, our main conclusions are not sensitive to this choice.  
Furthermore, our model is useful mainly at the tail end of reionization, where $\Gamma_{\rm HI}$ has been measured, and in this regime it is always true that $\langle R \rangle \gg \Delta s_{\rm cell}$.  

\subsection{Accuracy of $x_i$ and $\Gamma_{\rm HI}$ predictions}
\label{subsec:accuracy}

\begin{figure}
    \centering
    \includegraphics[scale=0.19]{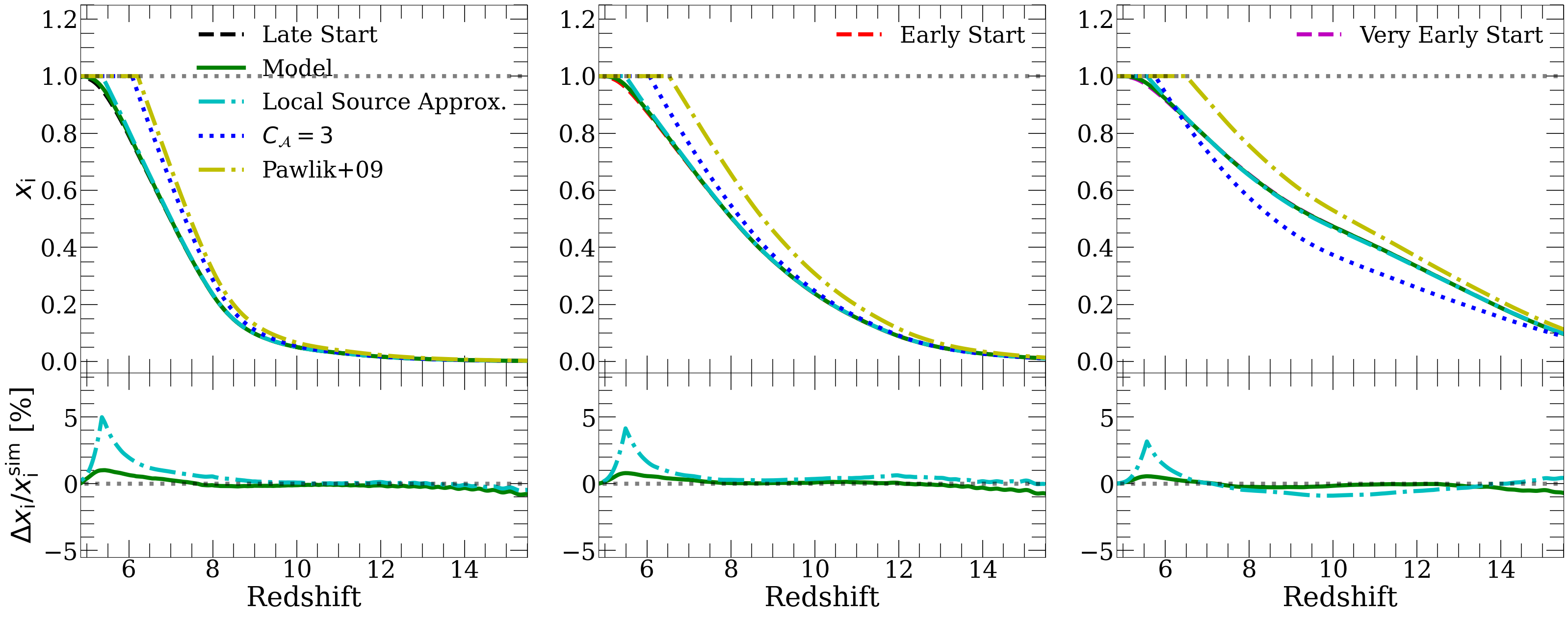}
    \caption{Mass-averaged ionized fraction predicted by our model (green solid) compared to simulation results (dashed lines).  From left to right, the columns show results for the \textsl{Late Start}, \textsl{Early Start}, and \textsl{Very Early Start} models, respectively.  The bottom row shows percentage differences between model and simulation.  The cyan dot-dashed curve shows a case where we assume the local source approximation (that is, set $d\Gamma_{\rm HI}/dt = 0$ in Equation~\ref{eq:first_eq}).  The dotted blue curve also assumes a constant $C_{\mathcal{A}} = 3$, and the yellow dot-dashed curve uses the Ref.~\cite{Pawlik2009} clumping factor.  The local source approximation causes up to a $\approx \%5$ error in $x_i$ because it neglects light travel time - correcting for this recovers $1\%$ or better accuracy.   }
    \label{fig:xi_sim_model_comparisons}
\end{figure}

In Figure~\ref{fig:xi_sim_model_comparisons}, we show our model prediction for $x_i$ in each simulation (different columns).  Each panel shows the simulated reionization history (dashed curve), and the model prediction (green solid), the model prediction assuming the local source approximation (that is, $d\Gamma_{\rm HI}/dt = 0$) in Equation~\ref{eq:first_eq} (dot-dashed cyan).  
We also show two scenarios where we replace $C_{\mathcal{A}}$ from the simulation with prescriptions from the literature.  
The blue dotted and yellow dot-dashed curves assume $C_{\mathcal{A}} = 3$ and the redshift-dependent clumping factor from Ref.~\cite{Pawlik2009}, respectively\footnote{We have checked functional form for the clumping factor from Ref.~\cite{Shull2011} gives similar results to the Ref.~\cite{Pawlik2009} one.  }.  
Ratios between the model and simulation are shown in the bottom row for the green and cyan curves only.  

We see that our model matches the simulated ionization history within $1\%$ or better at all redshifts, in all three scenarios.  
Ignoring the $d\Gamma_{\rm HI}/dt$ term, as is done to derive Equation~\ref{eq:madau99}, over-estimates $x_i$ by as much as $5\%$ at the tail end of reionization.  
The inaccuracy arises from neglecting the finite travel time of photons between emission and absorption.  
The deviation is largest at the end of reionization, when ionized bubbles are overlapping and $\Gamma_{\rm HI}$ is increasing rapidly.  
Ref.~\cite{Chen2020} saw a similar difference in accuracy when evaluating their analytical reionization model using the ionization clumping factor, $C_{\rm I}$ (their Equations 6 and 8), and $\dot{n}_{\rm ion}$ (their Equation 11).  
The reason for their finding is the same as ours.  
Their Equation 8 uses the absorption rate, there expressed as $\langle n_{\rm HI} \Gamma_{\rm HI}\rangle$, in the source term rather than $\dot{n}_{\rm ion}$, which has the net effect of accounting for the $d\Gamma_{\rm HI}/dt$ term.  

Replacing our simulation-calibrated $C_{\mathcal{A}}$ with commonly-used clumping factors leads to much larger deviations.  
Setting $C_{\mathcal{A}} = 3$ ends reionization too early by $\Delta z \approx 1$ in the \textsl{Late Start} case, as does the Ref.~\cite{Pawlik2009} model.  
For more gradual histories, the Ref.~\cite{Pawlik2009} model still finishes early by $\Delta z \approx 1$, while for $C_{\mathcal{A}} = 3$, the model ends later than the simulation in the \textsl{Very Early Start} case.  
The $C_{\mathcal{A}} = 3$ case neglects the evolution of small-scale IGM structure, which drives redshift evolution of $C_{\mathcal{A}}$.  
As such, it under-estimates absorption at low redshifts and over-estimates it at high redshift.  
The Ref.~\cite{Pawlik2009} clumping factor includes redshift evolution due to structure formation, but was derived from simulations in the ``pressure-smoothed'' limit that neglects the effects of gas clumping near the IGM Jeans scale (that is, mass scales $\lesssim 10^8$ $M_{\odot}$).  
As such, it under-estimates $C_{\mathcal{A}}$ at all redshifts (see Ref.~\cite{DAloisio2020}).   
We conclude that modeling uncertainty from $C_{\mathcal{A}}$ dominates over that from using the local source approximation.  
As such, the improvement in our $x_i$ prediction over that of the standard photon counting method is unlikely to be important for data interpretation until modeling uncertainty in $C_{\rm A}$ as been appreciably reduced.  
Rather, our ability to self-consistently model $\Gamma_{\rm HI}$ with $x_i$ is our most important improvement over prior work.  

\begin{figure}[h!]
    \centering
    \includegraphics[scale=0.19]{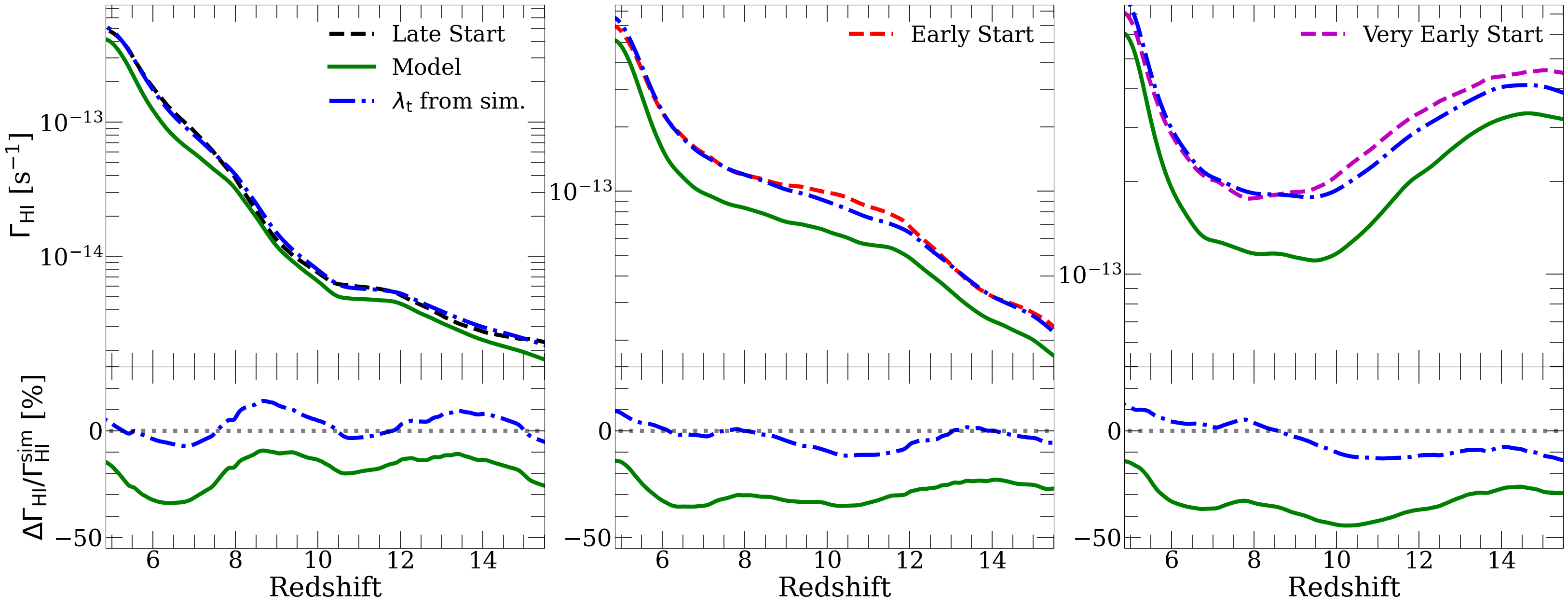}
    \caption{Our model prediction for the spatially-averaged $\Gamma_{\rm HI}$ (green solid), compared to the simulations (dashed curves).  The layout is the same as Figure~\ref{fig:xi_sim_model_comparisons}.  The blue dot-dashed curve shows the result of using $\lambda_t$, extracted from the simulation, and solving Equation~\ref{eq:first_step} instead of using Equations~\ref{eq:first_eq} and~\ref{eq:second_eq} (note that in Equation~\ref{eq:first_step}, $\kappa = \lambda_t^{-1}$).  }
    \label{fig:Gamma_sim_model_comparisons}
\end{figure}

In Figure~\ref{fig:Gamma_sim_model_comparisons} we assess the accuracy of our $\Gamma_{\rm HI}$ prediction.  As in Figure~\ref{fig:xi_sim_model_comparisons}, the green solid line shows the output of our model using the test setup described above.  
The blue dashed line shows the result of solving Equation~\ref{eq:second_step} using $\kappa = \lambda_t^{-1}$ calculated from the simulation (Equation~\ref{eq:lambda_def}), bypassing Equation~\ref{eq:second_eq}.  
We see that the model prediction is offset low by $20-40\%$ (depending on the reionization history), with $\pm 10\%$ fluctuations around this value.  
The accuracy improves slightly as reionization is approaching its end at $z \sim 5$, reaching to within $\sim 15\%$ of the truth for all the reionization histories.  
Using $\lambda_t$ directly removes this offset and recovers accuracy at the $10\%$ level or better at all redshifts.  
We will explore the cause of this offset in \S\ref{sec:inaccuracies}.  

We pause here to note that the bubble size distribution depends on the choice of starting points for sightlines.  
In this work, we have defined $dP/dR$ for randomly-placed sightlines within ionized regions, mirroring the original definition in Ref.~\cite{Mesinger2007}.  
However, it is also possible to define $dP/dR$ such that sightlines are anchored to halos (galaxies) of a given mass (brightness).  
Such definitions have been adopted in a number of works~\citep[e.g.][]{Roth2023,Neyer2024,Neyer2025}.  
Furthermore, observational constraints on the $\langle R \rangle$ likely probe something closer to one of these definitions, since they are derived from galaxy observations~\citep{Mason2020,Umeda2023}.  
We have experimented with such alternative definitions of $dP/dR$, and find that they nearly always give higher values of $\langle R \rangle$, since sightlines anchored to halos are further (on average) from the edges of ionized regions than randomly-placed sightlines.  
We can see from Equation~\ref{eq:second_eq} that a higher $\lambda_n$ would yield a larger $\Gamma_{\rm HI}$ at fixed $dx_i/dt$, suggesting that an alternative definition for $dP/dR$ might reduce the offset between model and simulation.  
We find that these alternative definitions yield modestly higher (by $\sim 10\%$) predictions for $\Gamma_{\rm HI}$, and leave $x_i$ unchanged.  
However, they do not raise $\Gamma_{\rm HI}$ enough to eliminate the observed offset, suggesting that the uncertainty in our model is larger than the effect of defining $dP/dR$ differently, and arises for some other reason.  
As such, we do not explore these different definitions in detail here.  

Our findings in this section establish two conclusions.  First, that our model captures the reionization history to better than $1\%$ at all redshifts, including at the tail end when the standard photon-counting approach errs at the $\sim 5\%$ level.  
This demonstrates that our formalism predicts ionized fraction as well as, or better than, existing photon counting models~\citep[e.g.][]{Chen2020}.  
Second, we have shown that we can predict $\Gamma_{\rm HI}$ with accuracy better than a factor of two at all redshifts, including to within $20-30\%$ at the end of reionization, where measurements are available.  
From Figure~\ref{fig:sim_summary}, we can see that this level of accuracy is comparable to or better than the uncertainty in recent measurements of $\Gamma_{\rm HI}$.  
This suggests that we should be able to extract useful constraints from existing data, despite the presence of some modeling error.  
We will consider this point in the next section, before searching for the source of this modeling error in \S\ref{sec:inaccuracies}.   

\section{Usefulness of the model for interpreting $\Gamma_{\rm HI}$ measurements}
\label{sec:usefulness}

Given the modeling uncertainty seen in Figure~\ref{fig:Gamma_sim_model_comparisons}, it is important to consider whether our model is useful for interpreting observational data.  
We will demonstrate in this section why we expect it to be very useful.  
To show this, we ask how strongly $\Gamma_{\rm HI}$ responds to modest changes in $\dot{n}_{\rm ion}$ and the reionization history.  
If differences in $\Gamma_{\rm HI}$ between such scenarios are much larger than the model uncertainty, we can conclude that the model should be useful for drawing conclusions about the properties of the galaxy population and reionization timeline. 

\begin{figure*}
    \centering
    \includegraphics[scale=0.25]{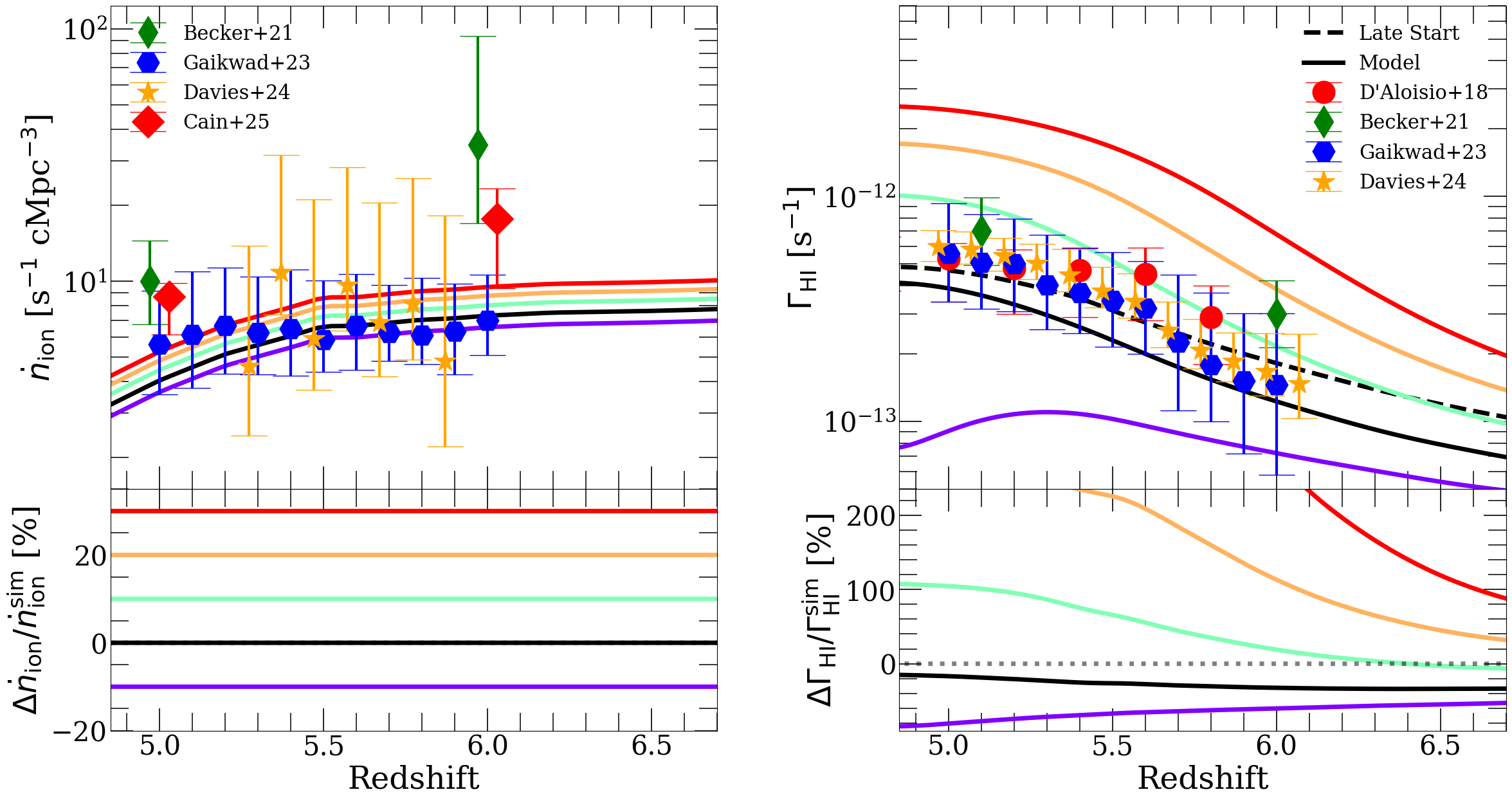}
    \caption{Simple illustration of our model's usefulness for leveraging measurements of $\Gamma_{\rm HI}$ at $5 < z < 6$ to constrain $\dot{n}_{\rm ion}$.  {\bf Left:} $\dot{n}_{\rm ion}$ from our \textsl{Late Start} simulation (black solid) alongside several re-scalings between different $-10\%$ and $+30\%$ (colored curves).  For reference, we show recent measurements of $\dot{n}_{\rm ion}$ in this redshift range by Refs.~\cite{Becker2021,Gaikwad2023,Davies2024,Cain2025}.  {\bf Right:} $\Gamma_{\rm HI}$ predicted by our model for each $\dot{n}_{\rm ion}$, compared to the truth from the \textsl{Late Start} simulation.  We show measurements from Refs.~\cite{DAloisio2018,Becker2021,Gaikwad2023,Davies2024}.  The dashed black curve on the right is the ``truth'' (that is, $\Gamma_{\rm HI}$ from the simulation).  }
    \label{fig:gamma_usefulness_plot}
\end{figure*}

In Figure~\ref{fig:gamma_usefulness_plot}, we re-scale $\dot{n}_{\rm ion}$ up and down in the \textsl{Late Start} model in increments of $10\%$ (left panels) and compute $\Gamma_{\rm HI}$ for each re-scaling (right panels), assuming that $C_{\mathcal{A}}$ and $\langle R \rangle (x_i)$ are held fixed.    
We show several recent sets of measurements in this redshift range for $\dot{n}_{
\rm ion}$ and $\Gamma_{\rm HI}$ (see caption).   
The former give a sense of how large a ``modest'' change in $\dot{n}_{\rm ion}$ is, and the latter show how tightly current measurements would constrain a simple 1-parameter model that re-scales $\dot{n}_{\rm ion}$.  
The black dashed line on the right indicates the $\Gamma_{\rm HI}$ from the simulation, while the solid black line is the model prediction without any re-scaling to $\dot{n}_{\rm ion}$, and the colored curves all re-scale $\dot{n}_{\rm ion}$ by some factor.   
By comparing the black curves, we see that the difference between the model prediction and the truth is comparable to or less than the size of $1\sigma$ error bars on recent $\Gamma_{\rm HI}$ measurements.   
By contrast, increasing $\dot{n}_{\rm ion}$ by just $10\%$ (much less than than the size of $\dot{n}_{\rm ion}$ measurement uncertainties) increases $\Gamma_{\rm HI}$ by a factor of $2-3$ at $z \leq 6$.  
Increasing $\Gamma_{\rm HI}$ by $20\%$ (yellow) or decreasing it by $10\%$ (magenta) pushes the prediction outside the $1\sigma$ error bars of most measurements.  
We see that $\Gamma_{\rm HI}$ is a very sensitive probe of $\dot{n}_{\rm ion}$ at the end of reionization.  
This finding is broadly consistent with the results of previous work~\citep{Keating2019,Ocvirk2021,Cain2023}.   
The sensitivity arises because of the sharp increase in the MFP caused by the disappearance of the last neutral regions.  
As neutral islands disappear,they stop regulating the growth of $\Gamma_{\rm HI}$.  
Ionizing photons produced in excess of the absorption rate in ionized gas subsequently drives a rapid 
increase in both $\lambda_i$ and $\Gamma_{\rm HI}$.  
As we have noted earlier, this strong sensitivity to $\dot{n}_{\rm ion}$ is missed if $\Gamma_{\rm HI}$ is not modeled self-consistently with $x_i$.  
This fact serves not only to motivate our formalism, but also validate its usefulness for reionization studies, as seen above.  

\begin{figure*}
    \centering
    \includegraphics[scale=0.25]{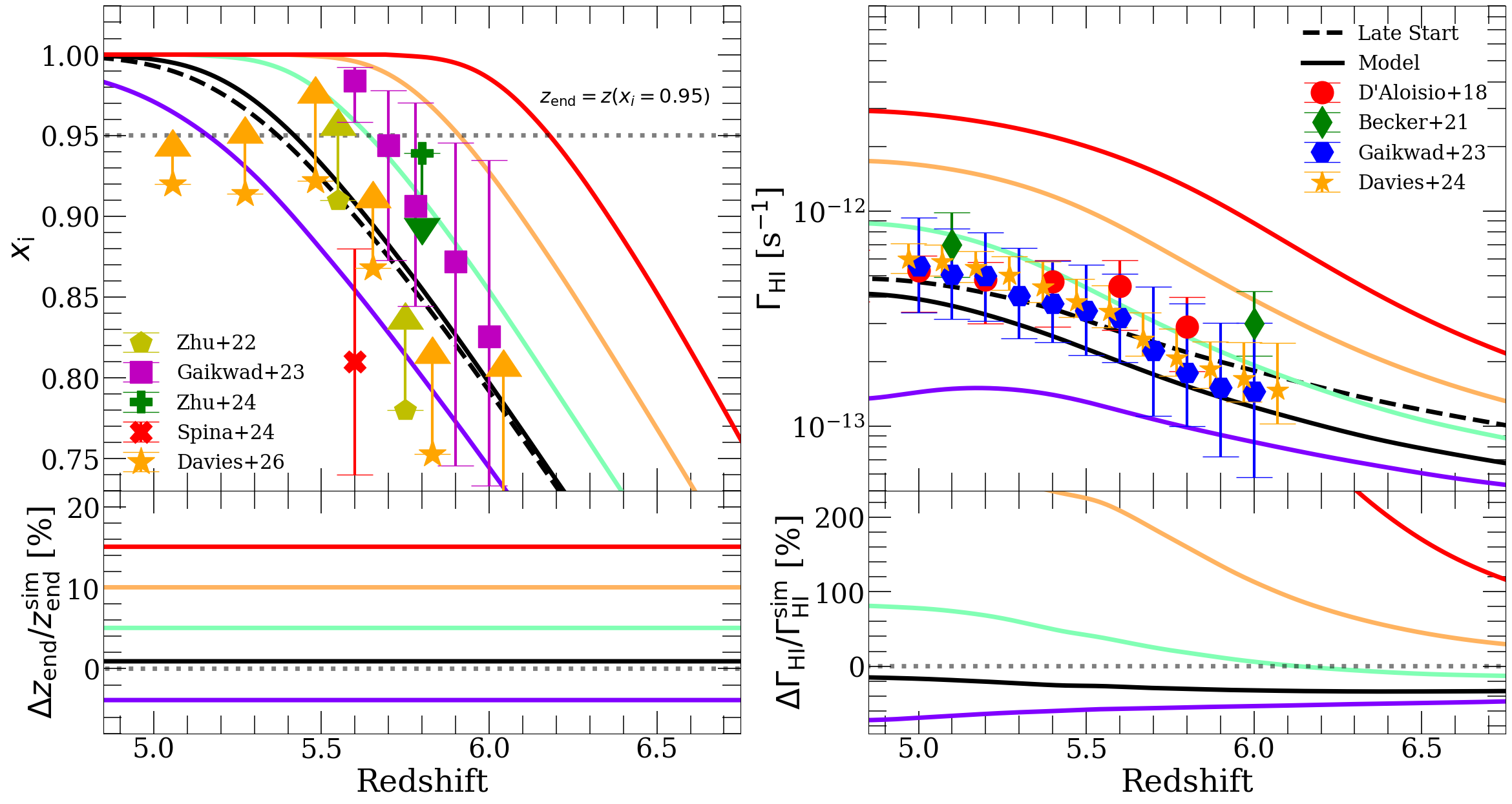}
    \caption{The same as Figure~\ref{fig:gamma_usefulness_plot}, but instead re-scaling $\dot{n}_{\rm ion}$ to achieve shifts in the redshift at which the ionized fraction is $95\%$.  Ionized fraction constraints are from Refs.~\cite{Zhu2022,Gaikwad2023,Zhu2024,Spina2024,Davies2026}.  }
    \label{fig:gamma_usefulness_plot_ion}
\end{figure*}

In Figure~\ref{fig:gamma_usefulness_plot_ion}, we carry out a similar exercise, this time shifting the endpoint of reionization ($z_{\rm end}$, defined to be the redshift when $x_i = 0.95$) by offsets between $-5\%$ and $+15\%$ from that of the \textsl{Late Start} model.  
We do this by re-scaling $\dot{n}_{\rm ion}$ by hand to achieve the desired shifts in $z_{\rm end}$.  
We show for comparison a subset of direct constraints on $x_i$ from the Ly$\alpha$ forest at $z \lesssim 6$ (references in caption). 
The shifts in $\Gamma_{\rm HI}$ are similar to those seen in Figure~\ref{fig:gamma_usefulness_plot}, reinforcing our conclusions.  
A $+5\%$ ($\Delta z \approx 0.25$) shift in $z_{\rm end}$ more than offsets the modeling error, and $-5\%$ to $+10\%$ shifts push the prediction outside the $1\sigma$ error bars on $\Gamma_{\rm HI}$ measurements.  
These shifts are comparable to the range allowed by current measurements, as seen on the left.  
From these results, we see that our modeling uncertainty in $\Gamma_{\rm HI}$ maps to a $\lesssim 10\%$ bias in $\dot{n}_{\rm ion}$ and a $\lesssim 5\%$ bias in reionization endpoint, which are acceptable given uncertainties on current measurements.  
Placing constraints on these quantities to higher precision than this will require improvements to our formalism (see \S\ref{sec:inaccuracies}). 
 
Thus far, we have assumed that $C_{\mathcal{A}}(z)$ remains fixed - that is, independent of $\Gamma_{\rm HI}$ - when we vary $\dot{n}_{\rm ion}$ and the reionization history.  
There is good reason to expect that this is incorrect.  
Studies of small-scale intergalactic structures in the context of reionization have found that the MFP in ionized regions scales sub-linearly with $\Gamma_{\rm HI}$~\citep[e.g.][]{McQuinn2011,Emberson2013,Park2016,DAloisio2020,Nasir2021,Chan2023,Cain2026}, which implies that $C_{\mathcal{A}}$ should scale positively ($\lambda_i \propto \Gamma_{\rm HI}^\xi$ $\rightarrow$ $C_{\mathcal{A}} \propto \Gamma_{\rm HI}^{1-\xi}$, from Equations~\ref{eq:C_A} and~\ref{eq:kappa_defs}).  
This scaling reflects the dependence of the self-shielding density on $\Gamma_{\rm HI}$, with higher $\Gamma_{\rm HI}$ increasing the maximum density of ionized gas and thus the recombination rate~\citep{MiraldaEscude2000,Furlanetto2005,McQuinn2011,Munoz2016,Theuns2023}. 
We note that there is considerable theoretical uncertainty in $\xi$, with physically-motivated values typically ranging from $1/3$ to $1$, with $2/3$ being a commonly-adopted value~\citep{Davies2016,DAloisio2018,Becker2021,Zhu2023}.  
The fact that we model $\Gamma_{\rm HI}$  means that it is possible to include this dependency self-consistently in our formalism, which is a key improvement in its own right.  
Forthcoming efforts to constrain $\dot{n}_{\rm ion}$ using our formalism will need to take this dependence into account.  

\begin{figure}
    \centering
    \includegraphics[scale=0.24]{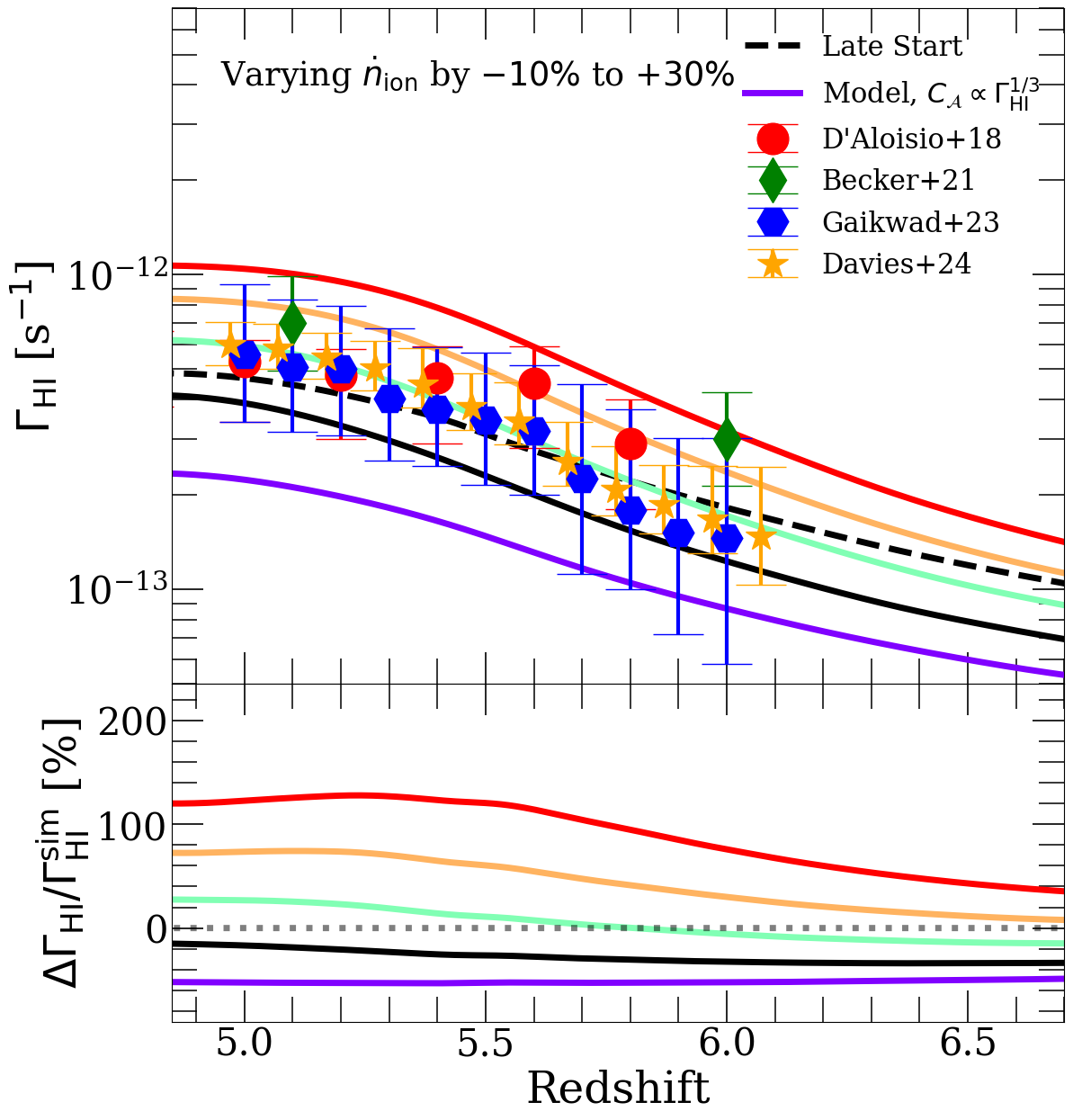}
    \includegraphics[scale=0.24]{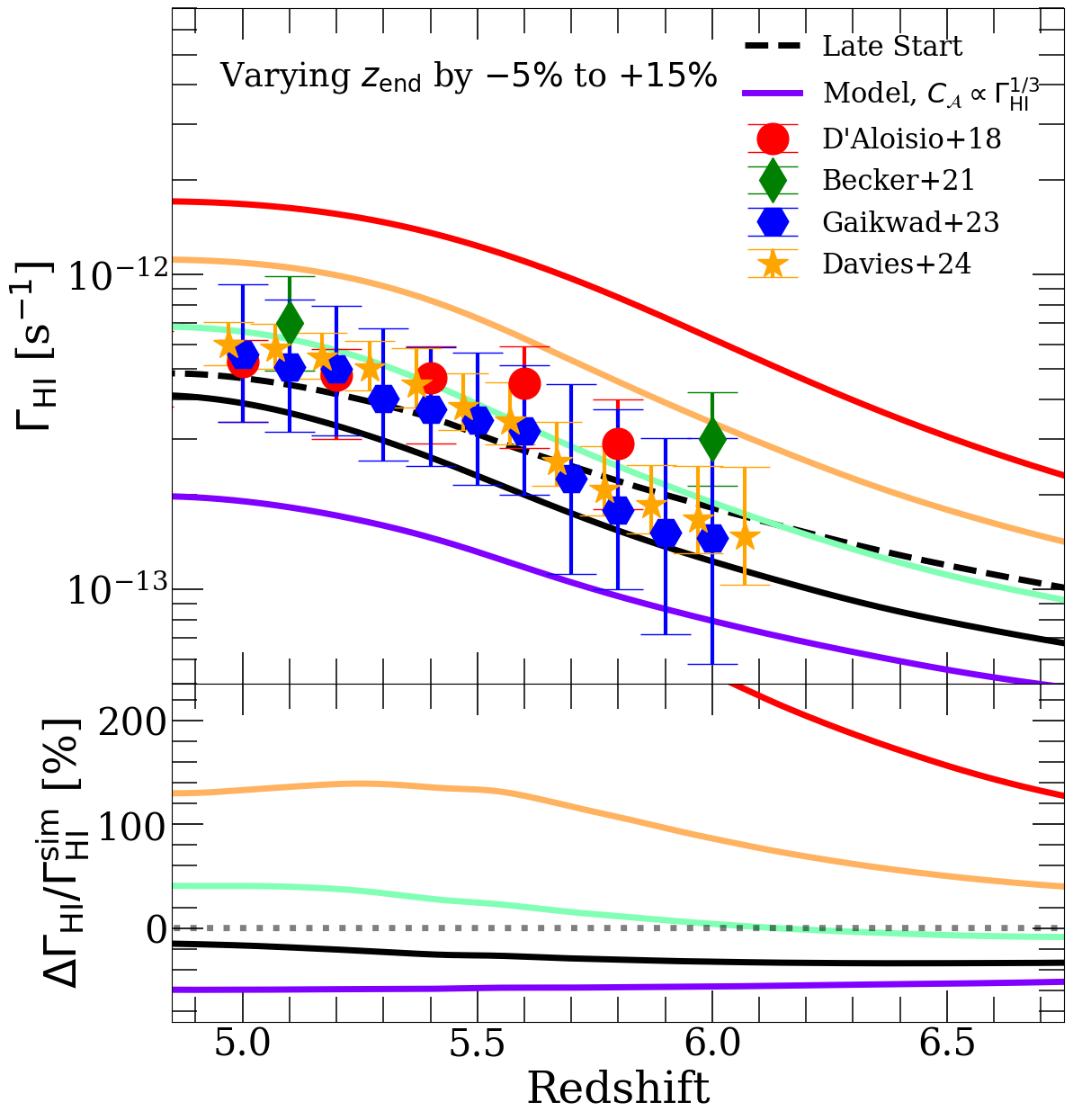}
    \caption{Effect of adding $\Gamma_{\rm HI}$ dependence to $C_{\mathcal{A}}$ (assuming $\xi = 2/3$) on the sensitivity of $\Gamma_{\rm HI}$ to $\dot{n}_{\rm ion}$ and $z_{\rm end}$.  {\bf Left}: The sensitivity to $\dot{n}_{\rm ion}$ decreases when $C_{\mathcal{A}}$ is allowed to increase with $\Gamma_{\rm HI}$, since the ionized IGM becomes more effective at regulating the growth of the UV background.  {\bf Right:} $\Gamma_{\rm HI}$ remains an equally sensitive probe of $z_{\rm end}$, because shifting $z_{\rm end}$ earlier by a fixed amount requires a larger boost to $\dot{n}_{\rm ion}$ when $C_{\rm A}$, which makes up for much of the loss of sensitivity of $\Gamma_{\rm HI}$ to $\dot{n}_{\rm ion}$.  }
    \label{fig:gamma_usefulness_CG}
\end{figure}

In Figure~\ref{fig:gamma_usefulness_CG}, we show again the right panels of Figures~\ref{fig:gamma_usefulness_plot} (left) and~\ref{fig:gamma_usefulness_plot_ion} (right), but this time assuming $\xi = 2/3$, such that $C_{\mathcal{A}} \propto \Gamma_{\rm HI}^{1/3}$.  
We then re-scale $C_{\rm A}$ from the FlexRT simulation by a factor of $[\Gamma_{\rm HI}(z)/\Gamma_{\rm HI}^0(z)]^{1/3}$, where $\Gamma_{\rm HI}^0(z)$ is our model prediction without any re-scaling of $\dot{n}_{\rm ion}$.  
In the left panel, we see that the same re-scalings of $\dot{n}_{\rm ion}$ we used in Figure~\ref{fig:gamma_usefulness_plot} map to a considerably smaller spread in $\Gamma_{\rm HI}$ when $\xi = 2/3$.  
Indeed, the model with a $30\%$ boost in $\dot{n}_{\rm ion}$ is still marginally consistent (at $1\sigma$) with some of the measurements.  
Physically, the weaker dependence of $\Gamma_{\rm HI}$ on $\dot{n}_{\rm ion}$ occurs because absorptions by Lyman Limit systems increase as $\Gamma_{\rm HI}$ increases, which partially regulates the growth of the UV background~\citep{Munoz2016,Theuns2023}.  
This suggests that, if $\xi$ is appreciably less than $1$, the uncertainty in our model will map to a larger systematic bias in $\dot{n}_{\rm ion}$ than Figure~\ref{fig:gamma_usefulness_plot} suggests. 
On the right, we see that the sensitivity of $\Gamma_{\rm HI}$ to $z_{\rm end}$ is weakened less than its sensitivity to $\dot{n}_{\rm ion}$ when $\xi = 2/3$.  
This owes to the fact that higher (lower) $C_{\mathcal{A}}$ slows down (speeds up) the pace of reionization, in turn requiring a larger change in $\dot{n}_{\rm ion}$ to recover the shame shift in $z_{\rm end}$.  
It follows that $\xi$ has a smaller effect on the sensitivity of $\Gamma_{\rm HI}$ to $x_i$, such that constraints on $z_{\rm end}$ inferred using our model should be only weakly sensitive to $\xi$.  
This is good news for applications of our formalism to analyses of cosmological data sets, which are generally only interested in the timeline of reionization (which sets $\tau$) and not so much on $\dot{n}_{\rm ion}$.

The dependence of $C_{\mathcal{A}}$ on $\Gamma_{\rm HI}$ can be treated in forthcoming analyses in a number of ways. 
One option is to assume a prescription for self-shielding of ionized gas, which can be predicted analytically~\citep[e.g.][]{MiraldaEscude2000} or extracted from numerical simulations~\citep{Rahmati2013,Chardin2018}, and use a model for the small-scale density PDF to calculate the recombination rate.  
One could also directly predict $\xi$ using an analytical model for $\lambda_i$, such as the ``halo model'' developed in Ref.~\cite{Theuns2023}.  
Simulations can also be used to directly predict $\lambda_i$ and its dependence on various parameters, including $\Gamma_{\rm HI}$.  
Recently, Ref.~\cite{Tohfa2026} developed an emulator for the mean free path in ionized regions (our $\lambda_i$) based on the suite of high-resolution hydrodynamical/radiative transfer simulations presented in Ref.~\cite{Cain2026}.  
Their suite includes simulations with different values of $\Gamma_{\rm HI}$, reionization redshift $z_{\rm reion}$, and local over-density $\delta$ at their $2$ $h^{-1}$Mpc box scale.  
Using their emulator, one could write $C_{\mathcal{A}}(\Gamma_{\rm HI},z_{\rm reion},\delta) \propto \Gamma_{\rm HI}/\lambda_i(\Gamma_{\rm HI}, z_{\rm re}, \delta)$.  
Since the distribution of $z_{\rm re}$ depends on the reionization history, their model for $C_{\mathcal{A}}$ could be self-consistently coupled to our predictions for $x_i$ and $\Gamma_{\rm HI}$.  
We plan to carry out this exercise in subsequent work. 

We have showed that modest changes in $\dot{n}_{\rm ion}$ and $z_{\rm end}$ map to changes in $\Gamma_{\rm HI}$ much larger than our modeling uncertainty at the end of reionization.  
We have further found that our model accuracy is comparable to or better than uncertainties on current measurements.  
As such, we expect our formalism to be useful for incorporating existing constraints on $\Gamma_{\rm HI}$ (and other observables derivative of high-$z$ quasar spectra) into high-dimensional analyses of high-redshift galaxy properties and cosmological data sets.  
However, if future analysis of quasar data can measure $\Gamma_{\rm HI}$ to $10\%$ accuracy or better, the $20-30\%$ discrepancy between our $\Gamma_{\rm HI}$ prediction and simulations will become dominant over measurement uncertainty.  
It is therefore important to understand the origin of this discrepancy and seek pathways for future improvements - a task we undertake in the next section.  
We note that readers interested only in the main results and conclusions of this paper may skip the (somewhat technical) next section and jump straight to \S\ref{sec:conc}.  

\section{Origin of $\Gamma_{\rm HI}$ offset \& pathways for improvement}
\label{sec:inaccuracies}

As we noted in the discussion surrounding Figure~\ref{fig:Gamma_sim_model_comparisons}, plugging the total MFP ($\lambda_t$) from the simulations directly into Equation~\ref{eq:first_step} to solve for $\Gamma_{\rm HI}$ removes the discrepancy between the model prediction and simulations.  
This suggests that the discrepancy has something to do with how we model $\lambda_n$ (and thereby $\lambda_t$) in \S\ref{subsec:opacity_neutral}.  
We will therefore return to the derivation in that section, and the assumptions underlying it.  
In that section, we made two important assumptions: (1) that the average transmission along sightlines with fixed $R$ can be modeled by an exponential transmission profile with attenuation length equal to $\lambda_i$ for any value of $R$ (Equation~\ref{eq:first_assumption}) and (2) that the ionized bubble size distribution is well-approximated by an exponential in $R$ (Equation~\ref{eq:dPdR_exp}).  
We will test the second assumption first.  

\begin{figure}[h!]
    \centering
    \includegraphics[scale=0.25]{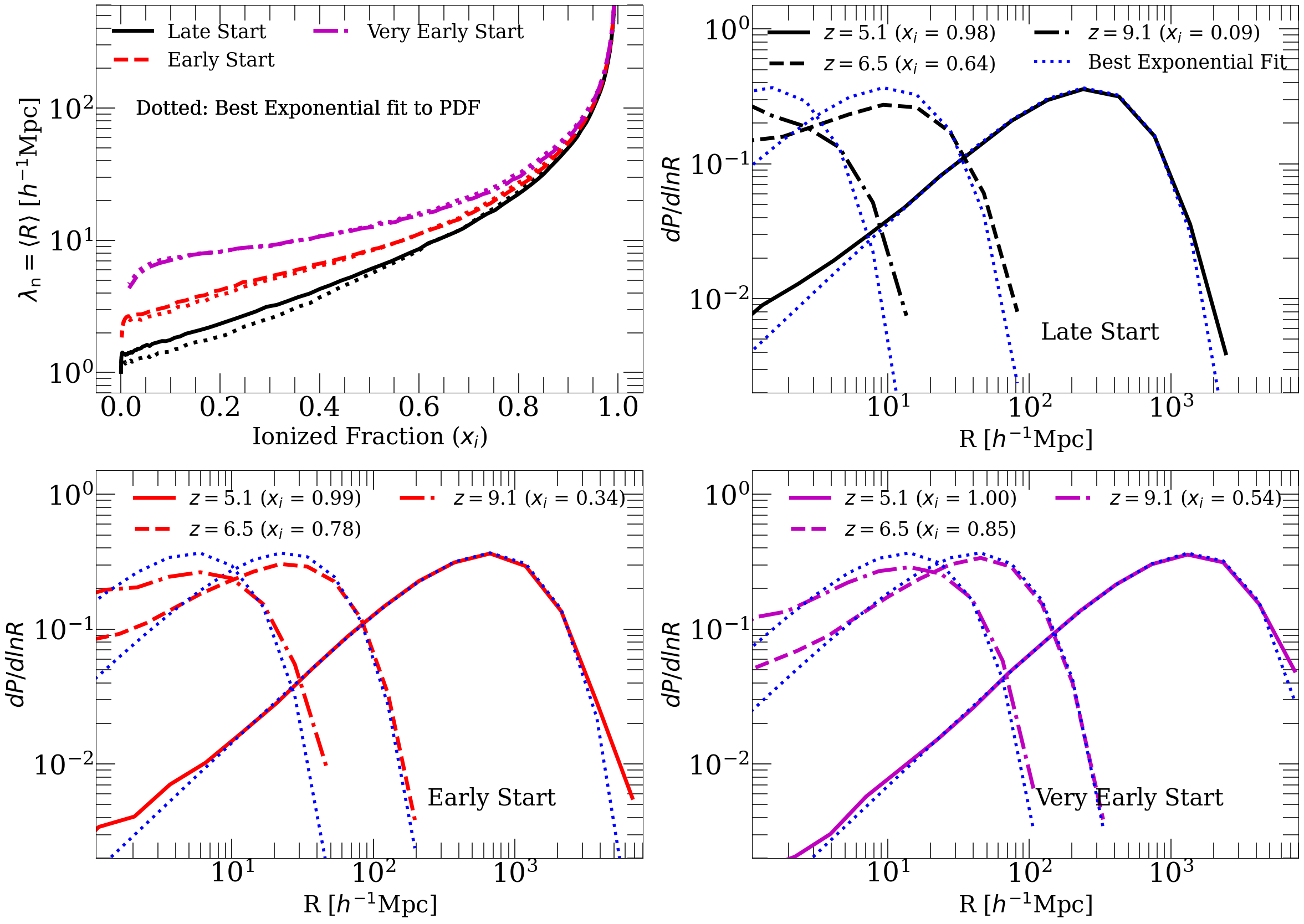}
    \caption{Bubble size distributions in our simulations.  {\bf Top Left:} $\langle R \rangle$ vs. $x_i$ for each simulation.  The dotted curves show the $\langle R \rangle$ obtained from fitting the bubble distribution to an exponential (Equation~\ref{eq:dPdR_exp}).  {\bf Remaining panels:} examples of the bubble size distribution at several redshifts for each reionization history.  The legend shows the $x_i$ at each redshift, and the blue dotted curves denote the best fit to Equation~\ref{eq:dPdR_exp}.  }
    \label{fig:bubble_plot}
\end{figure}

The top-left panel of Figure~\ref{fig:bubble_plot} shows $\lambda_n = \langle R \rangle$ as a function of $x_i$ (the input to Equation~\ref{eq:second_eq}) in our reionization scenarios.   
The other panels show $dP/d\ln R$ at several redshifts, with the corresponding $x_i(z)$ indicated in the legend.  
The blue dotted lines show the best fit to Equation~\ref{eq:dPdR_exp}, and the dotted lines in the top left show the $\langle R \rangle$ recovered from these fits.  
We see that Equation~\ref{eq:dPdR_exp} fits $dP/dR$ reasonably well, especially at lower redshifts when the ionized bubbles are largest\footnote{We note that within the last few percent of reionization, when $\langle R \rangle \sim$ our box size, likely do not capture $dP/dR$ accurately.  This is probably fine for two reasons: (1) the same issue affects FlexRT's internal prediction of $\Gamma_{\rm HI}$ and $x_i$ also, so our comparison remains value (2) absorption within ionized gas dominates in this regime, such that errors in $dP/dR$ should have little impact on $\Gamma_{\rm HI}$ anyway.  } (and $\Gamma_{\rm HI}$ measurements exist).  
At high redshifts (small $x_i$), the fit deviates modestly, but the best-fit $\langle R \rangle$ remains within $20\%$ of the truth, even when $\langle R \rangle \rightarrow \Delta s_{\rm cell}$.  
This deviation is only a few percent at $x_i > 0.5$. 
It therefore seems unlikely that our assumption about $dP/dR$ is the origin of the problem.  

As it turns out, the culprit is actually the first assumption, which underlies Equation~\ref{eq:first_assumption}.  
For the stacked transmission profiles of IGM sightlines to be well-described by an exponential profile, the sightline-to-sightline fluctuations in $\lambda_i(s)$ must be uncorrelated at a fixed distance $s$.  
This will be true, for example, if these fluctuations are dominated by the density field, which is (mostly) uncorrelated between many randomly positioned and oriented sightlines.  
We expect this to be true in general{\it as long as the placement of sightlines is not correlated in any way to the spatial fluctuations of the fields that set $\lambda_i$}.  Besides density, three other fields set the MFP in ionized gas - $\Gamma_{\rm HI}$, and to a lesser degree, temperature and $z_{\rm reion}$~\citep{DAloisio2020,Cain2026}.  
Unfortunately, as we will see, our sightline placement (at fixed $R$) is necessarily spatially correlated with all three of these variables.  

\begin{figure}[h!]
    \centering
    \includegraphics[scale=0.285]{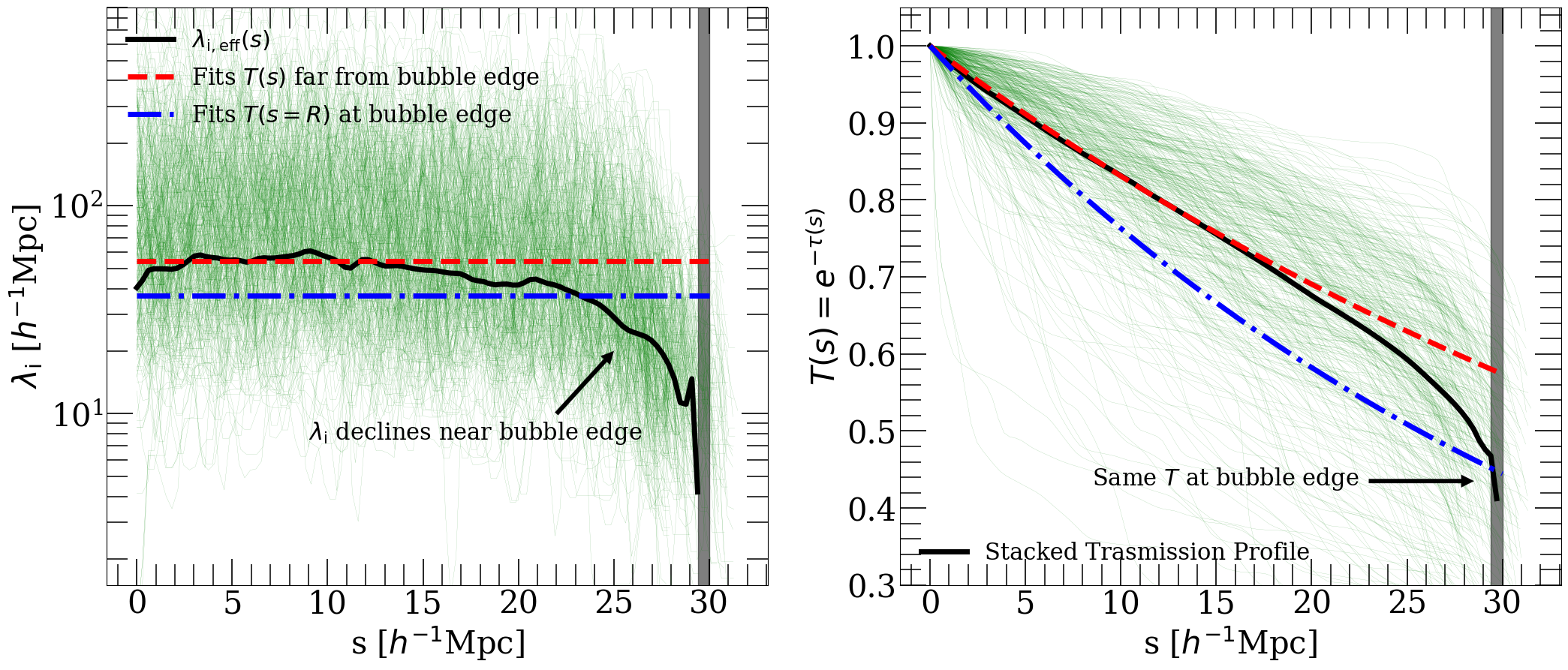}
    \caption{Direct test of the assumption underlying Equation~\ref{eq:first_assumption}.  {\bf Left:} $\lambda_i(s)$ along many sightlines of length $R = 30$ $h^{-1}$Mpc (thin green lines) and their effective MFP $\lambda_{\rm i, eff}(s)$ (Equation~\ref{eq:lam_eff}).  {\bf Right:} transmission profiles for each sightline, and their stacked average.  The red dashed line is a pure exponential transmission profile that fits $T(s)$ well within the first $\approx 15$ $h^{-1}$Mpc of the stack.  The blue dot-dashed line is also a pure exponential, but instead matches the transmission at $s = R = 30$ $h^{-1}$Mpc (the bubble edge).  The drop off in the average $\lambda_i$ and transmission close the bubble edge violates the assumption underlying Equation~\ref{eq:first_assumption}.  Matching the correct transmission at $s = R$ requires under-estimating $\lambda_i$, and thus $\Gamma_{\rm HI}$, $s < R$.  }
    \label{fig:sample_sightlines}
\end{figure}

In Figure~\ref{fig:sample_sightlines}, we show $\lambda_i(s)$ (left) and the integrated transmission profile, $T(s) = \exp[-\tau(s)]$ (right) for many individual sightlines with a fixed $R = 30$ $h^{-1}$Mpc (thin green curves).  
The gray shaded line on the right denotes the bubble edge at $s = R$.  
The black solid curve on the right is the stack of all these transmission profiles.   
On the left, it shows the effective $\lambda_i(s)$ for the stack obtained from differentiating the stacked transmission profile, given by
\begin{equation}
    \label{eq:lam_eff}
    \lambda_{\rm i, eff}(s) \equiv \kappa_{\rm i, eff}(s)^{-1} \equiv -\left[\frac{d}{ds} \ln T(s)\right]^{-1}
\end{equation}
where $\kappa_{\rm i, eff}$ is the analogously-defined effective absorption coefficient.  
The red dashed line shows a pure exponential transmission profile (with constant $\lambda_i(s)$) that fits well the transmission profile and $\lambda_{\rm i, eff}(s)$ from the simulated stack at $s \lesssim R/2$.  
These are regions well interior to ionized bubbles, where we can see that fluctuations between different sightlines are uncorrelated and average down to roughly an exponential transmission profile.  
However, near the bubble edge, this is no longer true.  
We see a sharp decline in $\lambda_{\rm eff}(s)$, accelerating near the bubble edge, which translates to a drop in transmission well below the red dashed line.  
This happens because all sightlines see a decrease in $\lambda_i(s)$ as as $s \rightarrow R$.  
The main culprit driving this decline is $\Gamma_{\rm HI}$.  
Close to the edges of ionized bubbles, ionizing flux from distance sources has been attenuated significantly by the intervening IGM, such that $\Gamma_{\rm HI}$ declines as the bubble edge is approached {\it from any direction}\footnote{Indeed, spatial fluctuation in $\Gamma_{\rm HI}$ that trace the structure of ionized bubbles are crucial to understanding spatial correlations between the Ly$\alpha$ forest and the sources of reionization~\citep{Gangolli2024,Kakiichi2025,Garaldi2025,Kashino2026}.  }.  
Indeed, spatial fluctuations in $\Gamma_{\rm HI}$ that trace the structure of the ionized bubbles likely play an important role in the $5 < z < 6$ Ly$\alpha$ forest~\citep{Davies2016,DAloisio2018}.  
Gas closest to the bubble edges is also most recently ionized, meaning it has more small-scale structure than average IGM gas and a still-smaller MFP~\citep{Park2016,DAloisio2020,Chan2023}.  
{\bf Since all sightlines have the same $R$, they all approach the edge of the bubble they inhabit at the same $s$, rendering this decrease in $\lambda_i$ coherent across all such sightlines and violating the assumption underlying Equation~\ref{eq:first_assumption}.  }

The result in Figure~\ref{fig:sample_sightlines} further explains why our model {\it under-estimates} $\Gamma_{\rm HI}$.  The blue dot-dashed curve shows another pure exponential profile, but this time with attenuation length selected to match the stacked transmission at the bubble edge, $s = R$.  
Indeed, it is $T(s = R)$ that describes the rate at which photons actually reach the edges of ionized bubbles and contribute to reionization.  
The stacked transmission profile is everywhere else higher than the blue dot-dashed curve, as is $\lambda_{\rm i, eff}(s)$ for most on the sightline distance on the left.  
This shows that {\bf a pure exponential transmission profile that gives the correct $T(s = R)$ at the bubble edge (and thus the correct rate of reionization) will under-estimate $T(s)$ for all $s < R$.}  Since we can see from Figure~\ref{fig:xi_sim_model_comparisons} that our model {\bf does} predict the correct reionization history, it therefore must be under-estimating IGM transmission within ionized bubbles, and thus under-predicting both $\lambda_i$ and $\Gamma_{\rm HI}$.  

It may seem counter-intuitive that our model recovers the correct reionization history despite under-estimating $\lambda_i$.  
We can understand why this is by again considering Equations~\ref{eq:first_eq} and~\ref{eq:kappa_defs}.  
Since the $d\Gamma_{\rm HI}/dt$ term in Equation~\ref{eq:first_eq} is small compared to the $dx_i/dt$ term, we expect to recover the correct reionization history as long as the terms on the RHS are correct (which they are by construction in our test).  
We can further see, from Equation~\ref{eq:kappa_defs}, that 
\begin{equation}
    \label{eq:abs_ratio}
    \frac{\lambda_i}{\lambda_n} \propto \frac{dx_i}{dt} / C_{\mathcal{A}}
\end{equation}
In other words, the reionization history depends only on $C_{\mathcal{A}}$ and the {\it ratio} of $\lambda_i$ and $\lambda_n$.  
This suggests that $\lambda_i$ and $\lambda_n$ in our model must be too low by the same factor, such that we recover the correct reionization history.  
This also explains why using $\lambda_t$ extracted from the simulation rather than predicting it based on $\lambda_n$ and $C_{\mathcal{A}}$ recovered the correct $\Gamma_{\rm HI}$ history to within $10\%$ in Figure~\ref{fig:Gamma_sim_model_comparisons}.  

Given these findings, we should stop to consider when we should expect our model to be most (and least) accurate.  
One scenario in which it should be quite accurate is that in which $\lambda_i$ is independent of $\Gamma_{\rm HI}$, temperature, and $z_{\rm reion}$, and depends only on the local density.  
Then, one can safely assume that sightline-to-sightline variations in $\lambda_i(s)$ are (largely) uncorrelated, since density fluctuations at the scale of ionized bubbles are small, especially late in reionization.  
The opposite limit, $\lambda_i \propto \Gamma_{\rm HI}$ (that is, $\xi = 1$), is most problematic.  
We note that in the simulations upon which FlexRT's sub-grid model is based, the clumping factor is sensitive to $\Gamma_{\rm HI}$ only weakly in regions that have been ionized for more than a few $10s$ of Myr~\citep{DAloisio2020}.  
This suggests that our tests in this paper likely err on the side of $\xi$ close to $1$ - so in that sense, they can likely be treated as conservative.  
We also expect the model to be accurate if $\lambda_i$ is large compared to $\lambda_n$ during most of reionization, because in that case spatial fluctuations in $\Gamma_{\rm HI}$ will be minimized.  
However, in the opposite limit - that of so-called ``absorption-dominated'' reionization~\citep{Davies2021b,Davies2024c} - we expect $\Gamma_{\rm HI}$ fluctuations to be maximized. 
If $\xi$ is also well above $0$, we expect our model to err badly in that limit.  
We show why this is the case in Appendix~\ref{app:stromgren} using a classic toy model - a Str\"omgren Sphere around an isolated source in a constant density and temperature medium.  
From Figure 4 of Ref.~\cite{Cain2024b}, we can see that our simulations are indeed absorption dominated - but only mildly so.  
This suggests that our model may fare worse than we report here in scenarios where the clumping factor is much larger than its canonical value of $\sim 3$, as suggested by some recent analyses of high-redshift quasar and galaxy data~\citep{Davies2024c,Austin2025}.  

Improving the model to recover more accurate predictions of $\Gamma_{\rm HI}$ therefore requires relaxing the assumption underlying Equation~\ref{eq:first_assumption}.  One approach to doing this would be to replace the constant $\overline{\lambda}_i$ on the RHS with a functional form that depends on both $s$ and $R$ that captures the decline in $\lambda_i(s)$ near bubble edges.  
This could be done, in principle, if the spatial fluctuations in $\Gamma_{\rm HI}$ could be modeled in a statistical way at spatial scales similar to those characterizing the bubbles.    
We speculate that it might be possible to develop a self-consistent treatment in this manner starting from a perturbation theory analysis based on the the position-dependent form of Equation~\ref{eq:cosmoRT_eq} (see Ref.~\cite{McQuinn2018} for an example of this in the context of modeling the EoR 21 cm signal).  
We leave this effort to future work.  

\section{Conclusions}
\label{sec:conc}

We have introduced a new analytical formalism that, for the first time, solves jointly for the ionized fraction and average photoionization rate during reionization.  
Our model has the potential to allow many-dimensional parameter analyses to take advantage of measurements from high-redshift quasar spectra beyond the IGM neutral fraction, potentially tightening constraints on the tail end of reionization.   
It requires (as input) the average ionized bubble size as a function of the ionization fraction, for which analytical and simulation-calibrated prescriptions already exist in the literature.  
We have tested our model against radiative transfer simulations of reionization, and come to the following conclusions: 

\begin{itemize}

    \item Our model reproduces the reionization history in radiative transfer simulations to within $1\%$ when its input parameters are calibrated to match the simulations.  It therefore matches or exceeds the accuracy of existing photon counting treatments.    

    \item Our prediction for $\Gamma_{\rm HI}$ is within a factor of $2$ of the simulations at all redshifts, with accuracy of $20-30\%$ at the end of reionizaiton.  This is comparable to or better than uncertainties on existing $\Gamma_{\rm HI}$ measurements at $z \lesssim 6$.  

    \item $\Gamma_{\rm HI}$ is highly sensitive the timing of the endpoint of reionization and the evolution of the ionizing emissivity of the galaxy population when it is modeled self-consistently with the reionization history.  We find that $10-20\%$ shifts in $\dot{n}_{\rm ion}$ and $5-10\%$ shifts in the reionization endpoint map to changes in $\Gamma_{\rm HI}$ larger than current measurement uncertainties if the clumping factor is independent of $\Gamma_{\rm HI}$.  Invoking a physically-motivated positive scaling of the clumping factor with $\Gamma_{\rm HI}$ weakens this sensitivity by a factor of $\sim 2$ for $\dot{n}_{\rm ion}$, but largely preservers the sensitivity to the reionization endpoint. 

    \item We investigated in detail the origin of the $20-40\%$ disagreement between our prediction for $\Gamma_{\rm HI}$ and simulations.  
    We found that it is caused by the fact that fluctuations in the mean free path within ionized bubbles are spatially correlated with the structure of the bubbles, which violates one of the key assumptions underlying our model.  
    We demonstrated that this issue explains the discrepancy in $\Gamma_{\rm HI}$, and suggested possible pathways for future improvement.  
    
\end{itemize}

Our formalism provides a timely improvement to computationally efficient photon-counting models, allowing them to take advantage of constraints on IGM-averaged properties inferred from high-redshift quasar spectra.  
We expect it to be useful for interpreting high-redshift galaxy observations and incorporating more astrophysical constraints into analyses of cosmological data sets.  
Further extensions to our formalism could include models for other EoR observables, such as the 21 cm power spectrum and patchy kSZ signals from the CMB~\citep[][]{Furlanetto2004,McQuinn2005,Georgiev2024}, as well as Ly$\alpha$ transmission and damping wing statistics from high-redshift galaxies~\citep{Mason2020,Umeda2023,Witstok2024,Kageura2025}, that can be connected analytically to the ionized bubble size distribution.  
This would provide even more constraining power to forthcoming analyses.

\acknowledgments

The authors thank Steven Furlanetto and Matthew McQuinn for helpful conversations and/or comments on the draft version of this manuscript.  CC acknowledges support from the Beus Center for Cosmic Foundations at Arizona State University for support during the preparation of this manuscript. AD acknowledges support from grant NSF AST-2045600. IG acknowledges support by GIF “German-Israeli Foundation for Scientific Research and Development". YZ acknowledges support from the NIRCam Science Team contract to the University of Arizona, NAS5-02105.  RAW acknowledges support from NASA JWST Interdisciplinary Scientist grants NAG5-12460, NNX14AN10G and 80NSSC18K0200 from GSFC.  

\bibliography{references.bib}

@string{aj = "AJ"}

@string{mnras = "MNRAS"}

@ARTICLE{Cohon2025,
       author = {{Cohon}, Joshua and {Cain}, Christopher and {Windhorst}, Rogier and {D'Aloisio}, Anson and {Carleton}, Timothy and {Zhu}, Yongda},
        title = "{A long time ago in an LAE far, far away: a signpost of early reionization or a nascent AGN at $z=13$?}",
      journal = {arXiv e-prints},
         year = 2025,
        month = aug,
          eid = {arXiv:2508.05739},
        pages = {arXiv:2508.05739},
          doi = {10.48550/arXiv.2508.05739},
archivePrefix = {arXiv},
       eprint = {2508.05739},
 primaryClass = {astro-ph.GA},
       adsurl = {https://ui.adsabs.harvard.edu/abs/2025arXiv250805739C}
}

@ARTICLE{Shull2011,
       author = {{Shull}, Michael and {Harness}, Anthony and {Trenti}, Michele and {Smith}, Britton},
        title = "{Critical Star-Formation Rates for Reionization: Full Reionization occurs at z = 7}",
      journal = {arXiv e-prints},
         year = 2011,
        month = aug,
          eid = {arXiv:1108.3334},
        pages = {arXiv:1108.3334},
          doi = {10.48550/arXiv.1108.3334},
archivePrefix = {arXiv},
       eprint = {1108.3334},
 primaryClass = {astro-ph.CO},
       adsurl = {https://ui.adsabs.harvard.edu/abs/2011arXiv1108.3334S}
}

@ARTICLE{Kashino2026,
       author = {{Kashino}, Daichi and {Lilly}, Simon J. and {Matthee}, Jorryt and {Mackenzie}, Ruari and {Eilers}, Anna-Christina and {Bordoloi}, Rongmon and {Simcoe}, Robert A. and {Naidu}, Rohan P. and {Yue}, Minghao and {Liu}, Bin},
        title = "{EIGER. VII. The Evolving Relationship between Galaxies and the Intergalactic Medium in the Final Stages of Reionization}",
      journal = {\apj},
         year = 2026,
        month = feb,
       volume = {997},
       number = {2},
          eid = {280},
        pages = {280},
          doi = {10.3847/1538-4357/ae2799},
archivePrefix = {arXiv},
       eprint = {2506.03121},
 primaryClass = {astro-ph.GA},
       adsurl = {https://ui.adsabs.harvard.edu/abs/2026ApJ...997..280K}
}

@ARTICLE{Kakiichi2025,
       author = {{Kakiichi}, Koki and {Jin}, Xiangyu and {Wang}, Feige and {Meyer}, Romain A. and {Garaldi}, Enrico and {Bosman}, Sarah E.~I. and {Davies}, Frederick B. and {Fan}, Xiaohui and {Trebitsch}, Maxime and {Yang}, Jinyi and {Ba{\~n}ados}, Eduardo and {Champagne}, Jaclyn B. and {Eilers}, Anna-Christina and {Hennawi}, Joseph F. and {Sun}, Fengwu and {Wu}, Yunjing and {Zou}, Siwei and {Kannan}, Rahul and {Smith}, Aaron and {Becker}, George D. and {D'Odorico}, Valentina and {Connor}, Thomas and {Liu}, Weizhe and {Protu{\v{s}}ov{\'a}}, Klaudia and {Walter}, Fabian and {Zhang}, Huanian},
        title = "{JWST ASPIRE: How Did Galaxies Complete Reionization? Evidence for Excess IGM Transmission around ${\rm [O\,{\scriptstyle III}]}$ Emitters during Reionization}",
      journal = {arXiv e-prints},
         year = 2025,
        month = mar,
          eid = {arXiv:2503.07074},
        pages = {arXiv:2503.07074},
          doi = {10.48550/arXiv.2503.07074},
archivePrefix = {arXiv},
       eprint = {2503.07074},
 primaryClass = {astro-ph.GA},
       adsurl = {https://ui.adsabs.harvard.edu/abs/2025arXiv250307074K}
}

@ARTICLE{Garaldi2025,
       author = {{Garaldi}, Enrico and {Bellscheidt}, Verena and {Smith}, Aaron and {Kannan}, Rahul},
        title = "{The galaxy-IGM connection in THESAN: observability and information content of the galaxy-Lyman- {\ensuremath{\alpha}} cross-correlation at z {\ensuremath{\geq}} 6}",
      journal = {The Open Journal of Astrophysics},
         year = 2025,
        month = dec,
       volume = {8},
        pages = {51666},
          doi = {10.33232/001c.151666},
archivePrefix = {arXiv},
       eprint = {2410.02850},
 primaryClass = {astro-ph.CO},
       adsurl = {https://ui.adsabs.harvard.edu/abs/2025OJAp....851666G}
}

@BOOK{Peebles1993,
       author = {{Peebles}, P.~J.~E.},
        title = "{Principles of Physical Cosmology}",
         year = 1993,
          doi = {10.1515/9780691206721},
       adsurl = {https://ui.adsabs.harvard.edu/abs/1993ppc..book.....P}
}

@ARTICLE{McQuinn2016b,
       author = {{McQuinn}, Matthew},
        title = "{The Evolution of the Intergalactic Medium}",
      journal = {\araa},
         year = 2016,
        month = sep,
       volume = {54},
        pages = {313-362},
          doi = {10.1146/annurev-astro-082214-122355},
archivePrefix = {arXiv},
       eprint = {1512.00086},
 primaryClass = {astro-ph.CO},
       adsurl = {https://ui.adsabs.harvard.edu/abs/2016ARA&A..54..313M}
}

@ARTICLE{Bolton2007,
       author = {{Bolton}, James S. and {Haehnelt}, Martin G.},
        title = "{The observed ionization rate of the intergalactic medium and the ionizing emissivity at z >= 5: evidence for a photon-starved and extended epoch of reionization}",
      journal = {\mnras},
         year = 2007,
        month = nov,
       volume = {382},
       number = {1},
        pages = {325-341},
          doi = {10.1111/j.1365-2966.2007.12372.x},
archivePrefix = {arXiv},
       eprint = {astro-ph/0703306},
 primaryClass = {astro-ph},
       adsurl = {https://ui.adsabs.harvard.edu/abs/2007MNRAS.382..325B}
}

@article{DAloisio2020,
	doi = {10.3847/1538-4357/ab9f2f},
	url = {https://doi.org/10.3847/1538-4357/ab9f2f},
	year = 2020,
	month = {aug},
	publisher = {American Astronomical Society},
	volume = {898},
	number = {2},
	pages = {149},
	author = {Anson D'Aloisio and Matthew McQuinn and Hy Trac and Christopher Cain and Andrei Mesinger},
	title = {Hydrodynamic Response of the Intergalactic Medium to Reionization},
	journal = {The Astrophysical Journal}
}

@ARTICLE{Cain2021,
       author = {{Cain}, Christopher and {D'Aloisio}, Anson and {Gangolli}, Nakul and {Becker}, George D.},
        title = "{A Short Mean Free Path at z = 6 Favors Late and Rapid Reionization by Faint Galaxies}",
      journal = {\apjl},
         year = 2021,
        month = aug,
       volume = {917},
       number = {2},
          eid = {L37},
        pages = {L37},
          doi = {10.3847/2041-8213/ac1ace},
archivePrefix = {arXiv},
       eprint = {2105.10511},
 primaryClass = {astro-ph.CO},
       adsurl = {https://ui.adsabs.harvard.edu/abs/2021ApJ...917L..37C}
}

@ARTICLE{Nasir2021,
       author = {{Nasir}, Fahad and {Cain}, Christopher and {D'Aloisio}, Anson and {Gangolli}, Nakul and {McQuinn}, Matthew},
        title = "{Hydrodynamic Response of the Intergalactic Medium to Reionization. II. Physical Characteristics and Dynamics of Ionizing Photon Sinks}",
      journal = {\apj},
         year = 2021,
        month = dec,
       volume = {923},
       number = {2},
          eid = {161},
        pages = {161},
          doi = {10.3847/1538-4357/ac2eb9},
archivePrefix = {arXiv},
       eprint = {2108.04837},
 primaryClass = {astro-ph.CO},
       adsurl = {https://ui.adsabs.harvard.edu/abs/2021ApJ...923..161N}
}

@ARTICLE{Cain2022b,
       author = {{Cain}, Christopher and {D'Aloisio}, Anson and {Gangolli}, Nakul and {McQuinn}, Matthew},
        title = "{The morphology of reionization in a dynamically clumpy universe}",
      journal = {\mnras},
         year = 2023,
        month = jun,
       volume = {522},
       number = {2},
        pages = {2047-2064},
          doi = {10.1093/mnras/stad1057},
archivePrefix = {arXiv},
       eprint = {2207.11266},
 primaryClass = {astro-ph.CO},
       adsurl = {https://ui.adsabs.harvard.edu/abs/2023MNRAS.522.2047C}
}

@article{McQuinn2011,
	doi = {10.1088/0004-637x/743/1/82},
	url = {https://doi.org/10.1088\%2F0004-637x\%2F743\%2F1\%2F82},
	year = {2011},
	month = {nov},
	publisher = {{IOP} Publishing},
	volume = {743},
	number = {1},
	pages = {82},
	author = {Matthew McQuinn and S. Peng Oh and Claude-Andr{\'{e}} Faucher-Gigu{\`{e}}re},
	title = {{ON} {LYMAN}-{LIMIT} {SYSTEMS} {AND} {THE} {EVOLUTION} {OF} {THE} {INTERGALACTIC} {IONIZING} {BACKGROUND}},
	journal = {The Astrophysical Journal},
}

@article{Emberson2013,
	doi = {10.1088/0004-637x/763/2/146},
	url = {https://doi.org/10.1088\%2F0004-637x\%2F763\%2F2\%2F146},
	year = {2013},
	month = {jan},
	publisher = {{IOP} Publishing},
	volume = {763},
	number = {2},
	pages = {146},
	author = {J. D. Emberson and Rajat M. Thomas and Marcelo A. Alvarez},
	title = {{THE} {OPACITY} {OF} {THE} {INTERGALACTIC} {MEDIUM} {DURING} {REIONIZATION}: {RESOLVING} {SMALL}-{SCALE} {STRUCTURE}},
	journal = {The Astrophysical Journal},
}

@ARTICLE{Reichardt2020,
       author = {{Reichardt}, C.~L. and {Patil}, S. and {Ade}, P.~A.~R. and {Anderson}, A.~J. and {Austermann}, J.~E. and {Avva}, J.~S. and {Baxter}, E. and {Beall}, J.~A. and {Bender}, A.~N. and {Benson}, B.~A. and {Bianchini}, F. and {Bleem}, L.~E. and {Carlstrom}, J.~E. and {Chang}, C.~L. and {Chaubal}, P. and {Chiang}, H.~C. and {Chou}, T.~L. and {Citron}, R. and {Moran}, C. Corbett and {Crawford}, T.~M. and {Crites}, A.~T. and {de Haan}, T. and {Dobbs}, M.~A. and {Everett}, W. and {Gallicchio}, J. and {George}, E.~M. and {Gilbert}, A. and {Gupta}, N. and {Halverson}, N.~W. and {Harrington}, N. and {Henning}, J.~W. and {Hilton}, G.~C. and {Holder}, G.~P. and {Holzapfel}, W.~L. and {Hrubes}, J.~D. and {Huang}, N. and {Hubmayr}, J. and {Irwin}, K.~D. and {Knox}, L. and {Lee}, A.~T. and {Li}, D. and {Lowitz}, A. and {Luong-Van}, D. and {McMahon}, J.~J. and {Mehl}, J. and {Meyer}, S.~S. and {Millea}, M. and {Mocanu}, L.~M. and {Mohr}, J.~J. and {Montgomery}, J. and {Nadolski}, A. and {Natoli}, T. and {Nibarger}, J.~P. and {Noble}, G. and {Novosad}, V. and {Omori}, Y. and {Padin}, S. and {Pryke}, C. and {Ruhl}, J.~E. and {Saliwanchik}, B.~R. and {Sayre}, J.~T. and {Schaffer}, K.~K. and {Shirokoff}, E. and {Sievers}, C. and {Smecher}, G. and {Spieler}, H.~G. and {Staniszewski}, Z. and {Stark}, A.~A. and {Tucker}, C. and {Vanderlinde}, K. and {Veach}, T. and {Vieira}, J.~D. and {Wang}, G. and {Whitehorn}, N. and {Williamson}, R. and {Wu}, W.~L.~K. and {Yefremenko}, V.},
        title = "{An Improved Measurement of the Secondary Cosmic Microwave Background Anisotropies from the SPT-SZ + SPTpol Surveys}",
      journal = {\apj},
         year = 2021,
        month = feb,
       volume = {908},
       number = {2},
          eid = {199},
        pages = {199},
          doi = {10.3847/1538-4357/abd407},
archivePrefix = {arXiv},
       eprint = {2002.06197},
 primaryClass = {astro-ph.CO},
       adsurl = {https://ui.adsabs.harvard.edu/abs/2021ApJ...908..199R}
}

@article{MiraldaEscude2000,
	doi = {10.1086/308330},
	url = {https://doi.org/10.1086\%2F308330},
	year = 2000,
	month = {feb},
	publisher = {{IOP} Publishing},
	volume = {530},
	number = {1},
	pages = {1--16},
	author = {Jordi Miralda-Escude and Martin Haehnelt and Martin J. Rees},
	title = {Reionization of the Inhomogeneous Universe},
	journal = {The Astrophysical Journal},
}

@article{Furlanetto2005,
    author = {Furlanetto, Steven R. and Oh, S. Peng},
    title = "{Taxing the rich: recombinations and bubble growth during reionization}",
    journal = {Monthly Notices of the Royal Astronomical Society},
    volume = {363},
    number = {3},
    pages = {1031-1048},
    year = {2005},
    month = {11},
    issn = {0035-8711},
    doi = {10.1111/j.1365-2966.2005.09505.x},
    url = {https://doi.org/10.1111/j.1365-2966.2005.09505.x},
    eprint = {http://oup.prod.sis.lan/mnras/article-pdf/363/3/1031/3943318/363-3-1031.pdf},
}

@ARTICLE{Furlanetto2004,
   author = {{Furlanetto}, S.~R. and {Zaldarriaga}, M. and {Hernquist}, L.
	},
    title = "{The Growth of H II Regions During Reionization}",
  journal = {The Astrophysical Journal},
   eprint = {astro-ph/0403697},
     year = 2004,
    month = sep,
   volume = 613,
    pages = {1-15},
      doi = {10.1086/423025},
   url = {https://ui.adsabs.harvard.edu/abs/2004ApJ...613....1F}
}

@ARTICLE{Gnedin2014,
       author = {{Gnedin}, Nickolay Y.},
        title = "{Cosmic Reionization on Computers. I. Design and Calibration of Simulations}",
      journal = {\apj},
         year = 2014,
        month = sep,
       volume = {793},
       number = {1},
          eid = {29},
        pages = {29},
          doi = {10.1088/0004-637X/793/1/29},
archivePrefix = {arXiv},
       eprint = {1403.4245},
 primaryClass = {astro-ph.CO},
       adsurl = {https://ui.adsabs.harvard.edu/abs/2014ApJ...793...29G}
}

@article{Shapiro2004,
    author = {Shapiro, Paul R. and Iliev, Ilian T. and Raga, Alejandro C.},
    title = "{Photoevaporation of cosmological minihaloes during reionization}",
    journal = {Monthly Notices of the Royal Astronomical Society},
    volume = {348},
    number = {3},
    pages = {753-782},
    year = {2004},
    month = {03},
    issn = {0035-8711},
    doi = {10.1111/j.1365-2966.2004.07364.x},
    url = {https://doi.org/10.1111/j.1365-2966.2004.07364.x},
    eprint = {http://oup.prod.sis.lan/mnras/article-pdf/348/3/753/4103465/348-3-753.pdf},
}

@article{Iliev2005,
    author = {Iliev, Ilian T. and Shapiro, Paul R. and Raga, Alejandro C.},
    title = "{Minihalo photoevaporation during cosmic reionization: evaporation times and photon consumption rates}",
    journal = {Monthly Notices of the Royal Astronomical Society},
    volume = {361},
    number = {2},
    pages = {405-414},
    year = {2005},
    month = {08},
    issn = {0035-8711},
    doi = {10.1111/j.1365-2966.2005.09155.x},
    url = {https://doi.org/10.1111/j.1365-2966.2005.09155.x},
    eprint = {http://oup.prod.sis.lan/mnras/article-pdf/361/2/405/18655918/361-2-405.pdf},
}

@ARTICLE{Becker2015,
   author = {{Becker}, G.~D. and {Bolton}, J.~S. and {Madau}, P. and {Pettini}, M. and 
	{Ryan-Weber}, E.~V. and {Venemans}, B.~P.},
    title = "{Evidence of patchy hydrogen reionization from an extreme Ly{$\alpha$} trough below redshift six}",
  journal = {MNRAS},
archivePrefix = "arXiv",
   eprint = {1407.4850},
     year = 2015,
    month = mar,
   volume = 447,
    pages = {3402-3419},
      doi = {10.1093/mnras/stu2646},
   url = {https://ui.adsabs.harvard.edu/abs/2015MNRAS.447.3402B}
}

@INPROCEEDINGS{Pawlik2009,
       author = {{Pawlik}, A.~H. and {Schaye}, J. and {van Scherpenzeel}, E.},
        title = "{Keeping the Universe Ionised: Photoheating and the High-redshift Clumping Factor of the Intergalactic Gas}",
    booktitle = {New Horizons in Astronomy: Frank N. Bash Symposium 2009},
         year = "2010",
       editor = {{Stanford}, L.~M. and {Green}, J.~D. and {Hao}, L. and {Mao}, Y.},
       series = {Astronomical Society of the Pacific Conference Series},
       volume = {432},
        month = "Oct",
        pages = {230},
archivePrefix = {arXiv},
       eprint = {0912.3034},
 primaryClass = {astro-ph.CO},
       url = {https://ui.adsabs.harvard.edu/abs/2010ASPC..432..230P}
}

@ARTICLE{Chardin2018,
       author = {{Chardin}, Jonathan and {Kulkarni}, Girish and {Haehnelt}, Martin G.},
        title = "{Self-shielding of hydrogen in the IGM during the epoch of reionization}",
      journal = {MNRAS},
         year = "2018",
        month = "Jul",
       volume = {478},
       number = {1},
        pages = {1065-1076},
          doi = {10.1093/mnras/sty992},
archivePrefix = {arXiv},
       eprint = {1707.06993},
 primaryClass = {astro-ph.CO},
       url = {https://ui.adsabs.harvard.edu/abs/2018MNRAS.478.1065C}
}

@ARTICLE{Park2016,
       author = {{Park}, Hyunbae and {Shapiro}, Paul R. and {Choi}, Jun-hwan and
         {Yoshida}, Naoki and {Hirano}, Shingo and {Ahn}, Kyungjin},
        title = "{The Hydrodynamic Feedback of Cosmic Reionization on Small-scale Structures and Its Impact on Photon Consumption During the Epoch of Reionization}",
      journal = {ApJ},
         year = "2016",
        month = "Nov",
       volume = {831},
       number = {1},
          eid = {86},
        pages = {86},
          doi = {10.3847/0004-637X/831/1/86},
archivePrefix = {arXiv},
       eprint = {1602.06472},
 primaryClass = {astro-ph.CO},
       url = {https://ui.adsabs.harvard.edu/abs/2016ApJ...831...86P}
}

@ARTICLE{Keating2019,
       author = {{Keating}, Laura C. and {Weinberger}, Lewis H. and {Kulkarni}, Girish and {Haehnelt}, Martin G. and {Chardin}, Jonathan and {Aubert}, Dominique},
        title = "{Long troughs in the Lyman-{\ensuremath{\alpha}} forest below redshift 6 due to islands of neutral hydrogen}",
      journal = {\mnras},
         year = 2020,
        month = jan,
       volume = {491},
       number = {2},
        pages = {1736-1745},
          doi = {10.1093/mnras/stz3083},
archivePrefix = {arXiv},
       eprint = {1905.12640},
 primaryClass = {astro-ph.CO},
       adsurl = {https://ui.adsabs.harvard.edu/abs/2020MNRAS.491.1736K}
}

@ARTICLE{Fan2006,
	author = {{Fan}, Xiaohui and {Strauss}, Michael A. and {Becker}, Robert H. and
	{White}, Richard L. and {Gunn}, James E. and {Knapp}, Gillian R. and
	{Richards}, Gordon T. and {Schneider}, Donald P. and {Brinkmann}, J. and
	{Fukugita}, Masataka},
	title = "{Constraining the Evolution of the Ionizing Background and the Epoch of Reionization with z\raisebox{-0.5ex}\textasciitilde6 Quasars. II. A Sample of 19 Quasars}",
	journal = {\aj},
	year = "2006",
	month = "Jul",
	volume = {132},
	number = {1},
	pages = {117-136},
	doi = {10.1086/504836},
	archivePrefix = {arXiv},
	eprint = {astro-ph/0512082},
	primaryClass = {astro-ph},
	url = {https://ui.adsabs.harvard.edu/abs/2006AJ....132..117F}
}

@ARTICLE{Planck2018,
       author = {{Planck Collaboration} and {Aghanim}, N. and {Akrami}, Y. and {Ashdown}, M. and {Aumont}, J. and {Baccigalupi}, C. and {Ballardini}, M. and {Banday}, A.~J. and {Barreiro}, R.~B. and {Bartolo}, N. and {Basak}, S. and {Battye}, R. and {Benabed}, K. and {Bernard}, J. -P. and {Bersanelli}, M. and {Bielewicz}, P. and {Bock}, J.~J. and {Bond}, J.~R. and {Borrill}, J. and {Bouchet}, F.~R. and {Boulanger}, F. and {Bucher}, M. and {Burigana}, C. and {Butler}, R.~C. and {Calabrese}, E. and {Cardoso}, J. -F. and {Carron}, J. and {Challinor}, A. and {Chiang}, H.~C. and {Chluba}, J. and {Colombo}, L.~P.~L. and {Combet}, C. and {Contreras}, D. and {Crill}, B.~P. and {Cuttaia}, F. and {de Bernardis}, P. and {de Zotti}, G. and {Delabrouille}, J. and {Delouis}, J. -M. and {Di Valentino}, E. and {Diego}, J.~M. and {Dor{\'e}}, O. and {Douspis}, M. and {Ducout}, A. and {Dupac}, X. and {Dusini}, S. and {Efstathiou}, G. and {Elsner}, F. and {En{\ss}lin}, T.~A. and {Eriksen}, H.~K. and {Fantaye}, Y. and {Farhang}, M. and {Fergusson}, J. and {Fernandez-Cobos}, R. and {Finelli}, F. and {Forastieri}, F. and {Frailis}, M. and {Fraisse}, A.~A. and {Franceschi}, E. and {Frolov}, A. and {Galeotta}, S. and {Galli}, S. and {Ganga}, K. and {G{\'e}nova-Santos}, R.~T. and {Gerbino}, M. and {Ghosh}, T. and {Gonz{\'a}lez-Nuevo}, J. and {G{\'o}rski}, K.~M. and {Gratton}, S. and {Gruppuso}, A. and {Gudmundsson}, J.~E. and {Hamann}, J. and {Handley}, W. and {Hansen}, F.~K. and {Herranz}, D. and {Hildebrandt}, S.~R. and {Hivon}, E. and {Huang}, Z. and {Jaffe}, A.~H. and {Jones}, W.~C. and {Karakci}, A. and {Keih{\"a}nen}, E. and {Keskitalo}, R. and {Kiiveri}, K. and {Kim}, J. and {Kisner}, T.~S. and {Knox}, L. and {Krachmalnicoff}, N. and {Kunz}, M. and {Kurki-Suonio}, H. and {Lagache}, G. and {Lamarre}, J. -M. and {Lasenby}, A. and {Lattanzi}, M. and {Lawrence}, C.~R. and {Le Jeune}, M. and {Lemos}, P. and {Lesgourgues}, J. and {Levrier}, F. and {Lewis}, A. and {Liguori}, M. and {Lilje}, P.~B. and {Lilley}, M. and {Lindholm}, V. and {L{\'o}pez-Caniego}, M. and {Lubin}, P.~M. and {Ma}, Y. -Z. and {Mac{\'\i}as-P{\'e}rez}, J.~F. and {Maggio}, G. and {Maino}, D. and {Mandolesi}, N. and {Mangilli}, A. and {Marcos-Caballero}, A. and {Maris}, M. and {Martin}, P.~G. and {Martinelli}, M. and {Mart{\'\i}nez-Gonz{\'a}lez}, E. and {Matarrese}, S. and {Mauri}, N. and {McEwen}, J.~D. and {Meinhold}, P.~R. and {Melchiorri}, A. and {Mennella}, A. and {Migliaccio}, M. and {Millea}, M. and {Mitra}, S. and {Miville-Desch{\^e}nes}, M. -A. and {Molinari}, D. and {Montier}, L. and {Morgante}, G. and {Moss}, A. and {Natoli}, P. and {N{\o}rgaard-Nielsen}, H.~U. and {Pagano}, L. and {Paoletti}, D. and {Partridge}, B. and {Patanchon}, G. and {Peiris}, H.~V. and {Perrotta}, F. and {Pettorino}, V. and {Piacentini}, F. and {Polastri}, L. and {Polenta}, G. and {Puget}, J. -L. and {Rachen}, J.~P. and {Reinecke}, M. and {Remazeilles}, M. and {Renzi}, A. and {Rocha}, G. and {Rosset}, C. and {Roudier}, G. and {Rubi{\~n}o-Mart{\'\i}n}, J.~A. and {Ruiz-Granados}, B. and {Salvati}, L. and {Sandri}, M. and {Savelainen}, M. and {Scott}, D. and {Shellard}, E.~P.~S. and {Sirignano}, C. and {Sirri}, G. and {Spencer}, L.~D. and {Sunyaev}, R. and {Suur-Uski}, A. -S. and {Tauber}, J.~A. and {Tavagnacco}, D. and {Tenti}, M. and {Toffolatti}, L. and {Tomasi}, M. and {Trombetti}, T. and {Valenziano}, L. and {Valiviita}, J. and {Van Tent}, B. and {Vibert}, L. and {Vielva}, P. and {Villa}, F. and {Vittorio}, N. and {Wandelt}, B.~D. and {Wehus}, I.~K. and {White}, M. and {White}, S.~D.~M. and {Zacchei}, A. and {Zonca}, A.},
        title = "{Planck 2018 results. VI. Cosmological parameters}",
      journal = {\aap},
         year = 2020,
        month = sep,
       volume = {641},
          eid = {A6},
        pages = {A6},
          doi = {10.1051/0004-6361/201833910},
archivePrefix = {arXiv},
       eprint = {1807.06209},
 primaryClass = {astro-ph.CO},
       adsurl = {https://ui.adsabs.harvard.edu/abs/2020A&A...641A...6P}
}

@ARTICLE{McQuinn2018,
	author = {{McQuinn}, Matthew and {D'Aloisio}, Anson},
	title = "{The observable 21cm signal from reionization may be perturbative}",
	journal = {JCAP},
	year = "2018",
	month = "Oct",
	volume = {2018},
	number = {10},
	eid = {016},
	pages = {016},
	doi = {10.1088/1475-7516/2018/10/016},
	archivePrefix = {arXiv},
	eprint = {1806.08372},
	primaryClass = {astro-ph.CO},
	url = {https://ui.adsabs.harvard.edu/abs/2018JCAP...10..016M}
}

@ARTICLE{Robertson2015,
       author = {{Robertson}, Brant E. and {Ellis}, Richard S. and
         {Furlanetto}, Steven R. and {Dunlop}, James S.},
        title = "{Cosmic Reionization and Early Star-forming Galaxies: A Joint Analysis of New Constraints from Planck and the Hubble Space Telescope}",
      journal = {\apjl},
         year = "2015",
        month = "Apr",
       volume = {802},
       number = {2},
          eid = {L19},
        pages = {L19},
          doi = {10.1088/2041-8205/802/2/L19},
archivePrefix = {arXiv},
       eprint = {1502.02024},
 primaryClass = {astro-ph.CO},
       url = {https://ui.adsabs.harvard.edu/abs/2015ApJ...802L..19R}
}

@article{Nasir2020,
   title={Observing the tail of reionization: neutral islands in the z - 5.5 Lyman-$\alpha$ forest},
   volume={494},
   ISSN={1365-2966},
   url={http://dx.doi.org/10.1093/mnras/staa894},
   DOI={10.1093/mnras/staa894},
   number={3},
   journal={Monthly Notices of the Royal Astronomical Society},
   publisher={Oxford University Press (OUP)},
   author={Nasir, Fahad and D’Aloisio, Anson},
   year={2020},
   month={Apr},
   pages={3080–3094}
}

@ARTICLE{Davies2016,
       author = {{Davies}, Frederick B. and {Furlanetto}, Steven R.},
        title = "{Large fluctuations in the hydrogen-ionizing background and mean free path following the epoch of reionization}",
      journal = {\mnras},
         year = "2016",
        month = "Aug",
       volume = {460},
       number = {2},
        pages = {1328-1339},
          doi = {10.1093/mnras/stw931},
archivePrefix = {arXiv},
       eprint = {1509.07131},
 primaryClass = {astro-ph.CO},
       url = {https://ui.adsabs.harvard.edu/abs/2016MNRAS.460.1328D}
}

@article{Davies2018,
	doi = {10.3847/1538-4357/aad6dc},
	url = {https://doi.org/10.3847%2F1538-4357%2Faad6dc},
	year = 2018,
	month = {sep},
	publisher = {American Astronomical Society},
	volume = {864},
	number = {2},
	pages = {142},
	author = {Frederick B. Davies and Joseph F. Hennawi and Eduardo Ba{\~{n}}ados and Zarija Luki{\'{c}} and Roberto Decarli and Xiaohui Fan and Emanuele P. Farina and Chiara Mazzucchelli and Hans-Walter Rix and Bram P. Venemans and Fabian Walter and Feige Wang and Jinyi Yang},
	title = {Quantitative Constraints on the Reionization History from the {IGM} Damping Wing Signature in Two Quasars at $z > 7$},
	journal = {The Astrophysical Journal}
}

@ARTICLE{Iliev2006,
       author = {{Iliev}, Ilian T. and {Ciardi}, Benedetta and {Alvarez}, Marcelo A. and
         {Maselli}, Antonella and {Ferrara}, Andrea and {Gnedin}, Nickolay Y. and
         {Mellema}, Garrelt and {Nakamoto}, Taishi and {Norman}, Michael L. and
         {Razoumov}, Alexei O. and {Rijkhorst}, Erik-Jan and
         {Ritzerveld}, Jelle and {Shapiro}, Paul R. and {Susa}, Hajime and
         {Umemura}, Masayuki and {Whalen}, Daniel J.},
        title = "{Cosmological radiative transfer codes comparison project - I. The static density field tests}",
      journal = {\mnras},
         year = "2006",
        month = "Sep",
       volume = {371},
       number = {3},
        pages = {1057-1086},
          doi = {10.1111/j.1365-2966.2006.10775.x},
archivePrefix = {arXiv},
       eprint = {astro-ph/0603199},
 primaryClass = {astro-ph},
       url = {https://ui.adsabs.harvard.edu/abs/2006MNRAS.371.1057I}
}

@ARTICLE{DAloisio2018,
       author = {{D'Aloisio}, Anson and {McQuinn}, Matthew and {Davies}, Frederick B. and
         {Furlanetto}, Steven R.},
        title = "{Large fluctuations in the high-redshift metagalactic ionizing background}",
      journal = {\mnras},
         year = "2018",
        month = "Jan",
       volume = {473},
       number = {1},
        pages = {560-575},
          doi = {10.1093/mnras/stx2341},
archivePrefix = {arXiv},
       eprint = {1611.02711},
 primaryClass = {astro-ph.CO},
       url = {https://ui.adsabs.harvard.edu/abs/2018MNRAS.473..560D}
}

@ARTICLE{Madau1999,
       author = {{Madau}, Piero and {Haardt}, Francesco and {Rees}, Martin J.},
        title = "{Radiative Transfer in a Clumpy Universe. III. The Nature of Cosmological Ionizing Sources}",
      journal = {\apj},
         year = 1999,
        month = apr,
       volume = {514},
       number = {2},
        pages = {648-659},
          doi = {10.1086/306975},
archivePrefix = {arXiv},
       eprint = {astro-ph/9809058},
 primaryClass = {astro-ph},
       url = {https://ui.adsabs.harvard.edu/abs/1999ApJ...514..648M}
}

@ARTICLE{Furlanetto2006,
       author = {{Furlanetto}, Steven R. and {Oh}, S. Peng and {Briggs}, Frank H.},
        title = "{Cosmology at low frequencies: The 21 cm transition and the high-redshift Universe}",
      journal = {\physrep},
         year = 2006,
        month = oct,
       volume = {433},
       number = {4-6},
        pages = {181-301},
          doi = {10.1016/j.physrep.2006.08.002},
archivePrefix = {arXiv},
       eprint = {astro-ph/0608032},
 primaryClass = {astro-ph},
       url = {https://ui.adsabs.harvard.edu/abs/2006PhR...433..181F}
}

@ARTICLE{Shimabukuro2020,
       author = {{Shimabukuro}, Hayato and {Mao}, Yi and {Tan}, Jianrong},
        title = "{Estimation of H II Bubble Size Distribution from 21 cm Power Spectrum with Artificial Neural Networks}",
      journal = {Research in Astronomy and Astrophysics},
         year = 2022,
        month = mar,
       volume = {22},
       number = {3},
          eid = {035027},
        pages = {035027},
          doi = {10.1088/1674-4527/ac4ca3},
archivePrefix = {arXiv},
       eprint = {2002.08238},
 primaryClass = {astro-ph.CO},
       adsurl = {https://ui.adsabs.harvard.edu/abs/2022RAA....22c5027S}
}

@ARTICLE{Kaur2020,
       author = {{Kaur}, Harman Deep and {Gillet}, Nicolas and {Mesinger}, Andrei},
        title = "{Minimum size of 21-cm simulations}",
      journal = {\mnras},
         year = 2020,
        month = jun,
       volume = {495},
       number = {2},
        pages = {2354-2362},
          doi = {10.1093/mnras/staa1323},
archivePrefix = {arXiv},
       eprint = {2004.06709},
 primaryClass = {astro-ph.CO},
       adsurl = {https://ui.adsabs.harvard.edu/abs/2020MNRAS.495.2354K}
}

@ARTICLE{Bosman2018,
       author = {{Bosman}, Sarah E.~I. and {Fan}, Xiaohui and {Jiang}, Linhua and
         {Reed}, Sophie and {Matsuoka}, Yoshiki and {Becker}, George and
         {Haehnelt}, Martin},
        title = "{New constraints on Lyman-{\ensuremath{\alpha}} opacity with a sample of 62 quasarsat z \&gt; 5.7}",
      journal = {\mnras},
         year = 2018,
        month = sep,
       volume = {479},
       number = {1},
        pages = {1055-1076},
          doi = {10.1093/mnras/sty1344},
archivePrefix = {arXiv},
       eprint = {1802.08177},
 primaryClass = {astro-ph.GA},
       url = {https://ui.adsabs.harvard.edu/abs/2018MNRAS.479.1055B}
}

@ARTICLE{Chen2025,
       author = {{Chen}, Huanqing and {Fan}, Jiawen and {Avestruz}, Camille},
        title = "{Cosmic Reionization On Computers: Biases and Uncertainties in the Measured Mean Free Path at the End Stage of Reionization}",
      journal = {The Open Journal of Astrophysics},
         year = 2025,
        month = jun,
       volume = {8},
          eid = {81},
        pages = {81},
          doi = {10.33232/001c.141394},
archivePrefix = {arXiv},
       eprint = {2410.05372},
 primaryClass = {astro-ph.CO},
       adsurl = {https://ui.adsabs.harvard.edu/abs/2025OJAp....8E..81C}
}

@ARTICLE{McQuinn2005,
       author = {{McQuinn}, Matthew and {Furlanetto}, Steven R. and {Hernquist}, Lars and
         {Zahn}, Oliver and {Zaldarriaga}, Matias},
        title = "{The Kinetic Sunyaev-Zel'dovich Effect from Reionization}",
      journal = {\apj},
         year = 2005,
        month = sep,
       volume = {630},
       number = {2},
        pages = {643-656},
          doi = {10.1086/432049},
archivePrefix = {arXiv},
       eprint = {astro-ph/0504189},
 primaryClass = {astro-ph},
       adsurl = {https://ui.adsabs.harvard.edu/abs/2005ApJ...630..643M}
}

@ARTICLE{Ocvirk2016,
       author = {{Ocvirk}, Pierre and {Gillet}, Nicolas and {Shapiro}, Paul R. and
         {Aubert}, Dominique and {Iliev}, Ilian T. and {Teyssier}, Romain and
         {Yepes}, Gustavo and {Choi}, Jun-Hwan and {Sullivan}, David and
         {Knebe}, Alexander and {Gottl{\"o}ber}, Stefan and {D'Aloisio}, Anson and
         {Park}, Hyunbae and {Hoffman}, Yehuda and {Stranex}, Timothy},
        title = "{Cosmic Dawn (CoDa): the First Radiation-Hydrodynamics Simulation of Reionization and Galaxy Formation in the Local Universe}",
      journal = {\mnras},
         year = 2016,
        month = dec,
       volume = {463},
       number = {2},
        pages = {1462-1485},
          doi = {10.1093/mnras/stw2036},
archivePrefix = {arXiv},
       eprint = {1511.00011},
 primaryClass = {astro-ph.GA},
       adsurl = {https://ui.adsabs.harvard.edu/abs/2016MNRAS.463.1462O}
}

@ARTICLE{Kulkarni2019,
       author = {{Kulkarni}, Girish and {Keating}, Laura C. and {Haehnelt}, Martin G. and
         {Bosman}, Sarah E.~I. and {Puchwein}, Ewald and {Chardin}, Jonathan and
         {Aubert}, Dominique},
        title = "{Large Ly {\ensuremath{\alpha}} opacity fluctuations and low CMB {\ensuremath{\tau}} in models of late reionization with large islands of neutral hydrogen extending to z < 5.5}",
      journal = {\mnras},
         year = 2019,
        month = may,
       volume = {485},
       number = {1},
        pages = {L24-L28},
          doi = {10.1093/mnrasl/slz025},
archivePrefix = {arXiv},
       eprint = {1809.06374},
 primaryClass = {astro-ph.CO},
       adsurl = {https://ui.adsabs.harvard.edu/abs/2019MNRAS.485L..24K}
}

@ARTICLE{Mason2018,
       author = {{Mason}, Charlotte A. and {Treu}, Tommaso and {de Barros}, Stephane and
         {Dijkstra}, Mark and {Fontana}, Adriano and {Mesinger}, Andrei and
         {Pentericci}, Laura and {Trenti}, Michele and {Vanzella}, Eros},
        title = "{Beacons into the Cosmic Dark Ages: Boosted Transmission of Ly{\ensuremath{\alpha}} from UV Bright Galaxies at z {\ensuremath{\gtrsim}} 7}",
      journal = {\apjl},
         year = 2018,
        month = apr,
       volume = {857},
       number = {2},
          eid = {L11},
        pages = {L11},
          doi = {10.3847/2041-8213/aabbab},
archivePrefix = {arXiv},
       eprint = {1801.01891},
 primaryClass = {astro-ph.CO},
       adsurl = {https://ui.adsabs.harvard.edu/abs/2018ApJ...857L..11M}
}

@ARTICLE{McGreer2015,
       author = {{McGreer}, Ian D. and {Mesinger}, Andrei and {D'Odorico}, Valentina},
        title = "{Model-independent evidence in favour of an end to reionization by z {\ensuremath{\approx}} 6}",
      journal = {\mnras},
         year = 2015,
        month = feb,
       volume = {447},
       number = {1},
        pages = {499-505},
          doi = {10.1093/mnras/stu2449},
archivePrefix = {arXiv},
       eprint = {1411.5375},
 primaryClass = {astro-ph.CO},
       adsurl = {https://ui.adsabs.harvard.edu/abs/2015MNRAS.447..499M}
}

@ARTICLE{Alvarez2012,
       author = {{Alvarez}, Marcelo A. and {Abel}, Tom},
        title = "{The Effect of Absorption Systems on Cosmic Reionization}",
      journal = {\apj},
         year = 2012,
        month = mar,
       volume = {747},
       number = {2},
          eid = {126},
        pages = {126},
          doi = {10.1088/0004-637X/747/2/126},
archivePrefix = {arXiv},
       eprint = {1003.6132},
 primaryClass = {astro-ph.CO},
       adsurl = {https://ui.adsabs.harvard.edu/abs/2012ApJ...747..126A}
}

@ARTICLE{Wyatt2026,
       author = {{Wyatt}, Michael M. and {Furlanetto}, Steven R. and {Minasyan}, Mary H.},
        title = "{The effect of a short mean free path on HII regions and 21 cm tomography during reionization}",
      journal = {\jcap},
         year = 2026,
        month = mar,
       volume = {2026},
       number = {3},
          eid = {007},
        pages = {007},
          doi = {10.1088/1475-7516/2026/03/007},
archivePrefix = {arXiv},
       eprint = {2510.09750},
 primaryClass = {astro-ph.CO},
       adsurl = {https://ui.adsabs.harvard.edu/abs/2026JCAP...03..007W}
}

@ARTICLE{Shin2008,
       author = {{Shin}, Min-Su and {Trac}, Hy and {Cen}, Renyue},
        title = "{Cosmological H II Bubble Growth during Reionization}",
      journal = {\apj},
         year = 2008,
        month = jul,
       volume = {681},
       number = {2},
        pages = {756-770},
          doi = {10.1086/588247},
archivePrefix = {arXiv},
       eprint = {0708.2425},
 primaryClass = {astro-ph},
       adsurl = {https://ui.adsabs.harvard.edu/abs/2008ApJ...681..756S}
}

@ARTICLE{Lin2016,
       author = {{Lin}, Yin and {Oh}, S. Peng and {Furlanetto}, Steven R. and
         {Sutter}, P.~M.},
        title = "{The distribution of bubble sizes during reionization}",
      journal = {\mnras},
         year = 2016,
        month = sep,
       volume = {461},
       number = {3},
        pages = {3361-3374},
          doi = {10.1093/mnras/stw1542},
archivePrefix = {arXiv},
       eprint = {1511.01506},
 primaryClass = {astro-ph.CO},
       adsurl = {https://ui.adsabs.harvard.edu/abs/2016MNRAS.461.3361L}
}

@ARTICLE{Chen2020,
       author = {{Chen}, Nianyi and {Doussot}, Aristide and {Trac}, Hy and {Cen}, Renyue},
        title = "{SCORCH. III. Analytical Models of Reionization with Varying Clumping Factors}",
      journal = {\apj},
         year = 2020,
        month = dec,
       volume = {905},
       number = {2},
          eid = {132},
        pages = {132},
          doi = {10.3847/1538-4357/abc890},
archivePrefix = {arXiv},
       eprint = {2004.07854},
 primaryClass = {astro-ph.CO},
       adsurl = {https://ui.adsabs.harvard.edu/abs/2020ApJ...905..132C}
}

@ARTICLE{Iliev2014,
       author = {{Iliev}, Ilian T. and {Mellema}, Garrelt and {Ahn}, Kyungjin and
         {Shapiro}, Paul R. and {Mao}, Yi and {Pen}, Ue-Li},
        title = "{Simulating cosmic reionization: how large a volume is large enough?}",
      journal = {\mnras},
         year = 2014,
        month = mar,
       volume = {439},
       number = {1},
        pages = {725-743},
          doi = {10.1093/mnras/stt2497},
archivePrefix = {arXiv},
       eprint = {1310.7463},
 primaryClass = {astro-ph.CO},
       adsurl = {https://ui.adsabs.harvard.edu/abs/2014MNRAS.439..725I}
}

@ARTICLE{Finkelstein2019,
       author = {{Finkelstein}, Steven L. and {D'Aloisio}, Anson and {Paardekooper}, Jan-Pieter and {Ryan}, Russell, Jr. and {Behroozi}, Peter and {Finlator}, Kristian and {Livermore}, Rachael and {Upton Sanderbeck}, Phoebe R. and {Dalla Vecchia}, Claudio and {Khochfar}, Sadegh},
        title = "{Conditions for Reionizing the Universe with a Low Galaxy Ionizing Photon Escape Fraction}",
      journal = {\apj},
         year = 2019,
        month = jul,
       volume = {879},
       number = {1},
          eid = {36},
        pages = {36},
          doi = {10.3847/1538-4357/ab1ea8},
archivePrefix = {arXiv},
       eprint = {1902.02792},
 primaryClass = {astro-ph.CO},
       adsurl = {https://ui.adsabs.harvard.edu/abs/2019ApJ...879...36F}
}

@ARTICLE{Worseck2014,
       author = {{Worseck}, G{\'a}bor and {Prochaska}, J. Xavier and {O'Meara}, John M. and {Becker}, George D. and {Ellison}, Sara L. and {Lopez}, Sebastian and {Meiksin}, Avery and {M{\'e}nard}, Brice and {Murphy}, Michael T. and {Fumagalli}, Michele},
        title = "{The Giant Gemini GMOS survey of z$_{em}$ > 4.4 quasars - I. Measuring the mean free path across cosmic time}",
      journal = {\mnras},
         year = 2014,
        month = dec,
       volume = {445},
       number = {2},
        pages = {1745-1760},
          doi = {10.1093/mnras/stu1827},
archivePrefix = {arXiv},
       eprint = {1402.4154},
 primaryClass = {astro-ph.CO},
       adsurl = {https://ui.adsabs.harvard.edu/abs/2014MNRAS.445.1745W}
}

@ARTICLE{Becker2013,
       author = {{Becker}, George D. and {Bolton}, James S.},
        title = "{New measurements of the ionizing ultraviolet background over 2 < z < 5 and implications for hydrogen reionization}",
      journal = {\mnras},
         year = 2013,
        month = dec,
       volume = {436},
       number = {2},
        pages = {1023-1039},
          doi = {10.1093/mnras/stt1610},
archivePrefix = {arXiv},
       eprint = {1307.2259},
 primaryClass = {astro-ph.CO},
       adsurl = {https://ui.adsabs.harvard.edu/abs/2013MNRAS.436.1023B}
}

@ARTICLE{Boera2019,
       author = {{Boera}, Elisa and {Becker}, George D. and {Bolton}, James S. and {Nasir}, Fahad},
        title = "{Revealing Reionization with the Thermal History of the Intergalactic Medium: New Constraints from the Ly{\ensuremath{\alpha}} Flux Power Spectrum}",
      journal = {\apj},
         year = 2019,
        month = feb,
       volume = {872},
       number = {1},
          eid = {101},
        pages = {101},
          doi = {10.3847/1538-4357/aafee4},
archivePrefix = {arXiv},
       eprint = {1809.06980},
 primaryClass = {astro-ph.CO},
       adsurl = {https://ui.adsabs.harvard.edu/abs/2019ApJ...872..101B}
}

@ARTICLE{Walther2019,
       author = {{Walther}, Michael and {O{\~n}orbe}, Jose and {Hennawi}, Joseph F. and {Luki{\'c}}, Zarija},
        title = "{New Constraints on IGM Thermal Evolution from the Ly{\ensuremath{\alpha}} Forest Power Spectrum}",
      journal = {\apj},
         year = 2019,
        month = feb,
       volume = {872},
       number = {1},
          eid = {13},
        pages = {13},
          doi = {10.3847/1538-4357/aafad1},
archivePrefix = {arXiv},
       eprint = {1808.04367},
 primaryClass = {astro-ph.CO},
       adsurl = {https://ui.adsabs.harvard.edu/abs/2019ApJ...872...13W}
}

@ARTICLE{Becker2011,
       author = {{Becker}, George D. and {Bolton}, James S. and {Haehnelt}, Martin G. and {Sargent}, Wallace L.~W.},
        title = "{Detection of extended He II reionization in the temperature evolution of the intergalactic medium}",
      journal = {\mnras},
         year = 2011,
        month = jan,
       volume = {410},
       number = {2},
        pages = {1096-1112},
          doi = {10.1111/j.1365-2966.2010.17507.x},
archivePrefix = {arXiv},
       eprint = {1008.2622},
 primaryClass = {astro-ph.CO},
       adsurl = {https://ui.adsabs.harvard.edu/abs/2011MNRAS.410.1096B}
}

@ARTICLE{Gaikwad2020,
       author = {{Gaikwad}, Prakash and {Rauch}, Michael and {Haehnelt}, Martin G. and {Puchwein}, Ewald and {Bolton}, James S. and {Keating}, Laura C. and {Kulkarni}, Girish and {Ir{\v{s}}i{\v{c}}}, Vid and {Ba{\~n}ados}, Eduardo and {Becker}, George D. and {Boera}, Elisa and {Zahedy}, Fakhri S. and {Chen}, Hsiao-Wen and {Carswell}, Robert F. and {Chardin}, Jonathan and {Rorai}, Alberto},
        title = "{Probing the thermal state of the intergalactic medium at z > 5 with the transmission spikes in high-resolution Ly {\ensuremath{\alpha}} forest spectra}",
      journal = {\mnras},
         year = 2020,
        month = jun,
       volume = {494},
       number = {4},
        pages = {5091-5109},
          doi = {10.1093/mnras/staa907},
archivePrefix = {arXiv},
       eprint = {2001.10018},
 primaryClass = {astro-ph.CO},
       adsurl = {https://ui.adsabs.harvard.edu/abs/2020MNRAS.494.5091G}
}

@ARTICLE{Gaikwad2023,
       author = {{Gaikwad}, Prakash and {Haehnelt}, Martin G. and {Davies}, Fredrick B. and {Bosman}, Sarah E.~I. and {Molaro}, Margherita and {Kulkarni}, Girish and {D'Odorico}, Valentina and {Becker}, George D. and {Davies}, Rebecca L. and {Nasir}, Fahad and {Bolton}, James S. and {Keating}, Laura C. and {Ir{\v{s}}i{\v{c}}}, Vid and {Puchwein}, Ewald and {Zhu}, Yongda and {Asthana}, Shikhar and {Yang}, Jinyi and {Lai}, Samuel and {Eilers}, Anna-Christina},
        title = "{Measuring the photoionization rate, neutral fraction, and mean free path of H I ionizing photons at 4.9 {\ensuremath{\leq}} z {\ensuremath{\leq}} 6.0 from a large sample of XShooter and ESI spectra}",
      journal = {\mnras},
         year = 2023,
        month = nov,
       volume = {525},
       number = {3},
        pages = {4093-4120},
          doi = {10.1093/mnras/stad2566},
archivePrefix = {arXiv},
       eprint = {2304.02038},
 primaryClass = {astro-ph.CO},
       adsurl = {https://ui.adsabs.harvard.edu/abs/2023MNRAS.525.4093G}
}

@article{Qin2021,
    author = {Qin, Yuxiang and Mesinger, Andrei and Bosman, Sarah E I and Viel, Matteo},
    title = "{Reionization and galaxy inference from the high-redshift Ly$\alpha$ forest}",
    journal = {Monthly Notices of the Royal Astronomical Society},
    volume = {506},
    number = {2},
    pages = {2390-2407},
    year = {2021},
    month = {06},
    issn = {0035-8711},
    doi = {10.1093/mnras/stab1833},
    url = {https://doi.org/10.1093/mnras/stab1833},
    eprint = {https://academic.oup.com/mnras/article-pdf/506/2/2390/39136191/stab1833.pdf},
}

@ARTICLE{Keating2020,
       author = {{Keating}, Laura C. and {Kulkarni}, Girish and {Haehnelt}, Martin G. and {Chardin}, Jonathan and {Aubert}, Dominique},
        title = "{Constraining the second half of reionization with the Ly {\ensuremath{\beta}} forest}",
      journal = {\mnras},
         year = 2020,
        month = sep,
       volume = {497},
       number = {1},
        pages = {906-915},
          doi = {10.1093/mnras/staa1909},
archivePrefix = {arXiv},
       eprint = {1912.05582},
 primaryClass = {astro-ph.CO},
       adsurl = {https://ui.adsabs.harvard.edu/abs/2020MNRAS.497..906K}
}

@ARTICLE{Choudhury2021,
       author = {{Choudhury}, T. Roy and {Paranjape}, Aseem and {Bosman}, Sarah E.~I.},
        title = "{Studying the Lyman {\ensuremath{\alpha}} optical depth fluctuations at z {\ensuremath{\sim}} 5.5 using fast semi-numerical methods}",
      journal = {\mnras},
         year = 2021,
        month = mar,
       volume = {501},
       number = {4},
        pages = {5782-5796},
          doi = {10.1093/mnras/stab045},
archivePrefix = {arXiv},
       eprint = {2003.08958},
 primaryClass = {astro-ph.CO},
       adsurl = {https://ui.adsabs.harvard.edu/abs/2021MNRAS.501.5782C}
}

@ARTICLE{Becker2021,
       author = {{Becker}, George D. and {D'Aloisio}, Anson and {Christenson}, Holly M. and {Zhu}, Yongda and {Worseck}, G{\'a}bor and {Bolton}, James S.},
        title = "{The mean free path of ionizing photons at 5 < z < 6: evidence for rapid evolution near reionization}",
      journal = {\mnras},
         year = 2021,
        month = dec,
       volume = {508},
       number = {2},
        pages = {1853-1869},
          doi = {10.1093/mnras/stab2696},
archivePrefix = {arXiv},
       eprint = {2103.16610},
 primaryClass = {astro-ph.CO},
       adsurl = {https://ui.adsabs.harvard.edu/abs/2021MNRAS.508.1853B}
}

@ARTICLE{Ocvirk2021,
       author = {{Ocvirk}, Pierre and {Lewis}, Joseph S.~W. and {Gillet}, Nicolas and {Chardin}, Jonathan and {Aubert}, Dominique and {Deparis}, Nicolas and {Th{\'e}lie}, {\'E}milie},
        title = "{Lyman-alpha opacities at z = 4-6 require low mass, radiatively-suppressed galaxies to drive cosmic reionization}",
      journal = {\mnras},
         year = 2021,
        month = nov,
       volume = {507},
       number = {4},
        pages = {6108-6117},
          doi = {10.1093/mnras/stab2502},
archivePrefix = {arXiv},
       eprint = {2105.01663},
 primaryClass = {astro-ph.CO},
       adsurl = {https://ui.adsabs.harvard.edu/abs/2021MNRAS.507.6108O}
}

@ARTICLE{Cain2024c,
       author = {{Cain}, Christopher and {D'Aloisio}, Anson},
        title = "{FLEXRT {\textemdash} A fast and flexible cosmological radiative transfer code for reionization studies. Part I. Code validation}",
      journal = {\jcap},
         year = 2024,
        month = dec,
       volume = {2024},
       number = {12},
          eid = {025},
        pages = {025},
          doi = {10.1088/1475-7516/2024/12/025},
archivePrefix = {arXiv},
       eprint = {2409.04521},
 primaryClass = {astro-ph.CO},
       adsurl = {https://ui.adsabs.harvard.edu/abs/2024JCAP...12..025C}
}

@ARTICLE{Davies2021b,
       author = {{Davies}, Frederick B. and {Bosman}, Sarah E.~I. and {Furlanetto}, Steven R. and {Becker}, George D. and {D'Aloisio}, Anson},
        title = "{The Predicament of Absorption-dominated Reionization: Increased Demands on Ionizing Sources}",
      journal = {\apjl},
         year = 2021,
        month = sep,
       volume = {918},
       number = {2},
          eid = {L35},
        pages = {L35},
          doi = {10.3847/2041-8213/ac1ffb},
archivePrefix = {arXiv},
       eprint = {2105.10518},
 primaryClass = {astro-ph.CO},
       adsurl = {https://ui.adsabs.harvard.edu/abs/2021ApJ...918L..35D}
}

@ARTICLE{Prochaska2009,
       author = {{Prochaska}, J. Xavier and {Worseck}, Gabor and {O'Meara}, John M.},
        title = "{A Direct Measurement of the Intergalactic Medium Opacity to H I Ionizing Photons}",
      journal = {\apjl},
         year = 2009,
        month = nov,
       volume = {705},
       number = {2},
        pages = {L113-L117},
          doi = {10.1088/0004-637X/705/2/L113},
archivePrefix = {arXiv},
       eprint = {0910.0009},
 primaryClass = {astro-ph.CO},
       adsurl = {https://ui.adsabs.harvard.edu/abs/2009ApJ...705L.113P}
}

@ARTICLE{Wu2021,
       author = {{Wu}, Xiaohan and {McQuinn}, Matthew and {Eisenstein}, Daniel},
        title = "{On the accuracy of common moment-based radiative transfer methods for simulating reionization}",
      journal = {\jcap},
         year = 2021,
        month = feb,
       volume = {2021},
       number = {2},
          eid = {042},
        pages = {042},
          doi = {10.1088/1475-7516/2021/02/042},
archivePrefix = {arXiv},
       eprint = {2009.07278},
 primaryClass = {astro-ph.CO},
       adsurl = {https://ui.adsabs.harvard.edu/abs/2021JCAP...02..042W}
}

@ARTICLE{HERA2021a,
       author = {{Abdurashidova}, Zara and {Aguirre}, James E. and {Alexander}, Paul and {Ali}, Zaki S. and {Balfour}, Yanga and {Beardsley}, Adam P. and {Bernardi}, Gianni and {Billings}, Tashalee S. and {Bowman}, Judd D. and {Bradley}, Richard F. and {Bull}, Philip and {Burba}, Jacob and {Carey}, Steve and {Carilli}, Chris L. and {Cheng}, Carina and {DeBoer}, David R. and {Dexter}, Matt and {de Lera Acedo}, Eloy and {Dibblee-Barkman}, Taylor and {Dillon}, Joshua S. and {Ely}, John and {Ewall-Wice}, Aaron and {Fagnoni}, Nicolas and {Fritz}, Randall and {Furlanetto}, Steven R. and {Gale-Sides}, Kingsley and {Glendenning}, Brian and {Gorthi}, Deepthi and {Greig}, Bradley and {Grobbelaar}, Jasper and {Halday}, Ziyaad and {Hazelton}, Bryna J. and {Hewitt}, Jacqueline N. and {Hickish}, Jack and {Jacobs}, Daniel C. and {Julius}, Austin and {Kern}, Nicholas S. and {Kerrigan}, Joshua and {Kittiwisit}, Piyanat and {Kohn}, Saul A. and {Kolopanis}, Matthew and {Lanman}, Adam and {La Plante}, Paul and {Lekalake}, Telalo and {Lewis}, David and {Liu}, Adrian and {MacMahon}, David and {Malan}, Lourence and {Malgas}, Cresshim and {Maree}, Matthys and {Martinot}, Zachary E. and {Matsetela}, Eunice and {Mesinger}, Andrei and {Molewa}, Mathakane and {Morales}, Miguel F. and {Mosiane}, Tshegofalang and {Murray}, Steven G. and {Neben}, Abraham R. and {Nikolic}, Bojan and {Nunhokee}, Chuneeta D. and {Parsons}, Aaron R. and {Patra}, Nipanjana and {Pascua}, Robert and {Pieterse}, Samantha and {Pober}, Jonathan C. and {Razavi-Ghods}, Nima and {Ringuette}, Jon and {Robnett}, James and {Rosie}, Kathryn and {Sims}, Peter and {Singh}, Saurabh and {Smith}, Craig and {Syce}, Angelo and {Thyagarajan}, Nithyanandan and {Williams}, Peter K.~G. and {Zheng}, Haoxuan and {HERA Collaboration}},
        title = "{First Results from HERA Phase I: Upper Limits on the Epoch of Reionization 21 cm Power Spectrum}",
      journal = {\apj},
         year = 2022,
        month = feb,
       volume = {925},
       number = {2},
          eid = {221},
        pages = {221},
          doi = {10.3847/1538-4357/ac1c78},
archivePrefix = {arXiv},
       eprint = {2108.02263},
 primaryClass = {astro-ph.CO},
       adsurl = {https://ui.adsabs.harvard.edu/abs/2022ApJ...925..221A}
}

@ARTICLE{HERA2021b,
       author = {{Abdurashidova}, Zara and {Aguirre}, James E. and {Alexander}, Paul and {Ali}, Zaki S. and {Balfour}, Yanga and {Barkana}, Rennan and {Beardsley}, Adam P. and {Bernardi}, Gianni and {Billings}, Tashalee S. and {Bowman}, Judd D. and {Bradley}, Richard F. and {Bull}, Philip and {Burba}, Jacob and {Carey}, Steve and {Carilli}, Chris L. and {Cheng}, Carina and {DeBoer}, David R. and {Dexter}, Matt and {de Lera Acedo}, Eloy and {Dillon}, Joshua S. and {Ely}, John and {Ewall-Wice}, Aaron and {Fagnoni}, Nicolas and {Fialkov}, Anastasia and {Fritz}, Randall and {Furlanetto}, Steven R. and {Gale-Sides}, Kingsley and {Glendenning}, Brian and {Gorthi}, Deepthi and {Greig}, Bradley and {Grobbelaar}, Jasper and {Halday}, Ziyaad and {Hazelton}, Bryna J. and {Heimersheim}, Stefan and {Hewitt}, Jacqueline N. and {Hickish}, Jack and {Jacobs}, Daniel C. and {Julius}, Austin and {Kern}, Nicholas S. and {Kerrigan}, Joshua and {Kittiwisit}, Piyanat and {Kohn}, Saul A. and {Kolopanis}, Matthew and {Lanman}, Adam and {La Plante}, Paul and {Lekalake}, Telalo and {Lewis}, David and {Liu}, Adrian and {Ma}, Yin-Zhe and {MacMahon}, David and {Malan}, Lourence and {Malgas}, Cresshim and {Maree}, Matthys and {Martinot}, Zachary E. and {Matsetela}, Eunice and {Mesinger}, Andrei and {Mirocha}, Jordan and {Molewa}, Mathakane and {Morales}, Miguel F. and {Mosiane}, Tshegofalang and {Mu{\~n}oz}, Julian B. and {Murray}, Steven G. and {Neben}, Abraham R. and {Nikolic}, Bojan and {Nunhokee}, Chuneeta D. and {Parsons}, Aaron R. and {Patra}, Nipanjana and {Pieterse}, Samantha and {Pober}, Jonathan C. and {Qin}, Yuxiang and {Razavi-Ghods}, Nima and {Reis}, Itamar and {Ringuette}, Jon and {Robnett}, James and {Rosie}, Kathryn and {Santos}, Mario G. and {Sikder}, Sudipta and {Sims}, Peter and {Smith}, Craig and {Syce}, Angelo and {Thyagarajan}, Nithyanandan and {Williams}, Peter K.~G. and {Zheng}, Haoxuan},
        title = "{HERA Phase I Limits on the Cosmic 21 cm Signal: Constraints on Astrophysics and Cosmology during the Epoch of Reionization}",
      journal = {\apj},
         year = 2022,
        month = jan,
       volume = {924},
       number = {2},
          eid = {51},
        pages = {51},
          doi = {10.3847/1538-4357/ac2ffc},
archivePrefix = {arXiv},
       eprint = {2108.07282},
 primaryClass = {astro-ph.CO},
       adsurl = {https://ui.adsabs.harvard.edu/abs/2022ApJ...924...51A}
}

@ARTICLE{Bosman2021,
       author = {{Bosman}, Sarah E.~I. and {Davies}, Frederick B. and {Becker}, George D. and {Keating}, Laura C. and {Davies}, Rebecca L. and {Zhu}, Yongda and {Eilers}, Anna-Christina and {D'Odorico}, Valentina and {Bian}, Fuyan and {Bischetti}, Manuela and {Cristiani}, Stefano V. and {Fan}, Xiaohui and {Farina}, Emanuele P. and {Haehnelt}, Martin G. and {Hennawi}, Joseph F. and {Kulkarni}, Girish and {Mesinger}, Andrei and {Meyer}, Romain A. and {Onoue}, Masafusa and {Pallottini}, Andrea and {Qin}, Yuxiang and {Ryan-Weber}, Emma and {Schindler}, Jan-Torge and {Walter}, Fabian and {Wang}, Feige and {Yang}, Jinyi},
        title = "{Hydrogen reionization ends by z = 5.3: Lyman-{\ensuremath{\alpha}} optical depth measured by the XQR-30 sample}",
      journal = {\mnras},
         year = 2022,
        month = jul,
       volume = {514},
       number = {1},
        pages = {55-76},
          doi = {10.1093/mnras/stac1046},
archivePrefix = {arXiv},
       eprint = {2108.03699},
 primaryClass = {astro-ph.CO},
       adsurl = {https://ui.adsabs.harvard.edu/abs/2022MNRAS.514...55B}
}

@ARTICLE{Mesinger2007,
       author = {{Mesinger}, Andrei and {Furlanetto}, Steven},
        title = "{Efficient Simulations of Early Structure Formation and Reionization}",
      journal = {\apj},
         year = 2007,
        month = nov,
       volume = {669},
       number = {2},
        pages = {663-675},
          doi = {10.1086/521806},
archivePrefix = {arXiv},
       eprint = {0704.0946},
 primaryClass = {astro-ph},
       adsurl = {https://ui.adsabs.harvard.edu/abs/2007ApJ...669..663M}
}

@ARTICLE{Furlanetto2009,
       author = {{Furlanetto}, Steven R. and {Mesinger}, Andrei},
        title = "{The ionizing background at the end of reionization}",
      journal = {\mnras},
         year = 2009,
        month = apr,
       volume = {394},
       number = {3},
        pages = {1667-1673},
          doi = {10.1111/j.1365-2966.2009.14449.x},
archivePrefix = {arXiv},
       eprint = {0809.4493},
 primaryClass = {astro-ph},
       adsurl = {https://ui.adsabs.harvard.edu/abs/2009MNRAS.394.1667F}
}

@ARTICLE{Zuo1992,
       author = {{Zuo}, Lin},
        title = "{Intensity correlation of ionizing background at high redshifts.}",
      journal = {\mnras},
         year = 1992,
        month = sep,
       volume = {258},
        pages = {45-56},
          doi = {10.1093/mnras/258.1.45},
       adsurl = {https://ui.adsabs.harvard.edu/abs/1992MNRAS.258...45Z}
}

@ARTICLE{Garaldi2022,
       author = {{Garaldi}, E. and {Kannan}, R. and {Smith}, A. and {Springel}, V. and {Pakmor}, R. and {Vogelsberger}, M. and {Hernquist}, L.},
        title = "{The THESAN project: properties of the intergalactic medium and its connection to reionization-era galaxies}",
      journal = {\mnras},
         year = 2022,
        month = feb,
          doi = {10.1093/mnras/stac257},
archivePrefix = {arXiv},
       eprint = {2110.01628},
 primaryClass = {astro-ph.CO},
       adsurl = {https://ui.adsabs.harvard.edu/abs/2022MNRAS.tmp..401G}
}

@ARTICLE{Lewis2022,
       author = {{Lewis}, Joseph S.~W. and {Ocvirk}, Pierre and {Sorce}, Jenny G. and {Dubois}, Yohan and {Aubert}, Dominique and {Conaboy}, Luke and {Shapiro}, Paul R. and {Dawoodbhoy}, Taha and {Teyssier}, Romain and {Yepes}, Gustavo and {Gottl{\"o}ber}, Stefan and {Rasera}, Yann and {Ahn}, Kyungjin and {Iliev}, Ilian T. and {Park}, Hyunbae and {Th{\'e}lie}, {\'E}milie},
        title = "{The short ionizing photon mean free path at z = 6 in Cosmic Dawn III, a new fully coupled radiation-hydrodynamical simulation of the Epoch of Reionization}",
      journal = {\mnras},
         year = 2022,
        month = nov,
       volume = {516},
       number = {3},
        pages = {3389-3397},
          doi = {10.1093/mnras/stac2383},
archivePrefix = {arXiv},
       eprint = {2202.05869},
 primaryClass = {astro-ph.CO},
       adsurl = {https://ui.adsabs.harvard.edu/abs/2022MNRAS.516.3389L}
}

@ARTICLE{Trac2021,
       author = {{Trac}, Hy and {Chen}, Nianyi and {Holst}, Ian and {Alvarez}, Marcelo A. and {Cen}, Renyue},
        title = "{AMBER: A Semi-numerical Abundance Matching Box for the Epoch of Reionization}",
      journal = {\apj},
         year = 2022,
        month = mar,
       volume = {927},
       number = {2},
          eid = {186},
        pages = {186},
          doi = {10.3847/1538-4357/ac5116},
archivePrefix = {arXiv},
       eprint = {2109.10375},
 primaryClass = {astro-ph.CO},
       adsurl = {https://ui.adsabs.harvard.edu/abs/2022ApJ...927..186T}
}

@ARTICLE{Kaurov2015,
       author = {{Kaurov}, Alexander A. and {Gnedin}, Nickolay Y.},
        title = "{Cosmic Reionization on Computers. III. The Clumping Factor}",
      journal = {\apj},
         year = 2015,
        month = sep,
       volume = {810},
       number = {2},
          eid = {154},
        pages = {154},
          doi = {10.1088/0004-637X/810/2/154},
archivePrefix = {arXiv},
       eprint = {1412.5607},
 primaryClass = {astro-ph.CO},
       adsurl = {https://ui.adsabs.harvard.edu/abs/2015ApJ...810..154K}
}

@ARTICLE{Xu2017,
       author = {{Xu}, Yidong and {Yue}, Bin and {Chen}, Xuelei},
        title = "{IslandFAST: A Semi-numerical Tool for Simulating the Late Epoch of Reionization}",
      journal = {\apj},
         year = 2017,
        month = aug,
       volume = {844},
       number = {2},
          eid = {117},
        pages = {117},
          doi = {10.3847/1538-4357/aa7b71},
archivePrefix = {arXiv},
       eprint = {1612.05703},
 primaryClass = {astro-ph.CO},
       adsurl = {https://ui.adsabs.harvard.edu/abs/2017ApJ...844..117X}
}

@ARTICLE{Kannan2022,
       author = {{Kannan}, R. and {Garaldi}, E. and {Smith}, A. and {Pakmor}, R. and {Springel}, V. and {Vogelsberger}, M. and {Hernquist}, L.},
        title = "{Introducing the THESAN project: radiation-magnetohydrodynamic simulations of the epoch of reionization}",
      journal = {\mnras},
         year = 2022,
        month = apr,
       volume = {511},
       number = {3},
        pages = {4005-4030},
          doi = {10.1093/mnras/stab3710},
archivePrefix = {arXiv},
       eprint = {2110.00584},
 primaryClass = {astro-ph.GA},
       adsurl = {https://ui.adsabs.harvard.edu/abs/2022MNRAS.511.4005K}
}

@ARTICLE{Gnedin2000,
       author = {{Gnedin}, Nickolay Y.},
        title = "{Effect of Reionization on Structure Formation in the Universe}",
      journal = {\apj},
         year = 2000,
        month = oct,
       volume = {542},
       number = {2},
        pages = {535-541},
          doi = {10.1086/317042},
archivePrefix = {arXiv},
       eprint = {astro-ph/0002151},
 primaryClass = {astro-ph},
       adsurl = {https://ui.adsabs.harvard.edu/abs/2000ApJ...542..535G}
}

@article{Calverley2011,
    author = {Calverley, Alexander P. and Becker, George D. and Haehnelt, Martin G. and Bolton, James S.},
    title = "{Measurements of the ultraviolet background at 4.6 < z < 6.4 using the quasar proximity effect*}",
    journal = {Monthly Notices of the Royal Astronomical Society},
    volume = {412},
    number = {4},
    pages = {2543-2562},
    year = {2011},
    month = {04},
    issn = {0035-8711},
    doi = {10.1111/j.1365-2966.2010.18072.x},
    url = {https://doi.org/10.1111/j.1365-2966.2010.18072.x},
    eprint = {https://academic.oup.com/mnras/article-pdf/412/4/2543/3356601/mnras0412-2543.pdf},
}

@article{Wyithe2011,
    author = {Wyithe, J. Stuart B. and Bolton, James S.},
    title = "{Near-zone sizes and the rest-frame extreme ultraviolet spectral index of the highest redshift quasars}",
    journal = {Monthly Notices of the Royal Astronomical Society},
    volume = {412},
    number = {3},
    pages = {1926-1936},
    year = {2011},
    month = {04},
    issn = {0035-8711},
    doi = {10.1111/j.1365-2966.2010.18030.x},
    url = {https://doi.org/10.1111/j.1365-2966.2010.18030.x},
    eprint = {https://academic.oup.com/mnras/article-pdf/412/3/1926/3610037/mnras0412-1926.pdf},
}

@ARTICLE{Zhu2022,
       author = {{Zhu}, Yongda and {Becker}, George D. and {Bosman}, Sarah E.~I. and {Keating}, Laura C. and {D'Odorico}, Valentina and {Davies}, Rebecca L. and {Christenson}, Holly M. and {Ba{\~n}ados}, Eduardo and {Bian}, Fuyan and {Bischetti}, Manuela and {Chen}, Huanqing and {Davies}, Frederick B. and {Eilers}, Anna-Christina and {Fan}, Xiaohui and {Gaikwad}, Prakash and {Greig}, Bradley and {Haehnelt}, Martin G. and {Kulkarni}, Girish and {Lai}, Samuel and {Pallottini}, Andrea and {Qin}, Yuxiang and {Ryan-Weber}, Emma V. and {Walter}, Fabian and {Wang}, Feige and {Yang}, Jinyi},
        title = "{Long Dark Gaps in the Ly{\ensuremath{\beta}} Forest at z < 6: Evidence of Ultra-late Reionization from XQR-30 Spectra}",
      journal = {\apj},
         year = 2022,
        month = jun,
       volume = {932},
       number = {2},
          eid = {76},
        pages = {76},
          doi = {10.3847/1538-4357/ac6e60},
archivePrefix = {arXiv},
       eprint = {2205.04569},
 primaryClass = {astro-ph.CO},
       adsurl = {https://ui.adsabs.harvard.edu/abs/2022ApJ...932...76Z}
}

@ARTICLE{Rosdahl2022,
       author = {{Rosdahl}, Joakim and {Blaizot}, J{\'e}r{\'e}my and {Katz}, Harley and {Kimm}, Taysun and {Garel}, Thibault and {Haehnelt}, Martin and {Keating}, Laura C. and {Martin-Alvarez}, Sergio and {Michel-Dansac}, L{\'e}o and {Ocvirk}, Pierre},
        title = "{LyC escape from SPHINX galaxies in the Epoch of Reionization}",
      journal = {\mnras},
         year = 2022,
        month = sep,
       volume = {515},
       number = {2},
        pages = {2386-2414},
          doi = {10.1093/mnras/stac1942},
archivePrefix = {arXiv},
       eprint = {2207.03232},
 primaryClass = {astro-ph.GA},
       adsurl = {https://ui.adsabs.harvard.edu/abs/2022MNRAS.515.2386R}
}

@ARTICLE{Haardt2012,
       author = {{Haardt}, Francesco and {Madau}, Piero},
        title = "{Radiative Transfer in a Clumpy Universe. IV. New Synthesis Models of the Cosmic UV/X-Ray Background}",
      journal = {\apj},
         year = 2012,
        month = feb,
       volume = {746},
       number = {2},
          eid = {125},
        pages = {125},
          doi = {10.1088/0004-637X/746/2/125},
archivePrefix = {arXiv},
       eprint = {1105.2039},
 primaryClass = {astro-ph.CO},
       adsurl = {https://ui.adsabs.harvard.edu/abs/2012ApJ...746..125H}
}

@ARTICLE{Munoz2016,
       author = {{Mu{\~n}oz}, Joseph A. and {Oh}, S. Peng and {Davies}, Frederick B. and {Furlanetto}, Steven R.},
        title = "{The flatness and sudden evolution of the intergalactic ionizing background}",
      journal = {\mnras},
         year = 2016,
        month = jan,
       volume = {455},
       number = {2},
        pages = {1385-1397},
          doi = {10.1093/mnras/stv2355},
archivePrefix = {arXiv},
       eprint = {1410.2249},
 primaryClass = {astro-ph.GA},
       adsurl = {https://ui.adsabs.harvard.edu/abs/2016MNRAS.455.1385M}
}

@ARTICLE{Zhu2023,
       author = {{Zhu}, Yongda and {Becker}, George D. and {Christenson}, Holly M. and {D'Aloisio}, Anson and {Bosman}, Sarah E.~I. and {Bakx}, Tom and {D'Odorico}, Valentina and {Bischetti}, Manuela and {Cain}, Christopher and {Davies}, Frederick B. and {Davies}, Rebecca L. and {Eilers}, Anna-Christina and {Fan}, Xiaohui and {Gaikwad}, Prakash and {Haehnelt}, Martin G. and {Keating}, Laura C. and {Kulkarni}, Girish and {Lai}, Samuel and {Ma}, Hai-Xia and {Mesinger}, Andrei and {Qin}, Yuxiang and {Satyavolu}, Sindhu and {Takeuchi}, Tsutomu T. and {Umehata}, Hideki and {Yang}, Jinyi},
        title = "{Probing Ultralate Reionization: Direct Measurements of the Mean Free Path over 5 < z < 6}",
      journal = {\apj},
         year = 2023,
        month = oct,
       volume = {955},
       number = {2},
          eid = {115},
        pages = {115},
          doi = {10.3847/1538-4357/aceef4},
archivePrefix = {arXiv},
       eprint = {2308.04614},
 primaryClass = {astro-ph.CO},
       adsurl = {https://ui.adsabs.harvard.edu/abs/2023ApJ...955..115Z}
}

@ARTICLE{Trebitsch2017,
       author = {{Trebitsch}, Maxime and {Blaizot}, J{\'e}r{\'e}my and {Rosdahl}, Joakim and {Devriendt}, Julien and {Slyz}, Adrianne},
        title = "{Fluctuating feedback-regulated escape fraction of ionizing radiation in low-mass, high-redshift galaxies}",
      journal = {\mnras},
         year = 2017,
        month = sep,
       volume = {470},
       number = {1},
        pages = {224-239},
          doi = {10.1093/mnras/stx1060},
archivePrefix = {arXiv},
       eprint = {1705.00941},
 primaryClass = {astro-ph.GA},
       adsurl = {https://ui.adsabs.harvard.edu/abs/2017MNRAS.470..224T}
}

@ARTICLE{Doussot2022,
       author = {{Doussot}, Aristide and {Semelin}, Beno{\^\i}t},
        title = "{A bubble size distribution model for the Epoch of Reionization}",
      journal = {\aap},
         year = 2022,
        month = nov,
       volume = {667},
          eid = {A118},
        pages = {A118},
          doi = {10.1051/0004-6361/202244108},
archivePrefix = {arXiv},
       eprint = {2208.14044},
 primaryClass = {astro-ph.CO},
       adsurl = {https://ui.adsabs.harvard.edu/abs/2022A&A...667A.118D}
}

@ARTICLE{Munoz2023,
       author = {{Mu{\~n}oz}, Julian B.},
        title = "{An effective model for the cosmic-dawn 21-cm signal}",
      journal = {\mnras},
         year = 2023,
        month = aug,
       volume = {523},
       number = {2},
        pages = {2587-2607},
          doi = {10.1093/mnras/stad1512},
archivePrefix = {arXiv},
       eprint = {2302.08506},
 primaryClass = {astro-ph.CO},
       adsurl = {https://ui.adsabs.harvard.edu/abs/2023MNRAS.523.2587M}
}

@ARTICLE{Sims2025,
       author = {{Sims}, Peter H. and {Bevins}, Harry T.~J. and {Fialkov}, Anastasia and {Anstey}, Dominic and {Handley}, Will J. and {Heimersheim}, Stefan and {de Lera Acedo}, Eloy and {Mondal}, Rajesh and {Barkana}, Rennan},
        title = "{Rapid and late cosmic reionization driven by massive galaxies: a joint analysis of constraints from 21-cm, Lyman line, and CMB data sets}",
      journal = {\mnras},
         year = 2025,
        month = dec,
       volume = {544},
       number = {4},
        pages = {3856-3882},
          doi = {10.1093/mnras/staf1864},
archivePrefix = {arXiv},
       eprint = {2504.09725},
 primaryClass = {astro-ph.CO},
       adsurl = {https://ui.adsabs.harvard.edu/abs/2025MNRAS.544.3856S}
}

@article{Chaubal2026,
    author = "Chaubal, P. and others",
    collaboration = "SPT-3G",
    title = "{SPT-3G D1: A Measurement of Secondary Cosmic Microwave Background Anisotropy Power}",
    eprint = "2601.20551",
    archivePrefix = "arXiv",
    primaryClass = "astro-ph.CO",
    reportNumber = "FERMILAB-PUB-26-0066-PPD",
    month = "1",
    year = "2026"
}

@ARTICLE{Davies2026,
       author = {{Davies}, Frederick B. and {Bosman}, Sarah E.~I. and {D'Odorico}, Valentina and {Campo}, Sofia and {Mesinger}, Andrei and {Qin}, Yuxiang and {Becker}, George D. and {Ba{\~n}ados}, Eduardo and {Chen}, Huanqing and {Cristiani}, Stefano and {Fan}, Xiaohui and {Gallerani}, Simona and {Haehnelt}, Martin G. and {Keating}, Laura C. and {Lai}, Samuel and {Ryan-Weber}, Emma and {Wang}, Feige and {Yang}, Jinyi and {Zhu}, Yongda},
        title = "{Updated dark pixel fraction constraints on reionization's end from the Lyman-series forests of XQR{\ensuremath{-}}30}",
      journal = {\mnras},
         year = 2026,
        month = jan,
       volume = {545},
       number = {2},
          eid = {staf1862},
        pages = {staf1862},
          doi = {10.1093/mnras/staf1862},
archivePrefix = {arXiv},
       eprint = {2510.25829},
 primaryClass = {astro-ph.CO},
       adsurl = {https://ui.adsabs.harvard.edu/abs/2026MNRAS.545f1862D}
}

@ARTICLE{Kageura2026,
       author = {{Kageura}, Yuta and {Ouchi}, Masami and {Naokawa}, Fumihiro and {Umeda}, Hiroya and {Matsumoto}, Akinori and {Harikane}, Yuichi and {Nakane}, Minami and {Thai}, Tran Thi},
        title = "{A New Constraint on the Optical Depth from the Reionization History Independent of CMB Large-Scale E-Mode Polarization}",
      journal = {arXiv e-prints},
         year = 2026,
        month = jan,
          eid = {arXiv:2601.09644},
        pages = {arXiv:2601.09644},
          doi = {10.48550/arXiv.2601.09644},
archivePrefix = {arXiv},
       eprint = {2601.09644},
 primaryClass = {astro-ph.CO},
       adsurl = {https://ui.adsabs.harvard.edu/abs/2026arXiv260109644K}
}

@ARTICLE{OMeara2013,
       author = {{O'Meara}, John M. and {Prochaska}, J. Xavier and {Worseck}, Gabor and {Chen}, Hsiao-Wen and {Madau}, Piero},
        title = "{The HST/ACS+WFC3 Survey for Lyman Limit Systems. II. Science}",
      journal = {\apj},
         year = 2013,
        month = mar,
       volume = {765},
       number = {2},
          eid = {137},
        pages = {137},
          doi = {10.1088/0004-637X/765/2/137},
archivePrefix = {arXiv},
       eprint = {1204.3093},
 primaryClass = {astro-ph.CO},
       adsurl = {https://ui.adsabs.harvard.edu/abs/2013ApJ...765..137O}
}

@ARTICLE{Fumagalli2013,
       author = {{Fumagalli}, Michele and {O'Meara}, John M. and {Prochaska}, J. Xavier and {Worseck}, Gabor},
        title = "{Dissecting the Properties of Optically Thick Hydrogen at the Peak of Cosmic Star Formation History}",
      journal = {\apj},
         year = 2013,
        month = sep,
       volume = {775},
       number = {1},
          eid = {78},
        pages = {78},
          doi = {10.1088/0004-637X/775/1/78},
archivePrefix = {arXiv},
       eprint = {1308.1101},
 primaryClass = {astro-ph.CO},
       adsurl = {https://ui.adsabs.harvard.edu/abs/2013ApJ...775...78F}
}

@ARTICLE{Gao2024,
       author = {{Gao}, Anning and {Prochaska}, J. Xavier and {Cai}, Zheng and {Zou}, Siwei and {Zhao}, Cheng and {Sun}, Zechang and {Ahlen}, S. and {Bianchi}, D. and {Brooks}, D. and {Claybaugh}, T. and {de la Macorra}, A. and {Dey}, Arjun and {Doel}, P. and {Forero-Romero}, J.~E. and {Gazta{\~n}aga}, E. and {Gontcho A Gontcho}, S. and {Gutierrez}, G. and {Honscheid}, K. and {Juneau}, S. and {Kremin}, A. and {Martini}, P. and {Meisner}, A. and {Miquel}, R. and {Moustakas}, J. and {Mu{\~n}oz-Guti{\'e}rrez}, A. and {Newman}, J.~A. and {P{\'e}rez-R{\`a}fols}, I. and {Rossi}, G. and {Sanchez}, E. and {Schubnell}, M. and {Sprayberry}, D. and {Tarl{\'e}}, G. and {Weaver}, B.~A. and {Zou}, H.},
        title = "{Measuring the Mean Free Path of H I Ionizing Photons at 3.2 {\ensuremath{\leq}} z {\ensuremath{\leq}} 4.6 with DESI Y1 Quasars}",
      journal = {\apjl},
         year = 2025,
        month = mar,
       volume = {981},
       number = {2},
          eid = {L27},
        pages = {L27},
          doi = {10.3847/2041-8213/adb48f},
archivePrefix = {arXiv},
       eprint = {2411.15838},
 primaryClass = {astro-ph.CO},
       adsurl = {https://ui.adsabs.harvard.edu/abs/2025ApJ...981L..27G}
}

@ARTICLE{Kim2021,
       author = {{Kim}, T.-S. and {Wakker}, B.~P. and {Nasir}, F. and {Carswell}, R.~F. and {Savage}, B.~D. and {Bolton}, J.~S. and {Fox}, A.~J. and {Viel}, M. and {Haehnelt}, M.~G. and {Charlton}, J.~C. and {Rosenwasser}, B.~E.},
        title = "{The evolution of the low-density H I> intergalactic medium from z = 3.6 to 0: data, transmitted flux, and H I> column density,}",
      journal = {\mnras},
         year = 2021,
        month = mar,
       volume = {501},
       number = {4},
        pages = {5811-5833},
          doi = {10.1093/mnras/staa3844},
archivePrefix = {arXiv},
       eprint = {2012.05861},
 primaryClass = {astro-ph.CO},
       adsurl = {https://ui.adsabs.harvard.edu/abs/2021MNRAS.501.5811K}
}

@ARTICLE{Rudie2013,
       author = {{Rudie}, Gwen C. and {Steidel}, Charles C. and {Shapley}, Alice E. and {Pettini}, Max},
        title = "{The Column Density Distribution and Continuum Opacity of the Intergalactic and Circumgalactic Medium at Redshift langzrang = 2.4}",
      journal = {\apj},
         year = 2013,
        month = jun,
       volume = {769},
       number = {2},
          eid = {146},
        pages = {146},
          doi = {10.1088/0004-637X/769/2/146},
archivePrefix = {arXiv},
       eprint = {1304.6719},
 primaryClass = {astro-ph.CO},
       adsurl = {https://ui.adsabs.harvard.edu/abs/2013ApJ...769..146R}
}

@ARTICLE{Cain2025c,
       author = {{Cain}, Christopher and {Van Engelen}, Alexander and {Croker}, Kevin S. and {Kramer}, Darby and {D'Aloisio}, Anson and {Lopez}, Garett},
        title = "{The Cosmic Microwave Background Optical Depth Constrains the Duration of Reionization}",
      journal = {The Astrophysical Journal Letters},
         year = 2025,
        month = jul,
       volume = {987},
       number = {2},
          eid = {L29},
        pages = {L29},
          doi = {10.3847/2041-8213/ade141},
archivePrefix = {arXiv},
       eprint = {2505.15899},
 primaryClass = {astro-ph.CO},
       adsurl = {https://ui.adsabs.harvard.edu/abs/2025ApJ...987L..29C}
}

@ARTICLE{Elbers2025,
       author = {{Elbers}, Willem},
        title = "{Rapid late-time reionization: constraints and cosmological implications}",
      journal = {arXiv e-prints},
         year = 2025,
        month = aug,
          eid = {arXiv:2508.21069},
        pages = {arXiv:2508.21069},
          doi = {10.48550/arXiv.2508.21069},
archivePrefix = {arXiv},
       eprint = {2508.21069},
 primaryClass = {astro-ph.CO},
       adsurl = {https://ui.adsabs.harvard.edu/abs/2025arXiv250821069E}
}

@ARTICLE{Chan2023,
       author = {{Chan}, Tsang Keung and {Ben{\'\i}tez-Llambay}, Alejandro and {Theuns}, Tom and {Frenk}, Carlos and {Bower}, Richard},
        title = "{The impact and response of mini-haloes and the interhalo medium on cosmic reionization}",
      journal = {\mnras},
         year = 2024,
        month = feb,
       volume = {528},
       number = {2},
        pages = {1296-1326},
          doi = {10.1093/mnras/stae114},
archivePrefix = {arXiv},
       eprint = {2305.04959},
 primaryClass = {astro-ph.CO},
       adsurl = {https://ui.adsabs.harvard.edu/abs/2024MNRAS.528.1296C}
}

@ARTICLE{Rosdahl2018,
       author = {{Rosdahl}, Joakim and {Katz}, Harley and {Blaizot}, J{\'e}r{\'e}my and {Kimm}, Taysun and {Michel-Dansac}, L{\'e}o and {Garel}, Thibault and {Haehnelt}, Martin and {Ocvirk}, Pierre and {Teyssier}, Romain},
        title = "{The SPHINX cosmological simulations of the first billion years: the impact of binary stars on reionization}",
      journal = {\mnras},
         year = 2018,
        month = sep,
       volume = {479},
       number = {1},
        pages = {994-1016},
          doi = {10.1093/mnras/sty1655},
archivePrefix = {arXiv},
       eprint = {1801.07259},
 primaryClass = {astro-ph.GA},
       adsurl = {https://ui.adsabs.harvard.edu/abs/2018MNRAS.479..994R}
}

@ARTICLE{Jin2023,
       author = {{Jin}, Xiangyu and {Yang}, Jinyi and {Fan}, Xiaohui and {Wang}, Feige and {Ba{\~n}ados}, Eduardo and {Bian}, Fuyan and {Davies}, Frederick B. and {Eilers}, Anna-Christina and {Farina}, Emanuele Paolo and {Hennawi}, Joseph F. and {Pacucci}, Fabio and {Venemans}, Bram and {Walter}, Fabian},
        title = "{(Nearly) Model-independent Constraints on the Neutral Hydrogen Fraction in the Intergalactic Medium at z   5-7 Using Dark Pixel Fractions in Ly{\ensuremath{\alpha}} and Ly{\ensuremath{\beta}} Forests}",
      journal = {\apj},
         year = 2023,
        month = jan,
       volume = {942},
       number = {2},
          eid = {59},
        pages = {59},
          doi = {10.3847/1538-4357/aca678},
archivePrefix = {arXiv},
       eprint = {2211.12613},
 primaryClass = {astro-ph.CO},
       adsurl = {https://ui.adsabs.harvard.edu/abs/2023ApJ...942...59J}
}

@ARTICLE{Robertson2013,
       author = {{Robertson}, Brant E. and {Furlanetto}, Steven R. and {Schneider}, Evan and {Charlot}, Stephane and {Ellis}, Richard S. and {Stark}, Daniel P. and {McLure}, Ross J. and {Dunlop}, James S. and {Koekemoer}, Anton and {Schenker}, Matthew A. and {Ouchi}, Masami and {Ono}, Yoshiaki and {Curtis-Lake}, Emma and {Rogers}, Alexander B. and {Bowler}, Rebecca A.~A. and {Cirasuolo}, Michele},
        title = "{New Constraints on Cosmic Reionization from the 2012 Hubble Ultra Deep Field Campaign}",
      journal = {\apj},
         year = 2013,
        month = may,
       volume = {768},
       number = {1},
          eid = {71},
        pages = {71},
          doi = {10.1088/0004-637X/768/1/71},
archivePrefix = {arXiv},
       eprint = {1301.1228},
 primaryClass = {astro-ph.CO},
       adsurl = {https://ui.adsabs.harvard.edu/abs/2013ApJ...768...71R}
}

@ARTICLE{FaucherGiguere2009,
       author = {{Faucher-Gigu{\`e}re}, Claude-Andr{\'e} and {Lidz}, Adam and {Zaldarriaga}, Matias and {Hernquist}, Lars},
        title = "{A New Calculation of the Ionizing Background Spectrum and the Effects of He II Reionization}",
      journal = {\apj},
         year = 2009,
        month = oct,
       volume = {703},
       number = {2},
        pages = {1416-1443},
          doi = {10.1088/0004-637X/703/2/1416},
archivePrefix = {arXiv},
       eprint = {0901.4554},
 primaryClass = {astro-ph.CO},
       adsurl = {https://ui.adsabs.harvard.edu/abs/2009ApJ...703.1416F}
}

@ARTICLE{Madau2017,
       author = {{Madau}, Piero},
        title = "{Cosmic Reionization after Planck and before JWST: An Analytic Approach}",
      journal = {\apj},
         year = 2017,
        month = dec,
       volume = {851},
       number = {1},
          eid = {50},
        pages = {50},
          doi = {10.3847/1538-4357/aa9715},
archivePrefix = {arXiv},
       eprint = {1710.07636},
 primaryClass = {astro-ph.GA},
       adsurl = {https://ui.adsabs.harvard.edu/abs/2017ApJ...851...50M}
}

@ARTICLE{Theuns2023,
       author = {{Theuns}, Tom and {Chan}, T.~K.},
        title = "{A halo model for cosmological Lyman-limit systems}",
      journal = {\mnras},
         year = 2024,
        month = jan,
       volume = {527},
       number = {1},
        pages = {689-705},
          doi = {10.1093/mnras/stad3176},
archivePrefix = {arXiv},
       eprint = {2310.09228},
 primaryClass = {astro-ph.CO},
       adsurl = {https://ui.adsabs.harvard.edu/abs/2024MNRAS.527..689T}
}

@article{Choudhury2018,
    author = {Choudhury, Tirthankar Roy and Paranjape, Aseem},
    title = {Photon number conservation and the large-scale 21 cm power spectrum in seminumerical models of reionization},
    journal = {Monthly Notices of the Royal Astronomical Society},
    volume = {481},
    number = {3},
    pages = {3821-3837},
    year = {2018},
    month = {09},
    issn = {0035-8711},
    doi = {10.1093/mnras/sty2551},
    url = {https://doi.org/10.1093/mnras/sty2551},
    eprint = {https://academic.oup.com/mnras/article-pdf/481/3/3821/25844366/sty2551.pdf},
}

@ARTICLE{Roth2023,
       author = {{Roth}, Joshua T. and {D'Aloisio}, Anson and {Cain}, Christopher and {Wilson}, Bayu and {Zhu}, Yongda and {Becker}, George D.},
        title = "{The effect of reionization on direct measurements of the mean free path}",
      journal = {\mnras},
         year = 2024,
        month = jun,
       volume = {530},
       number = {4},
        pages = {5209-5219},
          doi = {10.1093/mnras/stae1194},
archivePrefix = {arXiv},
       eprint = {2311.06348},
 primaryClass = {astro-ph.CO},
       adsurl = {https://ui.adsabs.harvard.edu/abs/2024MNRAS.530.5209R}
}

@ARTICLE{Neyer2024,
       author = {{Neyer}, Meredith and {Smith}, Aaron and {Kannan}, Rahul and {Vogelsberger}, Mark and {Garaldi}, Enrico and {Gal{\'a}rraga-Espinosa}, Daniela and {Borrow}, Josh and {Hernquist}, Lars and {Pakmor}, R{\"u}diger and {Springel}, Volker},
        title = "{The THESAN project: connecting ionized bubble sizes to their local environments during the Epoch of Reionization}",
      journal = {\mnras},
         year = 2024,
        month = jul,
       volume = {531},
       number = {3},
        pages = {2943-2957},
          doi = {10.1093/mnras/stae1325},
archivePrefix = {arXiv},
       eprint = {2310.03783},
 primaryClass = {astro-ph.GA},
       adsurl = {https://ui.adsabs.harvard.edu/abs/2024MNRAS.531.2943N}
}

@ARTICLE{Nasir2016,
       author = {{Nasir}, Fahad and {Bolton}, James S. and {Becker}, George D.},
        title = "{Inferring the IGM thermal history during reionization with the Lyman {\ensuremath{\alpha}} forest power spectrum at redshift z$\approx$ 5}",
      journal = {\mnras},
         year = 2016,
        month = dec,
       volume = {463},
       number = {3},
        pages = {2335-2347},
          doi = {10.1093/mnras/stw2147},
archivePrefix = {arXiv},
       eprint = {1605.04155},
 primaryClass = {astro-ph.CO},
       adsurl = {https://ui.adsabs.harvard.edu/abs/2016MNRAS.463.2335N}
}

@ARTICLE{Neyer2025,
       author = {{Neyer}, Meredith and {Smith}, Aaron and {Vogelsberger}, Mark and {{\'A}ngela Garc{\'\i}a}, Luz and {Kannan}, Rahul and {Garaldi}, Enrico and {Keating}, Laura},
        title = "{The THESAN project: Lyman-alpha emitters as probes of ionized bubble sizes}",
      journal = {arXiv e-prints},
         year = 2025,
        month = oct,
          eid = {arXiv:2510.18946},
        pages = {arXiv:2510.18946},
          doi = {10.48550/arXiv.2510.18946},
archivePrefix = {arXiv},
       eprint = {2510.18946},
 primaryClass = {astro-ph.GA},
       adsurl = {https://ui.adsabs.harvard.edu/abs/2025arXiv251018946N}
}

@ARTICLE{Satyavolu2023,
       author = {{Satyavolu}, Sindhu and {Kulkarni}, Girish and {Keating}, Laura C. and {Haehnelt}, Martin G.},
        title = "{Robustness of direct measurements of the mean free path of ionizing photons in the epoch of reionization}",
      journal = {\mnras},
         year = 2024,
        month = sep,
       volume = {533},
       number = {1},
        pages = {676-686},
          doi = {10.1093/mnras/stae1717},
archivePrefix = {arXiv},
       eprint = {2311.06344},
 primaryClass = {astro-ph.CO},
       adsurl = {https://ui.adsabs.harvard.edu/abs/2024MNRAS.533..676S}
}

@ARTICLE{Davies2024,
       author = {{Davies}, Frederick B. and {Bosman}, Sarah E.~I. and {Gaikwad}, Prakash and {Nasir}, Fahad and {Hennawi}, Joseph F. and {Becker}, George D. and {Haehnelt}, Martin G. and {D'Odorico}, Valentina and {Bischetti}, Manuela and {Eilers}, Anna-Christina and {Keating}, Laura C. and {Kulkarni}, Girish and {Lai}, Samuel and {Mazzucchelli}, Chiara and {Qin}, Yuxiang and {Satyavolu}, Sindhu and {Wang}, Feige and {Yang}, Jinyi and {Zhu}, Yongda},
        title = "{Constraints on the Evolution of the Ionizing Background and Ionizing Photon Mean Free Path at the End of Reionization}",
      journal = {\apj},
         year = 2024,
        month = apr,
       volume = {965},
       number = {2},
          eid = {134},
        pages = {134},
          doi = {10.3847/1538-4357/ad1d5d},
archivePrefix = {arXiv},
       eprint = {2312.08464},
 primaryClass = {astro-ph.CO},
       adsurl = {https://ui.adsabs.harvard.edu/abs/2024ApJ...965..134D}
}

@ARTICLE{Adams2024,
       author = {{Adams}, Nathan J. and {Conselice}, Christopher J. and {Austin}, Duncan and {Harvey}, Thomas and {Ferreira}, Leonardo and {Trussler}, James and {Juod{\v{z}}balis}, Ignas and {Li}, Qiong and {Windhorst}, Rogier and {Cohen}, Seth H. and {Jansen}, Rolf A. and {Summers}, Jake and {Tompkins}, Scott and {Driver}, Simon P. and {Robotham}, Aaron and {D'Silva}, Jordan C.~J. and {Yan}, Haojing and {Coe}, Dan and {Frye}, Brenda and {Grogin}, Norman A. and {Koekemoer}, Anton M. and {Marshall}, Madeline A. and {Pirzkal}, Nor and {Ryan}, Russell E. and {Maksym}, W. Peter and {Rutkowski}, Michael J. and {Willmer}, Christopher N.~A. and {Hammel}, Heidi B. and {Nonino}, Mario and {Bhatawdekar}, Rachana and {Wilkins}, Stephen M. and {Bradley}, Larry D. and {Broadhurst}, Tom and {Cheng}, Cheng and {Dole}, Herv{\'e} and {Hathi}, Nimish P. and {Zitrin}, Adi},
        title = "{EPOCHS. II. The Ultraviolet Luminosity Function from 7.5 < z < 13.5 Using 180 arcmin$^{2}$ of Deep, Blank Fields from the PEARLS Survey and Public JWST Data}",
      journal = {\apj},
         year = 2024,
        month = apr,
       volume = {965},
       number = {2},
          eid = {169},
        pages = {169},
          doi = {10.3847/1538-4357/ad2a7b},
archivePrefix = {arXiv},
       eprint = {2304.13721},
 primaryClass = {astro-ph.GA},
       adsurl = {https://ui.adsabs.harvard.edu/abs/2024ApJ...965..169A}
}

@ARTICLE{Donnan2024,
       author = {{Donnan}, C.~T. and {McLure}, R.~J. and {Dunlop}, J.~S. and {McLeod}, D.~J. and {Magee}, D. and {Arellano-C{\'o}rdova}, K.~Z. and {Barrufet}, L. and {Begley}, R. and {Bowler}, R.~A.~A. and {Carnall}, A.~C. and {Cullen}, F. and {Ellis}, R.~S. and {Fontana}, A. and {Illingworth}, G.~D. and {Grogin}, N.~A. and {Hamadouche}, M.~L. and {Koekemoer}, A.~M. and {Liu}, F. -Y. and {Mason}, C. and {Santini}, P. and {Stanton}, T.~M.},
        title = "{JWST PRIMER: A new multi-field determination of the evolving galaxy UV luminosity function at redshifts z$\approx$ 9 - 15}",
      journal = {\mnras},
         year = 2024,
        month = aug,
          doi = {10.1093/mnras/stae2037},
archivePrefix = {arXiv},
       eprint = {2403.03171},
 primaryClass = {astro-ph.GA},
       adsurl = {https://ui.adsabs.harvard.edu/abs/2024MNRAS.tmp.1993D}
}

@ARTICLE{Cain2023,
       author = {{Cain}, Christopher and {D'Aloisio}, Anson and {Lopez}, Garett and {Gangolli}, Nakul and {Roth}, Joshua T.},
        title = "{On the rise and fall of galactic ionizing output at the end of reionization}",
      journal = {\mnras},
         year = 2024,
        month = jun,
       volume = {531},
       number = {1},
        pages = {1951-1970},
          doi = {10.1093/mnras/stae1223},
archivePrefix = {arXiv},
       eprint = {2311.13638},
 primaryClass = {astro-ph.CO},
       adsurl = {https://ui.adsabs.harvard.edu/abs/2024MNRAS.531.1951C}
}

@ARTICLE{Asthana2024,
       author = {{Asthana}, Shikhar and {Haehnelt}, Martin G. and {Kulkarni}, Girish and {Aubert}, Dominique and {Bolton}, James S. and {Keating}, Laura C.},
        title = "{Late-end reionization with ATON-HE: towards constraints from Lyman-{\ensuremath{\alpha}} emitters observed with JWST}",
      journal = {\mnras},
         year = 2024,
        month = aug,
          doi = {10.1093/mnras/stae1945},
archivePrefix = {arXiv},
       eprint = {2404.06548},
 primaryClass = {astro-ph.CO},
       adsurl = {https://ui.adsabs.harvard.edu/abs/2024MNRAS.tmp.1922A}
}

@ARTICLE{Munoz2024,
       author = {{Mu{\~n}oz}, Julian B. and {Mirocha}, Jordan and {Chisholm}, John and {Furlanetto}, Steven R. and {Mason}, Charlotte},
        title = "{Reionization after JWST: a photon budget crisis?}",
      journal = {\mnras},
         year = 2024,
        month = nov,
       volume = {535},
       number = {1},
        pages = {L37-L43},
          doi = {10.1093/mnrasl/slae086},
archivePrefix = {arXiv},
       eprint = {2404.07250},
 primaryClass = {astro-ph.CO},
       adsurl = {https://ui.adsabs.harvard.edu/abs/2024MNRAS.535L..37M}
}

@ARTICLE{Morales2021,
       author = {{Morales}, Alexa M. and {Mason}, Charlotte A. and {Bruton}, Sean and {Gronke}, Max and {Haardt}, Francesco and {Scarlata}, Claudia},
        title = "{The Evolution of the Lyman-alpha Luminosity Function during Reionization}",
      journal = {\apj},
         year = 2021,
        month = oct,
       volume = {919},
       number = {2},
          eid = {120},
        pages = {120},
          doi = {10.3847/1538-4357/ac1104},
archivePrefix = {arXiv},
       eprint = {2101.01205},
 primaryClass = {astro-ph.GA},
       adsurl = {https://ui.adsabs.harvard.edu/abs/2021ApJ...919..120M}
}

@ARTICLE{Wold2022,
       author = {{Wold}, Isak G.~B. and {Malhotra}, Sangeeta and {Rhoads}, James and {Wang}, Junxian and {Hu}, Weida and {Perez}, Lucia A. and {Zheng}, Zhen-Ya and {Khostovan}, Ali Ahmad and {Walker}, Alistair R. and {Barrientos}, L. Felipe and {Gonz{\'a}lez-L{\'o}pez}, Jorge and {Harish}, Santosh and {Infante}, Leopoldo and {Jiang}, Chunyan and {Pharo}, John and {Moya-Sierralta}, Crist{\'o}bal and {Bauer}, Franz E. and {Galaz}, Gaspar and {Valdes}, Francisco and {Yang}, Huan},
        title = "{LAGER Ly{\ensuremath{\alpha}} Luminosity Function at z   7: Implications for Reionization}",
      journal = {\apj},
         year = 2022,
        month = mar,
       volume = {927},
       number = {1},
          eid = {36},
        pages = {36},
          doi = {10.3847/1538-4357/ac4997},
archivePrefix = {arXiv},
       eprint = {2105.12191},
 primaryClass = {astro-ph.GA},
       adsurl = {https://ui.adsabs.harvard.edu/abs/2022ApJ...927...36W}
}

@ARTICLE{Bolan2022,
       author = {{Bolan}, Patricia and {Lemaux}, Brian C. and {Mason}, Charlotte and {Brada{\v{c}}}, Maru{\v{s}}a and {Treu}, Tommaso and {Strait}, Victoria and {Pelliccia}, Debora and {Pentericci}, Laura and {Malkan}, Matthew},
        title = "{Inferring the intergalactic medium neutral fraction at z   6-8 with low-luminosity Lyman break galaxies}",
      journal = {\mnras},
         year = 2022,
        month = dec,
       volume = {517},
       number = {3},
        pages = {3263-3274},
          doi = {10.1093/mnras/stac1963},
archivePrefix = {arXiv},
       eprint = {2111.14912},
 primaryClass = {astro-ph.GA},
       adsurl = {https://ui.adsabs.harvard.edu/abs/2022MNRAS.517.3263B}
}

@ARTICLE{Umeda2023,
       author = {{Umeda}, Hiroya and {Ouchi}, Masami and {Nakajima}, Kimihiko and {Harikane}, Yuichi and {Ono}, Yoshiaki and {Xu}, Yi and {Isobe}, Yuki and {Zhang}, Yechi},
        title = "{JWST Measurements of Neutral Hydrogen Fractions and Ionized Bubble Sizes at z = 7-12 Obtained with Ly{\ensuremath{\alpha}} Damping Wing Absorptions in 27 Bright Continuum Galaxies}",
      journal = {\apj},
         year = 2024,
        month = aug,
       volume = {971},
       number = {2},
          eid = {124},
        pages = {124},
          doi = {10.3847/1538-4357/ad554e},
archivePrefix = {arXiv},
       eprint = {2306.00487},
 primaryClass = {astro-ph.GA},
       adsurl = {https://ui.adsabs.harvard.edu/abs/2024ApJ...971..124U}
}

@ARTICLE{Nakane2023,
       author = {{Nakane}, Minami and {Ouchi}, Masami and {Nakajima}, Kimihiko and {Harikane}, Yuichi and {Ono}, Yoshiaki and {Umeda}, Hiroya and {Isobe}, Yuki and {Zhang}, Yechi and {Xu}, Yi},
        title = "{Ly{\ensuremath{\alpha}} Emission at z = 7{\textendash}13: Clear Evolution of Ly{\ensuremath{\alpha}} Equivalent Width Indicating a Late Cosmic Reionization History}",
      journal = {\apj},
         year = 2024,
        month = may,
       volume = {967},
       number = {1},
          eid = {28},
        pages = {28},
          doi = {10.3847/1538-4357/ad38c2},
archivePrefix = {arXiv},
       eprint = {2312.06804},
 primaryClass = {astro-ph.GA},
       adsurl = {https://ui.adsabs.harvard.edu/abs/2024ApJ...967...28N}
}

@ARTICLE{Hsiao2023,
       author = {{Hsiao}, Tiger Yu-Yang and {Abdurro'uf} and {Coe}, Dan and {Larson}, Rebecca L. and {Jung}, Intae and {Mingozzi}, Matilde and {Dayal}, Pratika and {Kumari}, Nimisha and {Kokorev}, Vasily and {Vikaeus}, Anton and {Brammer}, Gabriel and {Furtak}, Lukas J. and {Adamo}, Angela and {Andrade-Santos}, Felipe and {Antwi-Danso}, Jacqueline and {Brada{\v{c}}}, Maru{\v{s}}a and {Bradley}, Larry D. and {Broadhurst}, Tom and {Carnall}, Adam C. and {Conselice}, Christopher J. and {Diego}, Jose M. and {Donahue}, Megan and {Eldridge}, Jan J. and {Fujimoto}, Seiji and {Henry}, Alaina and {Hernandez}, Svea and {Hutchison}, Taylor A. and {James}, Bethan L. and {Norman}, Colin and {Park}, Hyunbae and {Pirzkal}, Norbert and {Postman}, Marc and {Ricotti}, Massimo and {Rigby}, Jane R. and {Vanzella}, Eros and {Welch}, Brian and {Wilkins}, Stephen M. and {Windhorst}, Rogier A. and {Xu}, Xinfeng and {Zackrisson}, Erik and {Zitrin}, Adi},
        title = "{JWST NIRSpec Spectroscopy of the Triply Lensed z = 10.17 Galaxy MACS0647-JD}",
      journal = {\apj},
         year = 2024,
        month = sep,
       volume = {973},
       number = {1},
          eid = {8},
        pages = {8},
          doi = {10.3847/1538-4357/ad5da8},
archivePrefix = {arXiv},
       eprint = {2305.03042},
 primaryClass = {astro-ph.GA},
       adsurl = {https://ui.adsabs.harvard.edu/abs/2024ApJ...973....8H}
}

@ARTICLE{Simmonds2024,
       author = {{Simmonds}, C. and {Verhamme}, A. and {Inoue}, A.~K. and {Katz}, H. and {Garel}, T. and {De Barros}, S.},
        title = "{The impact of nebular Lyman-Continuum on ionizing photons budget and escape fractions from galaxies}",
      journal = {\mnras},
         year = 2024,
        month = may,
       volume = {530},
       number = {2},
        pages = {2133-2145},
          doi = {10.1093/mnras/stae1003},
archivePrefix = {arXiv},
       eprint = {2402.04052},
 primaryClass = {astro-ph.GA},
       adsurl = {https://ui.adsabs.harvard.edu/abs/2024MNRAS.530.2133S}
}

@ARTICLE{Chardin2015,
       author = {{Chardin}, Jonathan and {Haehnelt}, Martin G. and {Aubert}, Dominique and {Puchwein}, Ewald},
        title = "{Calibrating cosmological radiative transfer simulations with Ly {\ensuremath{\alpha}} forest data: evidence for large spatial UV background fluctuations at z {\ensuremath{\sim}} 5.6-5.8 due to rare bright sources}",
      journal = {\mnras},
         year = 2015,
        month = nov,
       volume = {453},
       number = {3},
        pages = {2943-2964},
          doi = {10.1093/mnras/stv1786},
archivePrefix = {arXiv},
       eprint = {1505.01853},
 primaryClass = {astro-ph.CO},
       adsurl = {https://ui.adsabs.harvard.edu/abs/2015MNRAS.453.2943C}
}

@ARTICLE{Mason2020,
       author = {{Mason}, Charlotte A. and {Gronke}, Max},
        title = "{Measuring the properties of reionized bubbles with resolved Ly{\ensuremath{\alpha}} spectra}",
      journal = {\mnras},
         year = 2020,
        month = nov,
       volume = {499},
       number = {1},
        pages = {1395-1405},
          doi = {10.1093/mnras/staa2910},
archivePrefix = {arXiv},
       eprint = {2004.13065},
 primaryClass = {astro-ph.GA},
       adsurl = {https://ui.adsabs.harvard.edu/abs/2020MNRAS.499.1395M}
}

@ARTICLE{Greig2024,
       author = {{Greig}, B. and {Mesinger}, A. and {Ba{\~n}ados}, E. and {Becker}, G.~D. and {Bosman}, S.~E.~I. and {Chen}, H. and {Davies}, F.~B. and {D'Odorico}, V. and {Eilers}, A. -C. and {Gallerani}, S. and {Haehnelt}, M.~G. and {Keating}, L. and {Lai}, S. and {Qin}, Y. and {Ryan-Weber}, E. and {Satyavolu}, S. and {Wang}, F. and {Yang}, J. and {Zhu}, Y.},
        title = "{IGM damping wing constraints on the tail end of reionization from the enlarged XQR-30 sample}",
      journal = {\mnras},
         year = 2024,
        month = may,
       volume = {530},
       number = {3},
        pages = {3208-3227},
          doi = {10.1093/mnras/stae1080},
archivePrefix = {arXiv},
       eprint = {2404.12585},
 primaryClass = {astro-ph.CO},
       adsurl = {https://ui.adsabs.harvard.edu/abs/2024MNRAS.530.3208G}
}

@ARTICLE{Davies2024c,
       author = {{Davies}, Frederick B. and {Bosman}, Sarah E.~I. and {Furlanetto}, Steven R.},
        title = "{The Predicament of Absorption-dominated Reionization II: Observational Estimate of the Clumping Factor at the End of Reionization}",
      journal = {arXiv e-prints},
         year = 2024,
        month = jun,
          eid = {arXiv:2406.18186},
        pages = {arXiv:2406.18186},
          doi = {10.48550/arXiv.2406.18186},
archivePrefix = {arXiv},
       eprint = {2406.18186},
 primaryClass = {astro-ph.CO},
       adsurl = {https://ui.adsabs.harvard.edu/abs/2024arXiv240618186D}
}
\bibliographystyle{JHEP}

\appendix

\section{Multi-frequency generalization of the formalism}
\label{app:multi_freq}

Here, we take up the multi-frequency derivation of Equation~\ref{eq:first_step}, noting that all subsequent steps in our formalism can be generalized trivially (as we will see shortly).  
We start by restoring frequency dependence to Equation~\ref{eq:cosmoRT_eq_nu0}: 
\begin{equation}
    \label{eq:cosmoRT_eq_nu}
    \frac{dJ_{\nu}}{dt} = - c \kappa(\nu) J_{\nu} + \frac{c}{4\pi}\epsilon_{\nu}
\end{equation}
We then multiply both sides by $4 \pi \sigma_{\rm HI}(\nu)/h_p \nu c$, yielding
\begin{equation}
    \label{eq:cosmoRT_eq_nu_2}
    \frac{4 \pi \sigma_{\rm HI}(\nu)}{h_p \nu c}\frac{dJ_{\nu}}{dt} = - 4 \pi \frac{\sigma_{\rm HI}(\nu)\kappa(\nu) J_{\nu}}{h_p\nu} + \frac{\sigma_{\rm HI}(\nu)}{h_p \nu}\epsilon_{\nu}
\end{equation}
Next, we define
\begin{equation}
    \label{eq:ndots}
    \dot{n}_{\rm ion,\nu} \equiv \frac{\epsilon_\nu}{h_p\nu} \hspace{1.5cm}  \Gamma_{\rm HI,\nu} \equiv \frac{4 \pi \sigma_{\rm HI}(\nu) J_{\nu}}{h_p\nu}
\end{equation}
The first expression is the emitted photon spectrum, and the second is defined such that $\int d\nu \Gamma_{\rm HI,\nu} = \Gamma_{\rm HI}$ (that is, it is the integrand of Equation 1 of Ref.~\cite{Becker2013}).  
We then have
\begin{equation}
    \label{eq:cosmoRT_eq_nu_2}
    \frac{1}{c}\frac{d\Gamma_{\rm HI,\nu}}{dt} = -\Gamma_{\rm HI,\nu}\kappa(\nu)  + \dot{n}_{\rm ion,\nu}\sigma_{\rm HI}(\nu)
\end{equation}
Next, define the weighted frequency averages
\begin{equation}
    \label{eq:avg_sigma}
    \overline{\sigma}_{\rm HI} \equiv \frac{\int d\nu \dot{n}_{\rm ion,\nu}\sigma_{\rm HI}(\nu)}{\int d\nu \dot{n}_{\rm ion,\nu}}
    \hspace{1.5cm}
    \overline{\kappa} \equiv \frac{\int d\nu \Gamma_{\rm HI,\nu}\kappa(\nu)}{\int d\nu \Gamma_{\rm HI,\nu}}
\end{equation}
where the integrals are defined over some range of frequencies where photons ionize hydrogen.  
The first expression is the average HI cross-section weighted by the emitted photon spectrum, and the second is the average absorption coefficient weighted by $\Gamma_{\rm HI,\nu}$.  
Then we have, after integrating both sides over frequency and re-arranging: 
\begin{equation}
    \frac{1}{c\overline{\sigma}_{\rm HI}}\frac{d\Gamma_{\rm HI}}{dt} = - \frac{\Gamma_{\rm HI}\overline{\kappa}}{\overline{\sigma}_{\rm HI}} + \dot{n}_{\rm ion}
\end{equation}
which is identical to Equation~\ref{eq:first_step}, but with $\sigma_{\rm HI}$ and $\kappa$ replaced with their appropriately-defined frequency averages.  
The remaining steps in our formalism are valid given these substitutions.  
Note that our definition of $\overline{\kappa}$ is equivalent to the definition of frequency-averaged $\kappa$ given in Equation 2.4 of Ref.~\cite{Wu2021}.

\section{Breakdown of the model in the ``Str\"omgren limit''}
\label{app:stromgren}

In this appendix, we further illuminate the reason why Equation~\ref{eq:first_assumption} gives the wrong answer for $\lambda_n$ in the presence of $\lambda_i$ fluctuations that are correlated with the ionization topology. We do so by considering a simple, classic problem - a Str\"omgren sphere in a constant-density, isothermal medium~\citep{Iliev2006}.  Consider an isotropic ionizing source that emits $\dot{N}_s$ ionizing photons per second embedded in an initially neutral medium with hydrogen density $n_{\rm H}$ and temperature $T$.  If the source turns on at $t = 0$, the radius of the spherical HII region around the source is
\begin{equation}
    \label{appC:Rt}
    R(t) = R_s (1 - \exp[-t/t_{\rm rec}])^{1/3}
\end{equation}
where $R_s \equiv (3 \dot{N}_s/4 \pi n_{\rm H}t_{\rm rec})^{1/3}$, the recombination timescale is $t_{\rm rec} \equiv 1/\alpha_{B}(T)n_{\rm H}$, and $\alpha_{\rm B}$ is the case-B recombination coefficient of ionized hydrogen.  It can be shown that the number of photons in the HII region within the sphere is
\begin{equation}
    N_{\gamma} = \frac{R}{c}\left[\dot{N}_{\rm s} - \frac{\pi}{3}R^3 \frac{n_{\rm H}}{t_{\rm rec}}\right]
\end{equation}
If the sphere is embedded in a finite volume $V$, the average photoionization rate over the volume is given by (see Equation 2.1 of Ref.~\cite{Cain2024d}), 
\begin{equation}
    \label{appC:gamma}
    \Gamma_{\rm HI} = \frac{N_{\gamma}}{V} c \sigma_{\rm HI} = \frac{R \sigma_{\rm HI}}{V}\left[\dot{N}_{\rm s} - \frac{\pi}{3}R^3 \frac{n_{\rm H}}{t_{\rm rec}}\right]
\end{equation}
The ionized fraction in the volume is simply $x_i = \frac{4 \pi}{3} \frac{R^3}{V}$.  The absorption rate inside the ionized region is exactly balanced by the recombination rate averaged over $V$, which is $n_{\rm H}/t_{\rm rec} x_i$.  Thus, we have, from Equation~\ref{eq:kappa_defs}, 
\begin{equation}
    \frac{\Gamma_{\rm HI}}{\lambda_i \sigma_{\rm HI}} = n_{\rm H}/t_{\rm rec} x_i = \frac{R}{V \lambda_i}\left[\dot{N}_{\rm s} - \frac{\pi}{3}R^3 \frac{n_{\rm H}}{t_{\rm rec}}\right]
\end{equation}
Using our expression for $x_i$ above and solving for $\lambda_i$ yields
\begin{equation}
    \label{appC:lambdai}
    \lambda_i = \frac{3 t_{\rm rec}}{4 \pi R^2 n_{\rm H}}\left[\dot{N}_{\rm s} - \frac{\pi}{3}R^3 \frac{n_{\rm H}}{t_{\rm rec}}\right]
\end{equation}
Per the analogous Equation~\ref{eq:second_eq}, we can write
\begin{equation}
    \frac{\Gamma_{\rm HI}}{\sigma_{\rm HI} \lambda_{\rm n}} = n_{\rm H} \frac{dx_i}{dt} = \frac{4 \pi}{3}\frac{n_{\rm H} R_s^3}{V t_{\rm rec}}\left[1 - \left(\frac{R}{R_s}\right)^3\right]
\end{equation}
Using Equation~\ref{appC:gamma} and solving for $\lambda_n$ gives
\begin{equation}
    \label{appC:lambdan}
    \lambda_n = \frac{3 R t_{\rm rec}}{4 \pi n_{\rm H} R_s^3} \frac{\dot{N}_{\rm s} - \frac{\pi}{3}R^3 \frac{n_{\rm H}}{t_{\rm rec}}}{1 - \left(\frac{R}{R_s}\right)^3}
\end{equation}
Finally, one can write the total mean free path (including absorption in the ionized region and at its boundary) as the difference between the emission rate and the time derivative of the photon number density - that is,
\begin{equation}
    \label{appC:lambdat}
    \frac{\Gamma_{\rm HI}}{\sigma_{\rm HI} \lambda_t} = \frac{\dot{N}_{\rm s} - \dot{N}_{\gamma}}{V}
\end{equation}
It can be verified that Eq.~\ref{appC:lambdai},~\ref{appC:lambdan}, and~\ref{appC:lambdat} satisfy the condition $\lambda_t^{-1} = \lambda_{i}^{-1} + \lambda_{n}^{-1}$ (indeed, they must by reason of photon conservation!).  

For our purposes, it is instructive to consider the limiting cases of Equation~\ref{appC:lambdan} - namely, the $R \ll R_s$ and $R \rightarrow R_s$ limits.  In the former, the bubble is small and the recombination rate is negligible.  Thus, the total absorption rate is dominated by ionizations at the bubble edge.  In this limit, we may let $R^3 \rightarrow 0$ in Eq.~\ref{appC:lambdan}, which gives
\begin{equation}
    \lim_{R \ll R_s} \lambda_n = \frac{3 R t_{\rm rec} \dot{N}_{\rm s}}{4 \pi n_{\rm H} R_s^3} = R
\end{equation}
This is exactly the result of \S\ref{subsec:opacity_neutral}, since the average bubble size distribution (measured from the center\footnote{Note that the average distance to the edge of a sphere from {\it any} point within the sphere is $\frac{3}{4}R$, which would be the $\lambda_n$ we would estimate using the bubble size distribution as defined in this paper.  The correct answer is $R$ in this case because all photons are produced at the center of the bubble.  In the real universe, bubbles from individual sources overlap quickly into large ionized regions wherein the sources are not concentrated at the center - thus our random placement of sightlines is much better justified for the real IGM.  }) of a spherical bubble is simply its radius!  However, we can see that in the opposite limit, 
\begin{equation}
    \lim_{R \rightarrow R_s} \lambda_n = \infty
\end{equation}
due to the $1-(R/R_s)^3$ term in the denominator.  This must be the case physically, {\bf because no photons reach the bubble's edge in the limit that the bubble has stopped growing}.  One can further verify that in this limit, $\lambda_t = \lambda_i$, and that the total absorption rate is equal to the recombination rate.  The reason that $\lambda_n$ deviates maximally from $R$ in this limit is that $\lambda_i$ inside the bubble is very inhomogeneous, and is coherently so for all sightlines originating from the center of the sphere (since the source is assumed isotropic).  It drops off as a function of distance from the center of the sphere, and formally approaches $0$ at the edge when $R = R_s$.  {\bf Thus, the approximation that fluctuations in $\lambda_i$ along different sightlines are un-correlated fails maximally in this example.}  The real universe is somewhere between this extreme and the opposite limit, in which fluctuations in $\lambda_i$ are completely un-correlated with the ionization topology.  Fortunately, the degree to which Equation~\ref{eq:first_assumption} is inaccurate in the real universe is small enough that our model remains useful for interpreting observations, per our findings in \S\ref{sec:usefulness}.   

\end{document}